\documentclass[12pt]{article}
\usepackage{amsfonts}
\usepackage{bm}
\begin{document}

\renewcommand{\theequation}{\thesection.\arabic{equation}}

\title{Weakly-nonlocal Symplectic Structures, Whitham method,
and weakly-nonlocal Symplectic Structures of Hydrodynamic Type.}

\author{A.Ya. Maltsev}

\date{
\centerline{L.D.Landau Institute for Theoretical Physics,}
\centerline{119334 ul. Kosygina 2, Moscow, maltsev@itp.ac.ru}}

\maketitle

\begin{abstract}
 We consider the special type of the field-theoretical Symplectic 
structures called weakly nonlocal. The structures of this type
are in particular very common for the integrable systems like KdV
or NLS. We introduce here the special class of the weakly
nonlocal Symplectic structures which we call the weakly
nonlocal Symplectic structures of Hydrodynamic Type. We investigate
then the connection of such structures with the Whitham averaging
method and propose the procedure of "averaging" of the weakly
nonlocal Symplectic structures. The averaging procedure gives the 
weakly nonlocal Symplectic Structure of Hydrodynamic Type for the 
corresponding Whitham system. The procedure gives also the "action 
variables" corresponding to the wave numbers of $m$-phase solutions 
of initial system which give the additional conservation laws
for the Whitham system.

\end{abstract}

\section{Introduction.}

 We are going to consider weaky-nonlocal Symplectic Structures having
the form:

$$\Omega_{ij}(x,y) \,\, = \,\, \sum_{k \geq 0} 
\omega^{(k)}_{ij}(\bm{\varphi}, \bm{\varphi}_{x}, \dots)
\, \delta^{(k)}(x-y) \,\, +$$
\begin{equation}
\label{gensympstr}
+ \,\, \sum_{s=1}^{g} e_{s} \,
q_{i}^{(s)} (\bm{\varphi}, \bm{\varphi}_{x}, \dots) \,
\nu (x-y) \,\,
q_{j}^{(s)} (\bm{\varphi}, \bm{\varphi}_{y}, \dots)
\end{equation}

 We put here $\bm{\varphi} = (\varphi^{1}, \dots, \varphi^{n})$, 
$i,j = 1, \dots, n$ , $e_{s} = \pm 1$, 
$\nu (x-y) = 1/2 \, sign \, (x-y)$ and
$\omega^{(k)}_{ij}$ and $q_{i}^{(s)}$ are some local functions of
$\bm{\varphi}$ and it's derivatives at the same point. We assume
that both sums contain finite number of terms and all 
$\omega^{(k)}_{ij}$ and $q_{i}^{(s)}$ depend on finite number
of derivatives of $\bm{\varphi}$.

 The form (\ref{gensympstr}) can be written also in more
general form:

$$\Omega_{ij}(x,y) \,\, = \,\, \sum_{k \geq 0}
\omega^{(k)}_{ij}(\bm{\varphi}, \bm{\varphi}_{x}, \dots)
\, \delta^{(k)}(x-y) \,\, +$$

$$+ \,\, \sum_{s,p=1}^{g} \kappa_{sp} \,
q_{i}^{(s)} (\bm{\varphi}, \bm{\varphi}_{x}, \dots) \,
\nu (x-y) \,\,
q_{j}^{(p)} (\bm{\varphi}, \bm{\varphi}_{y}, \dots)$$
where $\kappa_{sp}$ is some constant symmetric bilinear
form. The form (\ref{gensympstr}) gives then the "diagonal"
representation of the nonlocal part in the appropriate
basis ${\bf q}^{(1)}, \dots , {\bf q}^{(g)}$. 

 The form (\ref{gensympstr}) will play the role of the 
"symplectic" 2-form on the space of functions 

$$\bm{\varphi}(x) = (\varphi^{1}(x), \dots, \varphi^{n}(x))
\,\,\, , \,\,\, -\infty < x < +\infty $$
with the appropriate behavior at infinity. We will put for 
simplicity $\varphi^{i}(x) \rightarrow 0$ or, more generally,
$\varphi^{i}(x) \rightarrow const$ for $x \rightarrow \pm\infty$
in this paper. Let us call the corresponding space the loop
space ${\cal L}_{0}$. We require that the expression 
(\ref{gensympstr}) gives the skew-symmetric closed 2-form on
the space ${\cal L}_{0}$ (let us not put here the requirement
of non-degeneracy).

 The weakly nonlocal Symplectic Structures (\ref{gensympstr})
were introduced in \cite{PhysD} where also the fact that the 
"negative" Symplectic Structures for KdV and NLS have this form
was proved.

 Let us say here also some words about the weakly nonlocal
structures in the theory of integrable systems. Namely, we mention 
the weakly nonlocal Hamiltonian and Symplectic Structures which
seem to be closely connected with local PDE's integrable in 
the sence of the inverse scattering method. We will call here 
(like in \cite{PhysD})
the Hamiltonian Structure on ${\cal L}_{0}$ weakly 
nonlocal if it has the form similar to (\ref{gensympstr}), i.e.
the Poisson brackets of fields $\varphi^{i}(x)$ and 
$\varphi^{j}(y)$ can be formally written as

$$\{\varphi^{i}(x) , \varphi^{j}(y)\} = \sum_{k \geq 0}
B^{ij}_{(k)} (\bm{\varphi}, \bm{\varphi}_{x}, \dots)
\, \delta^{(k)}(x-y) +$$
\begin{equation}
\label{genhamstr}
+ \sum_{s=1}^{g} e_{s} \,
S^{i}_{(s)} (\bm{\varphi}, \bm{\varphi}_{x}, \dots) \,
\nu (x-y) \,\,
S^{j}_{(s)} (\bm{\varphi}, \bm{\varphi}_{y}, \dots)
\end{equation}
$e_{s} = \pm 1$. 

 We can introduce also the Hamiltonian Operator ${\hat J}^{ij}$:

\begin{equation}
\label{genhamop}
{\hat J}^{ij} = \sum_{k \geq 0}
B^{ij}_{(k)} (\bm{\varphi}, \bm{\varphi}_{x}, \dots)
\, {\partial^{k} \over \partial x^{k}}  \,\, + \,\,
\sum_{s=1}^{g} e_{s} \,
S^{i}_{(s)} (\bm{\varphi}, \bm{\varphi}_{x}, \dots) \,
D^{-1} \,\,
S^{j}_{(s)} (\bm{\varphi}, \bm{\varphi}_{x}, \dots)
\end{equation}
where $D^{-1}$ is the integration operator defined in the 
skew-symmetric way:

$$D^{-1} \, \xi({x}) = {1 \over 2} \int_{-\infty}^{x} \xi({y}) \, dy 
- {1 \over 2} \int_{x}^{+\infty} \xi({y}) \, dy $$

 For the functional $H[\varphi]$ the corresponding dynamical
system can be written in the form:

$$\varphi^{i}_{t} \,\,\, = \,\,\, {\hat J}^{ij} \, 
{\delta H \over \delta \varphi^{j}(x)} \,\,\, = \,\,\,
\sum_{k \geq 0} B^{ij}_{(k)} (\bm{\varphi}, \bm{\varphi}_{x}, \dots)
\, {\partial^{k} \over \partial x^{k}} \,  
{\delta H \over \delta \varphi^{j}(x)} \, + $$
\begin{equation}
\label{hamdynsyst}
+ \, \sum_{s=1}^{g} e_{s} \,
S^{i}_{(s)} (\bm{\varphi}, \bm{\varphi}_{x}, \dots) \,
D^{-1} \,\,
\left[ S^{j}_{(s)} (\bm{\varphi}, \bm{\varphi}_{x}, \dots)
{\delta H \over \delta \varphi^{j}(x)} \right]
\end{equation}

 The operator (\ref{genhamop}) should also be skew-symmetric and
satisfy to Jacobi identity:

$${\delta J^{ij}(x,y) \over \delta \varphi^{k}(z)} \,\, + \,\,
{\delta J^{jk}(y,z) \over \delta \varphi^{i}(x)} \,\, + \,\,
{\delta J^{ki}(z,x) \over \delta \varphi^{j}(y)} \,\,
\equiv \,\, 0 $$
(in sence of distributions).

 It's not difficult to see that the functional

$$H \,\, = \,\, \int_{-\infty}^{+\infty} 
h (\bm{\varphi}, \bm{\varphi}_{x}, \dots) , dx$$
generates a local dynamical system

$$\varphi^{i}_{t} \,\, = \,\, 
S^{i} (\bm{\varphi}, \bm{\varphi}_{x}, \dots)$$
according to (\ref{hamdynsyst}) if it gives a conservation law
for all the dynamical systems

\begin{equation}
\label{tssyst}
\varphi^{i}_{t_{s}} \,\, = \,\,
S^{i}_{(s)} (\bm{\varphi}, \bm{\varphi}_{x}, \dots)
\end{equation}
i.e.

$$h_{t_{s}}\,\, \equiv \,\, \partial_{x} \,
Q_{s} (\bm{\varphi}, \bm{\varphi}_{x}, \dots)$$
for some functions $Q_{s} (\bm{\varphi}, \bm{\varphi}_{x}, \dots)$.

 As far as we know the first example of the Poisson bracket in
this form (actually with zero local part) was the Sokolov bracket
(\cite{Sokolov})

$$\{\varphi (x) , \varphi (y)\} \,\, = \,\,
\varphi_{x} \, \nu (x-y) \, \varphi_{y}$$
for the Krichever-Novikov equation (\cite{KN80}):

$$\varphi_{t} \,\, = \,\, \varphi_{xxx} \, - \, {3 \over 2}
{{\varphi_{xx}}^{2} \over \varphi_{x}} \, + \,
{h(\varphi) \over \varphi_{x}} \,\, = \,\,
\varphi_{x} \, D^{-1} \, \varphi_{x}
{\delta H \over \delta \varphi(x)} $$
where $h(\varphi) = c_{3} \varphi^{3} + c_{2} \varphi^{2} +
c_{1} \varphi + c_{0}$ and

$$H \,\, = \,\, \int_{-\infty}^{+\infty} \left(
{1 \over 2} {{\varphi_{xx}}^{2} \over {\varphi_{x}}^{2}} \, + 
\, {1 \over 3} {h(\varphi) \over {\varphi_{x}}^{2}} \right) \,
dx $$
 
 This equation appeared originally in  work \cite{KN80} 
describing the "rank 2" solutions of the KP system. 
In  pure algebra it describes  the deformations of the commuting
genus 1 pairs  OD operators of the rank 2 whose classification
was obtained in this work. As it was found later, 
the Krichever-Novikov equation is a unique
third order in $x$  completely integrable evolution equation 
which cannot be reduced to KdV by Miura type transformations.

 The Symplectic Structure corresponding to Sokolov bracket
is purely local:

$$\Omega (x,y) \,\, = \,\, {1 \over \varphi_{x}} \,
\delta^{\prime} (x-y) \, {1 \over \varphi_{y}} $$

 Let us mention that the local symplectic structures was 
considered by I.Dorfman and O.I.Mokhov (see Review \cite{MokhRev}).
  
 The hierarchy of the Poisson Structures having the general
form (\ref{genhamstr}) was first written in \cite{EnOrRub} for
KdV 

$$\varphi_{t} = 6 \varphi \varphi_{x} - \varphi_{xxx} $$
using the local bi-hamiltonian formalism 
(Gardner - Zakharov - Faddeev and Magri brackets) and the
corresponding Recursion operator in Lenard - Magri scheme.
Let us present here the pair of corresponding local Hamiltonian
Structures

$${\hat J}_{0} = \partial/\partial x $$
(Gardner - Zakharov - Faddeev bracket)
and
$${\hat J}_{1} = - {\partial}^{3}/{\partial x}^{3}
+ 2 ( \varphi \, \partial/\partial x + 
\partial/\partial x \, \varphi ) $$
(Magri bracket)   
and the first weakly non-local Hamiltonian operator:

$${\hat J}_{2} = \partial^{5}/{\partial x}^{5} - 
8 \varphi \partial^{3}/{\partial x}^{3}
- 12 \varphi_{x} \partial^{2}/{\partial x}^{2} - 
8 \varphi_{xx} \partial/\partial x
+ 16 \varphi^{2} \partial/\partial x - 2 \varphi_{xxx} +
16 \varphi \varphi_{x} - 4 \varphi_{x} D^{-1}
\varphi_{x} $$

 The operator ${\hat J}_{2}$ is obtained by the action of the
Recursion operator

$${\hat R} = - \partial^{2}/{\partial x}^{2} + 4 \varphi + 
2 \varphi_{x} \, D^{-1} $$
(such that ${\hat R} \, {\hat J}_{0} \, = \, {\hat J}_{1}$)
to the operator ${\hat J}_{1}$. The higher ("positive") 
Hamiltonian operators ${\hat J}_{n}$ can be obtained in the same
recursion scheme by the formula 
${\hat J}_{n} \, = \, {\hat R}^{n} \, {\hat J}_{0}$.
It was proved in \cite{EnOrRub} that all operators ${\hat J}_{n}$
for $n >1$ can be written in the form:

$$ {\hat J}_{n} \,\, = \,\, (local \,\, part) - \sum_{k=1}^{n-1}
S_{(k)}(\varphi,\varphi_{x},\dots) D^{-1}
S_{(n-k-1)}(\varphi,\varphi_{x},\dots)$$
where $S_{(1)}(\varphi,\varphi_{x},\dots) = 2\varphi_{x}$
and 

$$S_{(k)}(\varphi,\varphi_{x},\dots) \equiv {\hat R}
S_{(k-1)}(\varphi,\varphi_{x},\dots)$$
are higher KdV flows.

 The similar weakly non-local expressions for positive powers
of the Recursion operator for KdV were also considered in 
\cite{EnOrRub}. Let us present here the corresponding result:

$${\hat R}^{n} = (local \,\, part) + \sum_{k=1}^{n}
S_{(k)}(\varphi,\varphi_{x},\dots) D^{-1}
{\delta H_{(n-k)} \over \delta \varphi(x)} \,\,\, , \,\,
n\geq 0$$
where $S_{(k)} = \partial_{x} \delta H_{(k)}/\delta\varphi(x)$,
$H_{(0)} = \int \varphi dx$ and

$${\delta H_{(k)} \over \delta \varphi(x)} \equiv 
{\delta H_{(k-1)} \over \delta \varphi(x)} {\hat R} $$
are Euler-Lagrange derivatives of higher Hamiltonian
functions for KdV hierarchy. Let us mention also that in our
notations ${\hat R}$ acts from the left on the vectors and from 
the right on the 1-forms in the functional space ${\cal L}_{0}$.

 Using the results of \cite{EnOrRub} it was proved in \cite{PhysD}
that the "negative" Symplectic Structures (i.e. the inverse of
"negative" Hamiltonian operators) also have the weakly nonlocal
form. Let us formulate here the corresponding statement:

 All the "negative" Symplectic Structures 
${\hat \Omega}_{-n} = ({\hat J}_{-n})^{-1} \,\, , n \geq 0$ 
for KdV hierarchy can be written in the following form:

$$\Omega_{-n} = (local \,\, part) + \sum_{k=0}^{n}
{\delta H_{(k)} \over \delta \varphi(x)} D^{-1}
{\delta H_{(n-k)} \over \delta \varphi(x)}$$ 

 It was conjectured in \cite{PhysD} that this structure of 
"positive" Hamiltonian and "negative" Symplectic hierarchies 
should be very common for the wide class of integrable systems.
In particular, the similar statements about NLS equation

$$i \psi_{t} = - \psi_{xx} + 2 \kappa |\psi|^{2} \psi $$
were proved in \cite{PhysD}. Let us give here also the
corresponding statements for this case.

 Two basic Hamiltonian operators can be written here in the
form:

$${\hat J}_{0} \, = \, \left( \begin{array}{cc}
0 & i \cr -i & 0 \end{array} \right) \,\,\,\,\, , \,\,\,\,\,
{\hat J}_{1} \, = \, \left( \begin{array}{cc}
0 & \partial \cr \partial & 0 \end{array} \right) -
2 \kappa \left( \begin{array}{cc}
- \psi \partial^{-1} \psi & \psi \partial^{-1} {\bar \psi} \cr
{\bar \psi} \partial^{-1} \psi & 
- {\bar \psi} \partial^{-1} {\bar \psi} \end{array} \right)$$

 The Recursion operator ${\hat R}$ is defined again by formula
${\hat R} \, {\hat J}_{0} \, = \, {\hat J}_{1}$. For the 
"positive" Hamiltonian operators 
${\hat J}_{n} \, = \, {\hat R}^{n} \, {\hat J}_{0}$ and
"negative" Symplectic Structures
${\hat \Omega}_{-n} = ({\hat J}_{-n})^{-1}$, $n \geq 1$
the following statements will then be true (\cite{PhysD}):

 The "positive" Hamiltonian operators ${\hat J}_{n}$ and 
"negative" Symplectic Structures ${\hat \Omega}_{-n}$ in the 
hierarchy of Hamiltonian Structures for NLS can be written in
the form:

$${\hat J}_{n} \, = \, (local \,\, part) - \sum_{k=1}^{n} 
S_{(k-1)}(\psi,{\bar \psi},\dots) D^{-1}
S_{(n-k)}(\psi,{\bar \psi},\dots) $$
 
$$ {\hat \Omega}_{-n} \, = \, (local \,\, part) + \sum_{k=1}^{n}
{\delta H_{(k-1)} \over \delta(\psi,{\bar \psi})(x)} \, D^{-1} \,
{\delta H_{(n-k)} \over \delta(\psi,{\bar \psi})(x)} $$
where 

$$S_{(k)} \equiv {\hat J}_{0} 
{\delta H_{(k)} \over \delta(\psi,{\bar \psi})(x)} \,\,\, ,
\,\,\,\,\, H_{(0)} = \sqrt{2\kappa} \int \psi {\bar \psi} dx 
\,\,\, , 
\,\,\,\,\, {\rm and} \,\,\,\,\, 
{\delta H_{(k)} \over \delta(\psi,{\bar \psi})(x)} =
{\hat R} {\delta H_{(k-1)} \over \delta(\psi,{\bar \psi})(x)}$$
for any $k \geq 1$.\footnote{Actually, as was pointed out in
\cite{PhysD} the NLS equation has in fact three local 
Hamiltonian Structures 
(${\hat J}_{0}$, ${\hat J}_{1}$, ${\hat J}_{2}$)
in the variables
$r = \sqrt{\psi {\bar \psi}}$, 
$\theta = -i(\psi_{x}/\psi - {\bar \psi}_{x}/{\bar \psi})$
(i.e. $\psi = r \exp(i\int \theta dx)$).}

 The general investigations of the weakly-nonlocal structures
of integrable hierarchies were made in the very recent works.
Let us cite here the work \cite{Serg} (see also the references
therein) where the weakly-nonlocal form of the structures described
above was established for the integrable hierarchies under rather
general requirements. 

 It's possible to state that the 
weakly-nonlocal structures play indeed quite important role
in the theory of integrable systems.

 Let us say that the "positive" Symplectic Structures
${\hat \Omega}_{n} \, = \, {\hat J}_{n}^{-1}$ and the "negative"
Hamiltonian operators ${\hat J}_{-n}$, $(n \geq 1)$ will have
much more complicated form (not weakly nonlocal) both for
KdV and NLS hierarchies. 

Let us formulate the Theorem proved 
in \cite{malnloc2} connecting the non-local and local parts
for the general weakly-nonlocal Poisson brackets 
(\ref{genhamstr}). We will assume that the bracket 
(\ref{genhamstr}) is written in "irreducible" form, i.e.
the "vector-fields"

$${\bf S}_{(s)} (\bm{\varphi}, \bm{\varphi}_{x}, \dots) \, = \,
\left( S^{1}_{(s)} (\bm{\varphi}, \bm{\varphi}_{x}, \dots),
\dots S^{n}_{(s)} (\bm{\varphi}, \bm{\varphi}_{x}, \dots) 
\right)^{t} $$
are linearly independent (with constant coefficients).

\vspace{0.5cm}

 {\bf Theorem.}

{\it For any bracket (\ref{genhamstr}) the flows

\begin{equation}
\label{commflows}
\varphi^{i}_{t_{s}} \, = \, 
S^{i}_{(s)} (\bm{\varphi}, \bm{\varphi}_{x}, \dots)
\end{equation}
commute with each other and leave the bracket (\ref{genhamstr})
invariant.}

\vspace{0.5cm}

 The second statement means here that the Lie derivative of 
the tensor (\ref{genhamstr}) along the flows (\ref{commflows})
is zero on the functional space ${\cal L}_{0}$.

\vspace{0.5cm}

 However the general classification of weakly nonlocal Hamiltonian
Structures (\ref{genhamstr}) is rather difficult and is absent
by now. 

\vspace{0.5cm}

 Let us say now some words about very important class of weakly
nonlocal Hamiltonian and Symplectic Structures of Hydrodynamic 
Type (HT). These structures are closely connected with the
Systems of Hydrodynamic Type (HT-Systems), i.e. the systems of 
the form:

\begin{equation}
\label{HTsyst}
U^{\nu}_{T} \, = \, V^{\nu}_{\mu} ({\bf U}) \, U^{\mu}_{X} 
\,\,\,\,\, , \,\,\,\,\, \nu, \, \mu \, = \, 1, \dots, N
\end{equation}
where $V^{\nu}_{\mu} (U)$ is some $N \times N$ matrix
depending on the variables $U^{1}, \dots, U^{N}$.

 The Hamiltonian approach to systems (\ref{HTsyst}) was
started by B.A. Dubrovin and S.P. Novikov (\cite{dn1,dn2,dn3})
who introduced the local (homogeneous) Poisson brackets of
Hydrodynamic Type (Dubrovin - Novikov brackets). Let us give
here the corresponding definition.

{\bf Definition 1.}

{\it
 Dubrovin - Novikov bracket (DN-bracket) is a bracket on the
functional space \linebreak
$(U^{1}(X), \dots, U^{N}(X))$ having the form

\begin{equation}
\label{DNbr}
\{U^{\nu}(X), U^{\mu}(Y)\} \, = \, g^{\nu\mu}({\bf U}) \,
\delta^{\prime} (X-Y) \, + \, b^{\nu\mu}_{\lambda} ({\bf U}) \,
U^{\lambda}_{X} \, \delta (X-Y)
\end{equation}

}

 The corresponding Hamiltonian operator ${\hat J}^{\nu\mu}$
can be written as

$${\hat J}^{\nu\mu} \, = \,  g^{\nu\mu}({\bf U})
{\partial \over \partial x} + b^{\nu\mu}_{\lambda} ({\bf U}) \,
U^{\lambda}_{X} $$
and is homogeneous w.r.t. transformation $X \rightarrow aX$.

 Every functional $H$ of Hydrodynamic Type, i.e. the functional
having the form
 
$$H \, = \, \int_{-\infty}^{+\infty} h({\bf U}) \, dX $$
generates the system of Hydrodynamic Type (\ref{HTsyst}) 
according to the formula

\begin{equation}
\label{HThamsyst}
U^{\nu}_{T} \, = \, {\hat J}^{\nu\mu} \, 
{\delta H \over \delta U^{\mu}(X)} \, = \, g^{\nu\mu}({\bf U})
{\partial \over \partial x} {\partial h \over \partial U^{\mu}}
\, + \, b^{\nu\mu}_{\lambda} ({\bf U}) \,
{\partial h \over \partial U^{\mu}} \, U^{\lambda}_{X} 
\end{equation}
 
 The DN-bracket (\ref{DNbr}) is called non-degenerate if
$det \, ||g^{\nu\mu}({\bf U})|| \, \neq \, 0$.

 As was shown by B.A. Dubrovin and S.P. Novikov the theory
of DN-brackets is closely connected with Riemannian 
geometry (\cite{dn1,dn2,dn3}). In fact, it follows from the 
skew-symmetry of (\ref{DNbr}) that the coefficients 
$g^{\nu\mu}({\bf U})$ give in the non-degenerate case the
contravariant pseudo-Riemannian metric on the manifold
${\cal M}^{N}$ with coordinates $(U^{1}, \dots, U^{N})$
while the functions 
$\Gamma^{\nu}_{\mu\lambda}({\bf U}) \, = \,
- g_{\mu\alpha}({\bf U}) \, b^{\alpha\nu}_{\lambda}({\bf U})$
(where $g_{\nu\mu}({\bf U})$ is the corresponding metric
with lower indices) give the connection coefficients
compatible with metric $g_{\nu\mu}({\bf U})$. The validity
of Jacobi identity requires then that $g_{\nu\mu}({\bf U})$
is actually a flat metric on the manifold ${\cal M}^{N}$
and the functions $\Gamma^{\nu}_{\mu\lambda}({\bf U})$
give a symmetric (Levi-Civita) connection on ${\cal M}^{N}$
(\cite{dn1,dn2,dn3}).

 In the flat coordinates $n^{1}({\bf U}), \dots, n^{N}({\bf U})$
the non-degenerate DN-bracket can be written in constant
form:

$$\{n^{\nu}(X), n^{\mu}(Y)\} \, = \, e^{\nu} \,
\delta^{\nu\mu} \, \delta^{\prime} (X-Y) $$
where $e^{\nu} \, = \, \pm 1$.

 The functionals 

$$N^{\nu} \, = \, \int_{-\infty}^{+\infty} 
n^{\nu}(X) \, dX $$
are the annihilators of the bracket (\ref{DNbr}) and the 
functional

$$P \, = \, {1 \over 2} \int_{-\infty}^{+\infty}
\sum_{\nu = 1}^{N} e^{\nu} \, \left(n^{\nu}(X)\right)^{2}
\, dX $$
is the momentum functional generating the system
$U^{\nu}_{T} \, = \, U^{\nu}_{X}$ according to 
(\ref{HThamsyst}). 

 The Symplectic Structure corresponding to non-degenerate
DN-bracket has the weakly nonlocal form and can be written as

$$\Omega_{\nu\mu}(X,Y) \, = \, e^{\nu} \, \delta_{\nu\mu} \,
\nu (X-Y) $$
in coordinates $n^{\nu}$ or, more generally,

$$\Omega_{\nu\mu}(X,Y) \, = \, \sum_{\lambda = 1}^{N}
e^{\lambda} \, 
{\partial n^{\lambda} \over \partial U^{\nu}}(X) \,
\nu (X-Y) \, 
{\partial n^{\lambda} \over \partial U^{\mu}}(Y) $$
in arbitrary coordinates $U^{\nu}$.

 Let us mention also that the degenerate brackets (\ref{DNbr})
are more complicated but also have a nice differential 
geometric structure (\cite{grinberg}).

 The brackets (\ref{DNbr}) are closely connected with the
integration theory of systems of Hydrodynamic Type 
(\ref{HTsyst}). Namely, according to conjecture of 
S.P. Novikov, all the diagonalizable systems (\ref{HTsyst})
which are Hamiltonian with respect to DN-brackets (\ref{DNbr})
(with Hamiltonian function of Hydrodynamic Type) are
completely integrable. This conjecture was proved by 
S.P. Tsarev (\cite{tsarev}) who proposed a general procedure
("generalized Hodograph method") of integration of 
Hamiltonian diagonalizable systems (\ref{HTsyst}).

 In fact Tsarev's "generalized Hodograph method" permits
to integrate the wider class of diagonalizable systems 
(\ref{HTsyst}) (semi-Hamiltonian systems, \cite{tsarev})
which appeared to be Hamiltonian in more general 
(weakly nonlocal) Hamiltonian formalism. 

 The corresponding Poisson brackets (Mokhov - Ferapontov
bracket and Ferapontov bracket) are the weakly nonlocal
generalizations of DN-bracket (\ref{DNbr}) and are connected
with geometry of submanifolds in pseudo-Euclidean spaces.
Let us describe here the corresponding structures.

\vspace{0.5cm}

 The Mokhov - Ferapontov bracket 
(MF-bracket) has the form (\cite{mohfer1})

\begin{equation}
\label{MFbr}
\{U^{\nu}(X), U^{\mu}(Y)\} \, = \, g^{\nu\mu}({\bf U}) \,
\delta^{\prime} (X-Y) \, + \, b^{\nu\mu}_{\lambda} ({\bf U}) \,
U^{\lambda}_{X} \, \delta (X-Y) \, + 
\, c \, U^{\nu}_{X} \, \nu (X-Y) \, U^{\mu}_{Y}
\end{equation}

 As was proved in \cite{mohfer1} the expression (\ref{MFbr})
with $det \, ||g^{\nu\mu}({\bf U})|| \, \neq \, 0$ gives the 
Poisson bracket on the space $U^{\nu}(X)$ if and only if:

1) The tensor $g^{\nu\mu}({\bf U})$ represents the 
pseudo-Riemannian contravariant metric of constant curvature
$c$ on the manifold ${\cal M}^{N}$, i.e.

$$R^{\nu\mu}_{\lambda\eta}({\bf U}) \, = \, c \, \left(
\delta^{\nu}_{\lambda} \, \delta^{\mu}_{\eta} \, - \,
\delta^{\mu}_{\lambda} \, \delta^{\nu}_{\eta} \right) $$

2) The functions 
$\Gamma^{\nu}_{\mu\lambda}({\bf U}) \, = \,
- g_{\mu\alpha}({\bf U}) \, b^{\alpha\nu}_{\lambda}({\bf U})$
represent the Levi-Civita connection of metric 
$g_{\nu\mu}({\bf U})$.

\vspace{0.5cm}

 The Ferapontov bracket (F-bracket) is more general 
weakly nonlocal generalization of DN-bracket having the
form (\cite{fer1,fer2,fer3,fer4}):

$$\{U^{\nu}(X), U^{\mu}(Y)\} \, = \, g^{\nu\mu}({\bf U}) \,
\delta^{\prime} (X-Y) \, + \, b^{\nu\mu}_{\lambda}({\bf U}) \,
U^{\lambda}_{X} \, \delta (X-Y) \, + $$
\begin{equation}
\label{Fbr}
+ \, \sum_{k=1}^{g} e_{k} \, w^{\nu}_{(k)\lambda}({\bf U}) \,
U^{\lambda}_{X} \, \nu (X-Y) \, 
w^{\mu}_{(k)\delta}({\bf U}) \, U^{\delta}_{Y}
\end{equation}
$e_{k} = \pm 1$, $\nu, \mu = 1, \dots, N$.

 The expression (\ref{Fbr}) 
(with $det \, ||g^{\nu\mu}({\bf U})|| \, \neq \, 0$) gives the  
Poisson bracket on the space $U^{\nu}(X)$ if and only if
(\cite{fer1,fer4}):

1) Tensor $g^{\nu\mu}({\bf U})$ represents the metric
of the submanifold ${\cal M}^{N} \subset {\mathbb E}^{N+g}$
with flat normal connection in the pseudo-Euclidean space
${\mathbb E}^{N+g}$ of dimension ${N+g}$;

2) The functions
$\Gamma^{\nu}_{\mu\lambda}({\bf U}) \, = \,
- g_{\mu\alpha}({\bf U}) \, b^{\alpha\nu}_{\lambda}({\bf U})$
represent the Levi-Civita connection of metric
$g_{\nu\mu}({\bf U})$;

3) The set of affinors $\{w^{\nu}_{(k)\lambda}({\bf U})\}$
represents the full set of Weingarten operators corresponding
to $g$ linearly independent parallel vector fields in the
normal bundle, such that:

$$g_{\nu\tau}({\bf U}) \, w^{\tau}_{(k)\mu}({\bf U}) \, = \,
g_{\mu\tau}({\bf U}) \, w^{\tau}_{(k)\nu}({\bf U}) 
\,\,\,\,\, , \,\,\,\,\,
\nabla_{\nu} \, w^{\mu}_{(k)\lambda}({\bf U}) \, = \,
\nabla_{\lambda} \, w^{\mu}_{(k)\nu}({\bf U}) $$

$$R^{\nu\mu}_{\lambda\eta}({\bf U}) \, = \,
\sum_{k=1}^{g} e_{k} \, \left( 
w^{\nu}_{(k)\lambda}({\bf U}) \, 
w^{\mu}_{(k)\eta}({\bf U}) \, - \,
w^{\mu}_{(k)\lambda}({\bf U}) \,
w^{\nu}_{(k)\eta}({\bf U}) \right) $$

 Besides that the set of affinors $w_{(k)}$ is commutative
$[w_{(k)},w_{(k^{\prime})}]  = 0$.

 As was shown in \cite{fer2} the expression (\ref{Fbr}) can
be considered as the Dirac reduction of Dubrovin-Novikov 
bracket connected with metric in ${\mathbb E}^{N+g}$ to the
manifold ${\cal M}^{N}$ with flat normal connection. Let
us note also that MF-bracket can be considered as a case
of the F-bracket when ${\cal M}^{N}$ is a (pseudo)-sphere
${\cal S}^{N} \subset {\mathbb E}^{N+1}$ in a pseudo-Euclidean
space.

\vspace{0.5cm}

 The Symplectic Structures $\Omega_{\nu\mu}(X,Y)$ for
both (non-degenerate) MF-bracket and F-bracket have also
the weakly nonlocal form (\cite{PhysD,IntJMMS2}) 
and can be written in general coordinates $U^{\nu}$ as

$$\Omega_{\nu\mu}(X,Y) \, = \, \sum_{s=1}^{N+g} 
\epsilon_{s} \,
{\partial n^{s} \over \partial U^{\nu}}(X) \,
\nu (X-Y) \,
{\partial n^{s} \over \partial U^{\mu}}(Y) $$
where $\epsilon_{s} \, = \, \pm 1$ and the metric $G_{IJ}$
in the space ${\mathbb E}^{N+g}$ has the form
$G_{IJ} \, = \, diag (\epsilon_{1}, \dots, \epsilon_{N+g})$.
The functions $n^{1}({\bf U}), \dots, n^{N+g}({\bf U})$
are the "Canonical forms" on the manifold ${\cal M}^{N}$
and play the role of densities and annihilators of bracket
(\ref{Fbr}) and "Canonical Hamiltonian functions"
(see \cite{PhysD}) depending on the definition of phase
space. In fact, the functions $n^{s}({\bf U})$ are the
restrictions of flat coordinates of metric $G_{IJ}$
giving the DN-bracket in ${\mathbb E}^{N+g}$ on manifold
${\cal M}^{N}$. The mapping
${\cal M}^{N} \rightarrow {\mathbb E}^{N+g}$:

$$\left( U^{1}, \dots, U^{N}\right) \,\,\,\,\,\,\,
\rightarrow \,\,\,\,\,\,\, 
\left( n^{1}({\bf U}), \dots,  n^{N+g}({\bf U}) \right) $$
gives locally the embedding of ${\cal M}^{N}$ in 
${\mathbb E}^{N+g}$ as a submanifold with flat normal connection.

\vspace{0.5cm}

 All the brackets (\ref{DNbr}), (\ref{MFbr}), (\ref{Fbr}) 
are connected with Tsarev method of integration of systems
(\ref{HTsyst}). Namely, any diagonalizable system (\ref{HTsyst})
Hamiltonian w.r.t. the (non-degenerate) bracket
(\ref{DNbr}), (\ref{MFbr}) or (\ref{Fbr}) can be integrated
by "generalized Hodograph method".

 We will not describe here Tsarev method in details. However,
let us point out that "generalized Hodograph method" and the
HT Hamiltonian Structures were very useful for Whitham's 
systems obtained by the averaging of integrable PDE's
(\cite{whith,ffm,dn1,krichev1,krichev2,dn2,krichev3,dn3}).

\vspace{0.5cm}

 Let us discuss now the Whitham averaging method
(\cite{whith,ffm,dn1,krichev1,krichev2,dn2,krichev3,dn3,dm}).
We will restrict ourselves to the evolution systems 

\begin{equation}
\label{insyst}
\varphi^{i}_{t} \, = \, 
Q^{i}(\bm{\varphi}, \bm{\varphi}_{x}, \dots) 
\end{equation}
although the Whitham method can be applied also to more
general PDE systems.

 The $m$-phase Whitham averaging method is based on the 
existence of the finite-parametric family of solutions of 
(\ref{insyst}) having the form

\begin{equation}
\label{qpersol}
\varphi^{i} (x,t) \, = \, 
\Phi^{i} ({\bf k}({\bf U}) \, x \, + \, 
{\bm \omega}({\bf U}) \, t \, + \, {\bm \theta}_{0}, \,\,
U^{1}, \dots, U^{N})
\end{equation}
where ${\bf k} \, = \, (k^{1}, \dots, k^{m})$,
${\bm \omega} \, = \, (\omega^{1}, \dots, \omega^{m})$,
${\bm \theta} \, = \, (\theta^{1}, \dots, \theta^{m})$, 
and $\Phi^{i}({\bm \theta}, {\bf U})$ are the functions
$2\pi$-periodic w.r.t. each $\theta^{\alpha}$ and depending
on the finite set of additional parameters 
$U^{1}, \dots, U^{N}$. The solutions (\ref{qpersol}) are 
the quasiperiodic functions depending on $N + m$ parameters 
$U^{1}, \dots, U^{N}$ and
$\theta_{0}^{1}, \dots, \theta_{0}^{m}$. 

 In Whitham method the parameters $U^{1}, \dots, U^{N}$ and
$\theta_{0}^{1}, \dots, \theta_{0}^{m}$ become the 
slow-modulated functions of $x$ and $t$ to get the 
slow-modulated $m$-phase solution of (\ref{insyst}).
We introduce then the slow variables
$X = \epsilon \, x$, $T = \epsilon \, t$, 
$\epsilon \rightarrow 0$ and then try to find a solution
of system

\begin{equation}
\label{epsyst}
\epsilon \, \varphi^{i}_{T} \, = \,
Q^{i}(\bm{\varphi}, \epsilon \, \bm{\varphi}_{X}, \dots)
\end{equation}
having the form 

\begin{equation}
\label{sol}
\varphi^{i} (X,T) \, = \, \sum_{k=0}^{+\infty} \, 
\epsilon^{k} \,\,
\Phi^{i}_{(k)} 
\left({{\bf S}(X,T) \over \epsilon} \, +\, {\bm \theta}, \,
X, \, T \right)
\end{equation}
where $\Phi^{i}_{(k)} ({\bm \theta}, X, T)$ are 
$2\pi$-periodic w.r.t. each $\theta^{\alpha}$ and
${\bf S}(X,T)\, = \, (S^{1}(X,T), \dots, S^{m}(X,T))$
is a "phase" depending on the slow variables $X$ and $T$
(\cite{whith,luke,dm}).

 It follows then that $\Phi^{i}_{(0)} ({\bm \theta}, X, T)$
should always belong to the family of exact $m$-phase 
solutions of (\ref{insyst}) at any $X$ and $T$ and we have to 
find the functions $\Phi^{i}_{(k)} ({\bm \theta}, X, T)$,
$k \geq 1$ from the system (\ref{epsyst}). The existence of the
solution (\ref{sol}) implies some conditions on the parameters
${\bf U}(X,T)$, ${\bm \theta}_{0}(X,T)$ giving the zero
approximation of (\ref{sol}). In particular, the existence of
$\Phi^{i}_{(1)} ({\bm \theta}, X, T)$ implies the conditions
on ${\bf U}(X,T)$ having the form of the system (\ref{HTsyst}).
This system is called the Whitham system and describes the
evolution of the "averaged" characteristics of the solution
(\ref{sol}) in the main order. The solution of the Whitham
system (\ref{HTsyst}) is actually the main step in the
whole procedure. Let us mention also that the Whitham systems
for so-called "Integrable systems" like KdV can usually be
written in the diagonal form 
(\cite{whith,ffm,dn1,dn2,dn3, pavlov2}).

 The Lagrangian formalism of the Whitham
system and the averaging of Lagrangian function were
considered by Whitham (\cite{whith}) who pointed out that the
Whitham system admits the (local) Lagrangian formalism if the
initial system (\ref{insyst}) was Lagrangian.

 The Hamiltonian approach to the Whitham method was started
by B.A. Dubrovin and S.P. Novikov
in \cite{dn1} (see also \cite{dn2,dn3}) where the procedure
of "averaging" of local field-theoretical Poisson bracket
was proposed. The Dubrovin - Novikov procedure gives the
DN-bracket for the Whitham system (\ref{HTsyst}) in case
when the initial system (\ref{insyst}) is Hamiltonian
w.r.t. a local Poisson bracket

$$\{\varphi^{i}(x), \varphi^{j}(y)\} \, = \, \sum_{k \geq o} 
B_{(k)}^{ij} (\bm{\varphi}, \bm{\varphi}_{x}, \dots) \,
\delta^{(k)}(x-y)$$
with local Hamiltonian functional\footnote{The proof
of Jacobi identity for the averaged bracket was obtained
in \cite{engam}.}

$$H \, = \, \int_{-\infty}^{+\infty}
h (\bm{\varphi}, \bm{\varphi}_{x}, \dots) \, dx $$
 
 This procedure was generalized in \cite{malnloc1,malnloc2}
for the weakly nonlocal Hamiltonian structures. In this case
the procedure of construction of general F-bracket 
(or MF-bracket) for the Whitham system from the weakly 
non-local Poison bracket (\ref{genhamstr}) for initial
system (\ref{insyst}) was proposed.

 In this paper we will consider the Whitham averaging method
for PDE's having the weakly nonlocal Symplectic Structures
(\ref{gensympstr}) and construct the Symplectic Structures
of Hydrodynamic Type for the corresponding Whitham systems.
Let us say that the corresponding HT Symplectic Structures can 
in principle be more general than those connected with the Tsarev 
integration method. The theory of integration of corresponding 
HT systems (\ref{HTsyst}) should then be more complicated in 
general case.

 We call here the weakly nonlocal Symplectic Structure of
Hydrodynamic Type the Symplectic form 
$\Omega_{\nu\mu}(X,Y)$ having the form:

\begin{equation}  
\label{HTsympform}
\Omega_{\nu\mu}(X,Y) \, = \, \sum_{s,p=1}^{M}
\kappa_{sp} \, \omega^{(s)}_{\nu}({\bf U}(X)) \, \nu (X-Y) \,
\omega^{(p)}_{\mu}({\bf U}(Y))
\end{equation}
or in "diagonal" form

$$\Omega_{\nu\mu}(X,Y) \, = \, \sum_{s=1}^{M}
e_{s} \, \omega^{(s)}_{\nu}({\bf U}(X)) \, \nu (X-Y) \,
\omega^{(s)}_{\mu}({\bf U}(Y)) $$
in coordinates $U^{\nu}$ where $\kappa_{sp}$ is some quadratic 
form, $e_{s} = \pm 1$, and
$\omega^{(s)}_{\nu}({\bf U})$ are closed 1-forms on the 
manifold ${\cal M}^{N}$. Locally the forms 
$\omega^{(s)}_{\nu}({\bf U})$ can be represented as the gradients
of some functions $f^{(s)}({\bf U})$ such that

\begin{equation}
\label{HTconsform}
\Omega_{\nu\mu}(X,Y) \, = \, \sum_{s,p=1}^{M}
\kappa_{sp} \, {\partial f^{(s)} \over \partial U^{\nu}}(X) 
\, \nu (X-Y) \, {\partial f^{(p)} \over \partial U^{\mu}}(Y)
\end{equation}

 Generally speaking, we don't require here that the 
embedding ${\cal M}^{N} \subset {\mathbb E}^{M}$ given by
$(U^{1}, \dots, U^{N}) \rightarrow 
(f^{(1)}({\bf U}), \dots, f^{(M)}({\bf U}))$ gives the
submanifold with flat normal connection. Therefore, the
corresponding Hamiltonian operators will not necessary
have the weakly non-local form of  
the DN-brackets, MF-brackets or F-brackets.

 We propose here the procedure which permits to construct
the Symplectic Structure (\ref{HTsympform}) for the Whitham
system in case when the (local) initial system (\ref{insyst})
has the weakly nonlocal symplectic structure 
(\ref{gensympstr}) with some local Hamiltonian function

$$H \, = \, \int_{-\infty}^{+\infty}
h(\bm{\varphi}, \bm{\varphi}_{x}, \dots) \, dx $$

 In Chapter 2 we consider the general Symplectic forms
(\ref{gensympstr}) and the HT Symplectic forms 
(\ref{HTsympform}). In Chapter 3 we consider the general features 
of the Whitham method and introduce some conditions which we will
need for the next considerations. In Chapter 4 we introduce the
"extended" phase space and prove some technical Lemmas about
the "extended" Symplectic form necessary for the averaging
procedure of the forms (\ref{gensympstr}). In Chapter 5 we
give the procedure of averaging of the forms (\ref{gensympstr})
and prove that the Whitham system admits the Symplectic 
Structure of Hydrodynamic Type given by the corresponding
"averaged" Symplectic form. In Chapter 6 we give another 
variant of averaging of forms (\ref{gensympstr}) based on
the averaging of weakly nonlocal 1-forms and give the weakly
nonlocal Lagrangian formalism for the Whitham system.

\section{General weakly nonlocal Symplectic Forms and the
weakly nonlocal Symplectic Forms of Hydrodynamic Type.}
\setcounter{equation}{0}

 Let us consider first the general weakly nonlocal Symplectic 
Forms (\ref{gensympstr}). The nonlocal part of 
(\ref{gensympstr}) is skew-symmetric  and we should require
then also the skew-symmetry of the 
local part of (\ref{gensympstr}).
We will assume everywhere that (\ref{gensympstr}) is written
in "irreducible" form, i.e. the functions
${\bf q}^{(s)} (\bm{\varphi}, \bm{\varphi}_{x}, \dots)$ 
are linearly independent (with constant coefficients). Let us
prove here the following statement formulated in 
\cite{PhysD}.

\vspace{0.5cm}

{\bf Theorem 1.}

{\it
For any closed 2-form (\ref{gensympstr}) the functions
$q^{(s)}_{i} (\bm{\varphi}, \bm{\varphi}_{x}, \dots)$
represent the closed 1-forms on ${\cal L}_{0}$.
}

\vspace{0.5cm}

 Proof.

 Let us denote $\Omega^{\prime}_{ij}(x,y)$ the local part
of (\ref{gensympstr}). We have to check the closeness of 
 2-form (\ref{gensympstr}), i.e.

$$\left( d \Omega \right)_{ijk} (x,y,z) \, = \,
{\delta \Omega_{ij} (x,y) \over \delta \varphi^{k}(z)} +
{\delta \Omega_{jk} (y,z) \over \delta \varphi^{i}(x)} +
{\delta \Omega_{ki} (z,x) \over \delta \varphi^{j}(y)} \,
\equiv \, 0 $$
(in sence of distributions) on ${\cal L}_{0}$.

 We have then

$$\left( d \Omega \right)_{ijk} (x,y,z) \, = \,
\left( d \Omega^{\prime} \right)_{ijk} (x,y,z) \, + \,$$
$$+ \, \sum_{s=1}^{g} e_{s} \left[
{\delta q^{(s)}_{i} (x) 
\over \delta \varphi^{k}(z)} \, \nu (x-y) \,
q^{(s)}_{j} (y) \, + \,
q^{(s)}_{i} (x)
\, \nu (x-y) \, 
{\delta q^{(s)}_{j} (y)
\over \delta \varphi^{k}(z)} \right]\, + $$
$$+\, \sum_{s=1}^{g} e_{s} \left[
{\delta q^{(s)}_{j} (y)
\over \delta \varphi^{i}(x)} \, \nu (y-z) \,
q^{(s)}_{k} (z) \, + \,
q^{(s)}_{j} (y)
\, \nu (y-z) \,
{\delta q^{(s)}_{k} (z)
\over \delta \varphi^{i}(x)} \right]\, + $$  
\begin{equation}
\label{domega}
+ \, \sum_{s=1}^{g} e_{s} \left[
{\delta q^{(s)}_{k} (z)
\over \delta \varphi^{j}(y)} \, \nu (z-x) \,
q^{(s)}_{i} (x) \, + \,
q^{(s)}_{k} (z)
\, \nu (z-x) \,
{\delta q^{(s)}_{i} (x)
\over \delta \varphi^{j}(y)} \right]
\end{equation} 

 We use here the Leibnits identity and the relations 

\begin{equation}
\label{derrel}
{\delta \varphi^{i}(x) \over \delta \varphi^{j}(y)} 
\, = \, \delta^{i}_{j} \, \delta (x-y) \,\,\, , \,\,\,
{\delta \varphi_{x}^{i}(x) \over \delta \varphi^{j}(y)}
\, = \, \delta^{i}_{j} \, \delta^{\prime} (x-y) \,\,\, ,
\dots 
\end{equation}

 The expression 
$\left( d \Omega^{\prime} \right)_{ijk} (x,y,z)$ is then
purely local and all the nonlocality arises just in the 
remaining part of 
$\left( d \Omega \right)_{ijk} (x,y,z)$. Let us consider
now the values

$$d \Omega (\bm{\xi},\bm{\eta},\bm{\zeta}) \, = \,
\int_{-\infty}^{+\infty} \int_{-\infty}^{+\infty}
\int_{-\infty}^{+\infty} 
\left( d \Omega \right)_{ijk} (x,y,z) \,
\xi^{i}(x) \, \eta^{j}(y) \, \zeta^{k}(z) \, dx \, dy \, dz$$
where $\xi^{i}(x)$, $\eta^{i}(x)$,  $\zeta^{i}(x)$ are
the functions with finite supports such that the supports
of all $\zeta^{k}(x)$ do not intersect with the supports
of all $\xi^{i}(x)$, $\eta^{j}(x)$ and moreover all supports
of $\xi^{i}(x)$, $\eta^{j}(x)$ lie on the left from any support 
of $\zeta^{k}(x)$. Using (\ref{domega}) and (\ref{derrel})
it's easy to see then that we can write in this case

$$d \Omega (\bm{\xi},\bm{\eta},\bm{\zeta}) \, = \,$$

$$= \, \int_{-\infty}^{+\infty} \int_{-\infty}^{+\infty}
\int_{-\infty}^{+\infty} \sum_{s=1}^{g} e_{s} \left[
{\delta q^{(s)}_{j} (y)
\over \delta \varphi^{i}(x)} \, \nu (y-z) \,
q^{(s)}_{k} (z) \, + \,
q^{(s)}_{k} (z)
\, \nu (z-x) \,
{\delta q^{(s)}_{i} (x)
\over \delta \varphi^{j}(y)} \right] \, \times $$
$$\times \, \xi^{i}(x) \, \eta^{j}(y) \, \zeta^{k}(z) \, 
dx \, dy \, dz \, =$$

$$= {1 \over 2}  \int_{-\infty}^{+\infty} \!\!
\int_{-\infty}^{+\infty} \!\!
\int_{-\infty}^{+\infty} 
\sum_{s=1}^{g} e_{s} \left[
{\delta q^{(s)}_{i} (x) \over \delta \varphi^{j}(y)} \, - \,
{\delta q^{(s)}_{j} (y) \over \delta \varphi^{i}(x)} \right] \,
q^{(s)}_{k} (z) \,\,\, 
\xi^{i}(x) \, \eta^{j}(y) \, \zeta^{k}(z) \,
dx \, dy \, dz \, =$$

$$= {1 \over 2}
\sum_{s=1}^{g} e_{s} \left[ \int_{-\infty}^{+\infty} \!\!
q^{(s)}_{k} (z) \, \zeta^{k}(z) \, dz \right]  \times \! 
\int_{-\infty}^{+\infty} \!\! \int_{-\infty}^{+\infty} 
\left[
{\delta q^{(s)}_{i} (x) \over \delta \varphi^{j}(y)} \, - \,
{\delta q^{(s)}_{j} (y) \over \delta \varphi^{i}(x)} \right]
\xi^{i}(x) \, \eta^{j}(y) \, dx \, dy \, \equiv  0 $$

 Let us use now the fact that the functions
$q^{(s)}_{i} (x) \, = \, 
q^{(s)}_{i} (\bm{\varphi}, \bm{\varphi}_{x}, \dots)$ 
are the local translationally invariant (i.e. they do not
depend explicitly on $x$) expressions depending on 
$\bm{\varphi}(x)$ and their derivatives. Let us consider
the functions $\varphi^{i}(x)$ which can be represented as

$$\varphi^{i}(x) \,\, = \,\, {\tilde \varphi}^{i}(x)
\,\, + \,\, {\tilde {\tilde \varphi}}^{i}(x) $$
where

$$Supp \,\, \tilde{\bm{\varphi}}(x) \,\, \subset \,\,
\bigcup_{k} \, Supp \,\, \zeta^{k}(x) $$

$$Supp \,\, \tilde{\tilde{\bm{\varphi}}}(x) \,\, \subset \,\,
\left[ \bigcup_{i} \, Supp \,\, \xi^{i}(x) \right] \bigcup
\left[ \bigcup_{j} \, Supp \,\, \eta^{j}(x) \right] $$

 Denote

$$A^{(s)}[\tilde{\tilde{\bm{\varphi}}}, \bm{\xi}, \bm{\eta}] 
\, = \, \int_{-\infty}^{+\infty} 
\int_{-\infty}^{+\infty} \left[ 
{\delta q^{(s)}_{i} 
(\tilde{\tilde{\bm{\varphi}}},\tilde{\tilde{\bm{\varphi}}}_{x}, 
\dots) \over \delta \tilde{\tilde{\varphi}}^{j}(y)} \, - \,
{\delta q^{(s)}_{j}  
(\tilde{\tilde{\bm{\varphi}}},\tilde{\tilde{\bm{\varphi}}}_{y},
\dots) \over \delta \tilde{\tilde{\varphi}}^{j}(x)} \right]
\xi^{i}(x) \, \eta^{j}(x) \, dx \, dy $$

 We have then

$$\sum_{s=1}^{g} e_{s} \, 
A^{(s)}[\tilde{\tilde{\bm{\varphi}}}, \bm{\xi}, \bm{\eta}]
\times \int_{-\infty}^{+\infty} 
q^{(s)}_{k} (\tilde{\bm{\varphi}}, 
\tilde{\bm{\varphi}}_{z}, \dots) \, \zeta^{k}(z) \, dz \,
\equiv \, 0 $$
(for all $\tilde{\bm{\varphi}}(z), \zeta^{k}(z)$)

 It's easy to show that for linearly independent set
${\bf q}^{(s)} 
(\tilde{\bm{\varphi}},\tilde{\bm{\varphi}}_{z}, \dots)$
this system can have only trivial solution 
$A^{(s)}[\tilde{\tilde{\bm{\varphi}}}, \bm{\xi}, \bm{\eta}]
\equiv 0$ for any $\xi^{i}(x)$, $\eta^{j}(y)$ and
$\tilde{\tilde{\bm{\varphi}}}$ which is equivalent to
condition 
$\left(d \, q^{(s)} \right)_{ij} (x,y) = 0 $ for any
$q^{(s)}_{i} (\bm{\varphi}, \bm{\varphi}_{x}, \dots)$.

{\hfill Theorem 1 is proved.}

\vspace{0.5cm}

 We will put now
$q^{(s)}_{i} (\bm{\varphi}, \bm{\varphi}_{x}, \dots)
\, = \, \delta H^{(s)}/\delta \varphi^{i}(x)$
where $H^{(s)}$ are some "local" functionals

$$H^{(s)}[\bm{\varphi}] \, = \, \int_{-\infty}^{+\infty}
h^{(s)} (\bm{\varphi}, \bm{\varphi}_{x}, \dots) \, dx $$
and $\delta/\delta \varphi^{i}(x)$ is the Euler-Lagrange
derivative and consider the structures (\ref{gensympstr}) 
in the form

\begin{equation}
\label{conssympform}
\Omega_{ij} (x,y) \, = \, \sum_{k \geq 0} 
\omega^{(k)}_{ij} (\bm{\varphi}, \bm{\varphi}_{x}, \dots) 
\, \delta^{(k)}(x-y) \, + \,
\sum_{s=1}^{g} \, e_{s} \,
{\delta H^{(s)} \over \delta \varphi^{i}(x)} \, \nu (x-y) \,
{\delta H^{(s)} \over \delta \varphi^{j}(y)}
\end{equation}

 Let us consider now the weakly-nonlocal Symplectic
Structures of Hydrodynamic Type (\ref{HTsympform}).

\vspace{0.5cm}

{\bf Theorem 2.}

{\it The expression (\ref{HTsympform}) gives the closed
2-form on the space $\{{\bf U} (X)\}$ if and only if the
1-forms $\omega_{\nu}^{(s)}({\bf U})$ on ${\cal M}^{N}$ 
are closed, \footnote{We assume that (\ref{HTsympform})
is written in the "irreducible" form, i.e. the 1-forms
$\omega_{\nu}^{(s)}({\bf U})$ are linearly independent 
(with constant coefficients).} i.e. 

$${\partial \over \partial U^{\nu}} 
\omega_{\mu}^{(s)}({\bf U}) \, = \,
{\partial \over \partial U^{\mu}} 
\omega_{\nu}^{(s)}({\bf U})$$
}

\vspace{0.5cm}
 
 Let us say here that the statement analogous to Theorem 2
was first proved by O.I. Mokhov for the weakly nonlocal 
Symplectic operators of Hydrodynamic Type having the form
${\hat \Omega}_{ij} \, = \, 
a_{i}({\bf U}) D^{-1} b_{j} ({\bf U}) \, + \,
b_{i}({\bf U}) D^{-1} a_{j}({\bf U})$ (see \cite{MokhDis,MokhSt}).
Theorem 2 represents the not difficult generalization of this
statement for the arbitrary number of terms in the non-local
structure (\ref{HTsympform}).

\vspace{0.5cm}

Proof.

Let us use the "diagonal" form of (\ref{HTsympform}). 
It's easy to see that the form (\ref{HTsympform}) is
skew-symmetric. From Theorem 1 we get that the forms
$\omega_{\nu}^{(s)}({\bf U})$ should be closed on the
functional space $\{{\bf U} (X)\}$. We have then:

$${\delta \omega_{\mu}^{(s)}({\bf U}(Y)) \over
\delta U^{\nu}(X)} \, - \,
{\delta \omega_{\nu}^{(s)}({\bf U}(X)) \over
\delta U^{\mu}(Y)} \, = \, 
{\partial \omega_{\mu}^{(s)}({\bf U}) \over
\partial U^{\nu}}(Y) \,\, \delta (Y-X) \, - \,
{\partial \omega_{\nu}^{(s)}({\bf U}) \over
\partial U^{\mu}}(X) \,\, \delta (X-Y) \, =  $$
$$= \, \left[{\partial \omega_{\mu}^{(s)}({\bf U}) 
\over \partial U^{\nu}}(X) \, - \,
{\partial \omega_{\nu}^{(s)}({\bf U}) \over
\partial U^{\mu}}(X)\right] \, \delta (X-Y) \,
\equiv \, 0$$

 So we have

$${\partial \omega_{\mu}^{(s)}({\bf U}) \over
\partial U^{\nu}} \, - \,
{\partial \omega_{\nu}^{(s)}({\bf U}) \over 
\partial U^{\mu}} \, \equiv \, 0$$

 It's not difficult now to get by direct calculation that
$\left( d \Omega \right)_{\nu\mu\lambda} (X,Y,Z)$ can be
written in the form

$$\left( d \Omega \right)_{\nu\mu\lambda} (X,Y,Z) 
\,\,\,\,\, = \,\,\,\,\,
\sum_{s=1}^{M} \, e_{s} \, \omega_{\nu}^{(s)}(X) \,
\nu (X-Y) \, \delta (Y-Z) \, \left[
{\partial \omega_{\mu}^{(s)} \over \partial 
U^{\lambda}}(Z) \, - \, 
{\partial \omega_{\lambda}^{(s)} \over \partial
U^{\mu}}(Z) \right] \, +$$
$$+ \, \sum_{s=1}^{M} \, e_{s} \, \omega_{\mu}^{(s)}(Y) \,
\nu (Y-Z) \, \delta (Z-X) \, \left[
{\partial \omega_{\lambda}^{(s)} \over \partial
U^{\nu}}(X) \, - \,
{\partial \omega_{\nu}^{(s)} \over \partial
U^{\lambda}}(X) \right] \, +$$
$$+ \, \sum_{s=1}^{M} \, e_{s} \, 
\omega_{\lambda}^{(s)}(Z) \,
\nu (Z-X) \, \delta (X-Y) \, \left[
{\partial \omega_{\nu}^{(s)} \over \partial
U^{\mu}}(Y) \, - \,
{\partial \omega_{\mu}^{(s)} \over \partial
U^{\nu}}(Y) \right]$$

 So we get the second part of the Theorem.

{\hfill Theorem 2 is proved.}

\vspace{0.5cm}

 We can put locally 
$\omega_{\nu}^{(s)}({\bf U}) \, = \,
\partial f^{(s)} ({\bf U})/ \partial U^{\nu}$ on
${\cal M}^{N}$ and write the Symplectic structure
(\ref{HTsympform}) in a "conservative form"

\begin{equation}
\label{consHTform}
\Omega_{\nu\mu}(X,Y) \, = \, \sum_{s=1}^{M} \, e_{s} \, 
{\partial f^{(s)} \over \partial U^{\nu}}(X) \, 
\nu (X-Y) \, {\partial f^{(s)} \over \partial U^{\mu}}(Y)
\end{equation}

We will usually consider the form $\Omega_{\nu\mu}(X,Y)$
on the loop space ${\cal L}_{P_{0}}$ such that 
$P_{0} \in {\cal M}^{N}$ is some fixed point of
${\cal M}^{N}$ and the functions 
${\bf U}(X) \rightarrow {P_{0}}$ (quickly enough)
for $X \rightarrow \pm \infty$. The action of 
$\Omega_{\nu\mu}(X,Y)$ will be usually defined on the
"vector fields" $\xi^{\nu}(X)$ rapidly decreasing for
$X \rightarrow \pm \infty$. 

\vspace{0.5cm}

 The 2-form $\Omega_{\nu\mu}(X,Y)$ written in the form
(\ref{consHTform}) can be considered as the pullback 
of the form

$$\Xi_{IJ} (X,Y) \, = \, e_{I} \, \delta_{IJ} \,
\nu (X-Y) \,\,\, , \,\,\, I,J = 1, \dots, M $$
defined in the pseudo-Euclidean space ${\mathbb E}^{N}$
with the metric 
$G_{IJ} \, = \, diag \, (e_{1}, \dots, e_{M})$
for the mapping $\alpha$: 
${\cal M}^{N} \rightarrow {\mathbb E}^{N}$

$$(U^{1}, \dots, U^{N}) \, \rightarrow \,
(f^{(1)} ({\bf U}), \dots, f^{(M)} ({\bf U})) $$

\vspace{0.5cm}

{\bf Definition 2.}

{\it We call the Symplectic Form (\ref{HTsympform})
non-degenerate if $M \geq N$ and

$$ rank \,\, \left( 
\begin{array}{c}
\omega^{(1)}_{i} ({\bf U})
\cr
\dots
\cr
\omega^{(M)}_{i} ({\bf U})
\end{array} \right) \,\,\, = \,\,\, N$$

}

 Easy to see that the non-degeneracy of 
$\Omega_{\nu\mu}(X,Y)$ coincides with the condition of
regularity of $N$-dimensional submanifold
$\alpha({\cal M}^{N}) \subset {\mathbb E}^{N}$
in the space ${\mathbb E}^{N}$ for $M \geq N$.

\section{The families of $m$-phase solutions and the
Whitham method.}
\setcounter{equation}{0}

 We will consider now the Whitham averaging method for
the local systems

\begin{equation}
\label{dynsyst}
{\varphi}^{i}_{t} \,\,\, = \,\,\,
Q^{i} (\bm{\varphi}, \bm{\varphi}_{x}, \dots)
\end{equation}
having the weakly nonlocal Symplectic Structure
(\ref{conssympform}) with a "local" Hamiltonian
functional

\begin{equation}
\label{hamiltonian}
H \,\,\, = \,\,\, \int_{-\infty}^{+\infty}
h (\bm{\varphi}, \bm{\varphi}_{x}, \dots) \, dx
\end{equation}

 This means that

$$\int_{-\infty}^{+\infty} \Omega_{ij}(x,y) \,
{\varphi}^{j}_{t}(y) \, dy \,\,\, = \,\,\,
\int_{-\infty}^{+\infty} \Omega_{ij}(x,y) \,
Q^{j} (\bm{\varphi}, \bm{\varphi}_{y}, \dots)
\, dy \,\,\, \equiv \,\,\,
{\delta H \over \delta \varphi^{i}(x)} $$
on ${\hat {\cal W}}_{0}$ where
$\delta/\delta \varphi^{i}(x)$ is the Euler-Lagrange
derivative.

 This requires in particular that
the functionals $H^{(s)}[\bm{\varphi}]$ are the
conservation laws for the system (\ref{dynsyst})
such that

\begin{equation}
\label{hjrel}
h^{(s)}_{t} \,\,\, \equiv \,\,\, \partial_{x} \,
J^{(s)} (\bm{\varphi}, \bm{\varphi}_{x}, \dots)
\end{equation}
for some functions 
$J^{(s)} (\bm{\varphi}, \bm{\varphi}_{x}, \dots) $.
The functional $H \, [\bm{\varphi}]$ is defined then
actually up to the linear combination of
$H^{(s)}[\bm{\varphi}]$ depending on the boundary
conditions at infinity.

 We assume now that the system (\ref{dynsyst}) has a
finite-parametric family of quasiperiodic solutions

\begin{equation}
\label{qps}
{\varphi}^{i}(x,t) \,\,\, = \,\,\,
\Phi^{i} \left({\bf k}({\bf U}) \, x \, + \,
\bm{\omega}({\bf U}) \, t \, + \, \bm{\theta}_{0},
\, {\bf U} \right) \,\,\,\,\, , \,\,\,\,\,
i \, = \, 1, \dots, n
\end{equation}
where
$\bm{\theta} \, = \, (\theta^{1}, \dots, \theta^{m})$,
${\bf k} \, = \, (k^{1}, \dots, k^{m})$,
$\bm{\omega}\, = \, (\omega^{1}, \dots, \omega^{m})$
and $\Phi^{i} (\bm{\theta}, {\bf U})$ give the family
of $2\pi$-periodic w.r.t. each $\theta^{\alpha}$
functions depending on the additional parameters
${\bf U} \, = \, (U^{1}, \dots, U^{N})$.

 The functions $\Phi^{i} (\bm{\theta}, {\bf U})$
satisfy the system

\begin{equation}
\label{phasesyst}
g^{i} \left(\bm{\Phi}, \, \omega^{\alpha}({\bf U}) \,
\bm{\Phi}_{\theta^{\alpha}}, \dots \right)
\,\,\, = \,\,\, \omega^{\alpha}({\bf U}) \,
\Phi^{i}_{\theta^{\alpha}} \, - \,
Q^{i} \left(\bm{\Phi}, \, k^{\alpha}({\bf U}) \,
\bm{\Phi}_{\theta^{\alpha}}, \dots \right)  
\,\,\, = \,\,\, 0
\end{equation}
and we assume that the system (\ref{phasesyst}) has the
finite-parametric family $\Lambda$ of solutions
(for generic ${\bf k}$ and $\bm{\omega}$) on the
space of $2\pi$-periodic w.r.t. each $\theta^{\alpha}$ 
functions with parameters
${\bf U} \, = \, (U^{1}, \dots, U^{N})$ and the
"initial phase shifts"
$\bm{\theta}_{0} \, = \,
(\theta_{0}^{1}, \dots, \theta_{0}^{m})$.
We can choose then (in a smooth way) at every
$(U^{1}, \dots, U^{N})$
some function $\bm{\Phi}(\bm{\theta}, {\bf U})$ as
having zero initial phase shifts and represent the
$m$-phase solutions of system (\ref{dynsyst}) in the
form (\ref{qps}).

 In Whitham method we make a rescaling
$X = \epsilon \, x$, $T = \epsilon \, t$  
($\epsilon \rightarrow 0$) of both variables $x$ and
$t$ and try to find a function

\begin{equation}
\label{Sfunc}
{\bf S}(X,T) \, = \, \left(
S^{1}(X,T), \dots, S^{m}(X,T) \right)
\end{equation}
and $2\pi$-periodic functions

\begin{equation}
\label{epsexp}
\Psi^{i} (\bm{\theta},X,T,\epsilon) \,\,\, = \,\,\,
\sum_{k \geq 0} \, \Psi_{(k)}^{i} (\bm{\theta},X,T) \,\,
\epsilon^{k}
\end{equation}
such that the functions

\begin{equation}
\label{whithsol}
\phi^{i} (\bm{\theta},X,T,\epsilon) \,\,\, = \,\,\,
\Psi^{i} \left({{\bf S}(X,T) \over
\epsilon} \, + \, \bm{\theta},X,T,\epsilon \right) 
\end{equation}
satisfy the system
 
\begin{equation}
\label{epssyst}
\epsilon \, \phi_{T}^{i} 
\,\,\, = \,\,\, Q^{i} \left(\bm{\phi},
\epsilon \, \bm{\phi}_{X}, \dots \right)
\end{equation}
at every $X$, $T$ and $\bm{\theta}$.

 It is easy to see that the function
$\bm{\Psi}_{(0)}(\bm{\theta},X,T)$ satisfies the
system (\ref{phasesyst}) at every $X$ and $T$ with

$$k^{\alpha} \,\,\, = \,\,\, S^{\alpha}_{X}
\,\,\,\,\, , \,\,\,\,\,
\omega^{\alpha} \,\,\, = \,\,\, S^{\alpha}_{T} $$
and so belongs at every $(X,T)$ to the family $\Lambda$.
We can write then

$$\Psi^{i}_{(0)}(\bm{\theta},X,T) \,\,\, = \,\,\,
\Phi^{i} (\bm{\theta} \, + \, \bm{\theta}_{0}(X,T),  
{\bf U}(X,T))$$

 We can introduce then the functions $U^{\nu}(X,T)$,
$\theta^{\alpha}_{0}(X,T)$ as the parameters characterizing
the main term in (\ref{epsexp}) which should satisfy to
condition

\begin{equation}
\label{comcond}
\left[k^{\alpha}({\bf U})\right]_{T} \,\,\, = \,\,\,
\left[\omega^{\alpha}({\bf U})\right]_{X}
\end{equation}
 
 We have to define now the functions
$\Psi_{(1)}^{i} (\bm{\theta},X,T)$ from the liner system

\begin{equation}
\label{firstappr}
{\hat L}^{i}_{j} \,
\Psi_{(1)}^{j} (\bm{\theta},X,T) \,\,\, = \,\,\,
f^{i}_{(1)} (\bm{\theta},X,T)
\end{equation}
where

$${\hat L}^{i}_{j} \,\,\,\,\, = \,\,\,\,\,
{\hat L}^{i}_{(X,T) \, j} \,\,\,\,\, = \,\,\,\,\,
\delta^{i}_{j} \, \omega^{\alpha} (X,T)
{\partial \over \partial \theta^{\alpha}} \, - \,
{\partial Q^{i} \over \partial \varphi^{j}}
\left(\bm{\Psi}_{(0)} (\bm{\theta},X,T), \dots
\right) \, - \,$$
\begin{equation}
\label{linop}  
- \,\,\, {\partial Q^{i} \over \partial \varphi^{j}_{x}}
\left(\bm{\Psi}_{(0)} (\bm{\theta},X,T), \dots
\right) \, k^{\alpha} (X,T) \,
{\partial \over \partial \theta^{\alpha}} \, - \, \dots
\end{equation}
is the linearization of system (\ref{phasesyst}) and
${\bf f}_{(1)}(\bm{\theta},X,T)$ is discrepancy
given by

$$f^{i}_{(1)} (\bm{\theta},X,T) \,\,\, = \,\,\,
- \, \Psi^{i}_{(0)T} (\bm{\theta},X,T) \, + \,  
{\partial Q^{i} \over \partial \varphi^{j}_{x}}
\left(\bm{\Psi}_{(0)} (\bm{\theta},X,T), \dots
\right) \, \Psi^{j}_{(0)X} (\bm{\theta},X,T) \, +  $$
\begin{equation}
\label{discreap}
+ \, {\partial Q^{i} \over \partial \varphi^{j}_{xx}}
\left(\bm{\Psi}_{(0)} (\bm{\theta},X,T), \dots
\right) \, \left( 2k^{\alpha}(X,T) \,
\Psi^{j}_{(0)\theta^{\alpha}X} \, + \,
k^{\alpha}_{X} \Psi^{j}_{(0)\theta^{\alpha}} \right)
\, + \, \dots
\end{equation}  
where

$${\partial \over \partial T} \,\,\, = \,\,\, 
U^{\nu}_{T} \, {\partial \over \partial U^{\nu}}
\,\,\, + \,\,\,
\theta^{\alpha}_{(0)T} \,
{\partial \over \partial \theta^{\alpha}}
\,\,\,\,\, , \,\,\,\,\,
{\partial \over \partial X} \,\,\, = \,\,\,
U^{\nu}_{X} \, {\partial \over \partial U^{\nu}}
\,\,\, + \,\,\,
\theta^{\alpha}_{(0)X} \,
{\partial \over \partial \theta^{\alpha}} $$   
for the functions

$$\Psi^{i}_{(0)} (\bm{\theta},X,T) \,\,\, = \,\,\,
\Phi^{i} (\bm{\theta} \, + \, \bm{\theta}_{0}(X,T),
{\bf U}(X,T))$$

 We will assume that $k^{\alpha}$ and $\omega^{\alpha}$
can be considered (locally) as the independent parameters
on the family $\Lambda$ and the total family of solutions
of (\ref{phasesyst}) depends (for generic $k^{\alpha}$,
$\omega^{\alpha}$) on $N \, = \, 2m \, + \, r$,
($r \geq 0$) parameters $U^{\nu}$ and $m$ initial
phases $\theta_{(0)}^{\alpha}$.

 Easy to see that the functions
$\bm{\Phi}_{\theta^{\alpha}} (\bm{\theta} \, + \,
\bm{\theta}_{0}(X,T), {\bf U}(X,T))$
and \linebreak
$\nabla_{\bm{\xi}} \, \bm{\Phi}_{\theta^{\alpha}}
(\bm{\theta} \, + \, \bm{\theta}_{0}(X,T), {\bf U}(X,T))$
where $\bm{\xi}$ is any vector in space of parameters
$U^{\nu}$ tangential to the surface
${\bf k} \, = \, const$, $\bm{\omega} \, = \, const$
belong to the kernel of operator
${\hat L}^{i}_{(X,T) \, j}$.

 Let us put now some "regularity" conditions on the family
(\ref{qps}) of quasiperiodic solutions of (\ref{dynsyst})

\vspace{0.5cm}
 
{\bf Definition 3.}

{\it We call the family (\ref{qps}) the full regular   
family of $m$-phase solutions of (\ref{dynsyst}) if:

1) The functions
$\bm{\Phi}_{\theta^{\alpha}} (\bm{\theta}, {\bf U})$,
$\bm{\Phi}_{U^{\nu}} (\bm{\theta}, {\bf U})$ are linearly
independent (almost everywhere) on the set $\Lambda$;

2) The $m \, + \, r$ linearly independent functions
$\bm{\Phi}_{\theta^{\alpha}} (\bm{\theta}, {\bf U})$,
$\nabla_{\bm{\xi}} \, \bm{\Phi} (\bm{\theta}, {\bf U})$  
($\nabla_{\bm{\xi}} {\bf k} \, = \, 0$,
$\nabla_{\bm{\xi}} \bm{\omega} \, = \, 0$)
give the full kernel of the operator
${\hat L}^{i}_{[{\bf U}] \, j}$ 
(here $\bm{\theta}_{0} \, = \, 0$) for generic
${\bf k}$ and $\bm{\omega}$.
 
3) There are exactly $m \, + \, r$ linearly independent  
"right eigen-vectors"
$\bm{\kappa}^{(q)}_{[{\bf U}]}(\bm{\theta})$,
$q = 1, \dots, m+r$ of the operator
${\hat L}^{i}_{[{\bf U}] \, j}$ (for generic
${\bf k}$ and $\bm{\omega}$) corresponding to zero
eigen values i.e.

$$\int_{0}^{2\pi}\!\!\!\dots\int_{0}^{2\pi}
\kappa_{[{\bf U}]i}^{(q)}(\bm{\theta}) \,
{\hat L}^{i}_{[{\bf U}] \, j} \,
\psi^{j} (\bm{\theta}) \,
{d^{m} \theta \over (2\pi)^{m}} \,\,\,
\equiv \,\,\, 0$$
for any periodic $\psi^{j} (\bm{\theta})$.
}

\vspace{0.5cm}

 We have then to put the $m \, + \, r$ conditions of
orthogonality of the discrepancy
${\bf f}_{(1)}(\bm{\theta},X,T)$
to the functions
$\bm{\kappa}^{(q)}_{[{\bf U}](X,T)}
(\bm{\theta}\, + \, \bm{\theta}_{0}(X,T))$

\begin{equation}
\label{ortcond}
\int_{0}^{2\pi}\!\!\!\dots\int_{0}^{2\pi}
\kappa^{(q)}_{[{\bf U}(X,T)]\, i}
(\bm{\theta}\, + \, \bm{\theta}_{0}(X,T)) \,
f^{i}_{(1)} (\bm{\theta},X,T) \,
{d^{m} \theta \over (2\pi)^{m}} \,\,\, = \,\,\, 0
\end{equation}
at every $X$, $T$ to be able to solve the system
(\ref{firstappr}) on the space of periodic w.r.t.
each $\theta^{\alpha}$ functions.

 The system (\ref{ortcond}) together with (\ref{comcond})
gives
$m \, + \, (m \, + \, r) \, = \, 2m \, + \, r \, = \, N$
conditions at each $X$ and $T$ on the parameters of zero
approximation $\bm{\Psi}_{(0)} (\bm{\theta},X,T)$
necessary for the construction of the first
$\epsilon$-term in the solution (\ref{epsexp}).
Let us prove now the following Lemma about the orthogonality
conditions (\ref{ortcond}):

\vspace{0.5cm}

{\bf Lemma 1.}  

{\it Under all the assumptions of regularity formulated above
the orthogonality conditions (\ref{ortcond}) do not contain
the functions $\theta^{\alpha}_{0}(X,T)$ and give just the restriction
on the functions $U^{\nu}(X,T)$ having the form

$$C^{(q)}_{\nu} ({\bf U}) \, U^{\nu}_{T} \,\, - \,\,
D^{(q)}_{\nu} ({\bf U}) \, U^{\nu}_{X} \,\, = \,\, 0 $$
(with some functions $C^{(q)}_{\nu} ({\bf U})$,  
$D^{(q)}_{\nu} ({\bf U})$).
}
 
\vspace{0.5cm}

 Proof.

 Let us write down the part
${\tilde {\bf f}}_{(1)} (\bm{\theta},X,T)$ of  
${\bf f}_{(1)} (\bm{\theta},X,T)$ which contains the derivatives
$\theta^{\alpha}_{0T}(X,T)$ and $\theta^{\alpha}_{0X}(X,T)$.
We have from (\ref{discreap})

$${\tilde f}^{i}_{(1)} (\bm{\theta},X,T) \,\,\, = \,\,\,
- \, \Psi^{i}_{(0)\theta^{\beta}} (\bm{\theta},X,T) \,
\theta^{\beta}_{0T} \,\, + \,\,
{\partial Q^{i} \over \partial \varphi^{j}_{x}}
\left(\bm{\Psi}_{(0)} (\bm{\theta},X,T), \dots \right) \,  
\Psi^{j}_{(0)\theta^{\beta}} (\bm{\theta},X,T) \,
\theta^{\beta}_{0X} \,\, + $$
$$+ \,\, {\partial Q^{i} \over \partial \varphi^{j}_{xx}}
\left(\bm{\Psi}_{(0)} (\bm{\theta},X,T), \dots \right) \,
2 k^{\alpha} (X,T) \,
\Psi^{j}_{(0)\theta^{\alpha}\theta^{\beta}} (\bm{\theta},X,T) \,
\theta^{\beta}_{0X} \,\, + \,\, \dots $$
 
 We can write then

$${\tilde f}^{i}_{(1)} (\bm{\theta},X,T) \,\,\, = \,\,\,
\left[ - \, {\partial \over \partial \omega^{\beta}} \,
g^{i} \left( \bm{\Phi} (\bm{\theta} \, + \, \bm{\theta}_{0},
{\bf U}), \dots \right) \,\, + \,\,
{\hat L}^{i}_{j} \, {\partial \over \partial \omega^{\beta}} \,
\Phi^{j} (\bm{\theta} \, + \, \bm{\theta}_{0},
{\bf U}), \dots ) \right] \, \theta^{\beta}_{0T} \,\, +$$   
$$+ \,\, \left[ {\partial \over \partial k^{\beta}} \,
g^{i} \left( \bm{\Phi} (\bm{\theta} \, + \, \bm{\theta}_{0},
{\bf U}), \dots \right) \,\, - \,\,
{\hat L}^{i}_{j} \, {\partial \over \partial k^{\beta}} \,
\Phi^{j} (\bm{\theta} \, + \, \bm{\theta}_{0},
{\bf U}), \dots ) \right] \, \theta^{\beta}_{0X} $$
where the constraints $g^{i}$ and the operator
${\hat L}^{i}_{(X,T)j}$ were introduced in (\ref{phasesyst})
and (\ref{linop}) respectively. 

 The derivatives $\partial g^{i}/\partial \omega^{\beta}$
and $\partial g^{i}/\partial k^{\beta}$ are identically zero
on $\Lambda$ according to (\ref{phasesyst}). We have then

$$\int_{0}^{2\pi}\!\!\!\dots\int_{0}^{2\pi}
\kappa^{(q)}_{[{\bf U}(X,T)]\, i}
(\bm{\theta}\, + \, \bm{\theta}_{0}(X,T)) \,
{\tilde f}^{i}_{(1)} (\bm{\theta},X,T) \,
{d^{m} \theta \over (2\pi)^{m}} \,\,\, \equiv \,\,\, 0 $$
since all $\bm{\kappa}^{(q)} (\bm{\theta},X,T)$ are the     
right eigen-vectors of ${\hat L}$ with zero eigen-values.

 It is easy to see also that all $\bm{\theta}_{0}(X,T)$
in the arguments of $\bm{\Phi}$ and $\bm{\kappa}^{(q)}$   
will disappear after the integration so we get the statement
of the Lemma.

{\hfill Lemma 1 is proved.}

\vspace{0.5cm}

 {\bf Remark.}

 As follows from the proof of Lemma 1 we will always have in
particular

$$\int_{0}^{2\pi}\!\!\!\dots\int_{0}^{2\pi}
\kappa^{(q)}_{[{\bf U}(X,T)]\, i}
(\bm{\theta}\, + \, \bm{\theta}_{0}(X,T)) \,
\Phi^{i}_{\theta^{\beta}} (\bm{\theta} \, + \,
\bm{\theta}_{0}(X,T), {\bf U}(X,T)) \,
{d^{m} \theta \over (2\pi)^{m}} \,\,\, \equiv \,\,\, 0 $$
for the case of full regular family of quasiperiodic solutions
(\ref{qps}).

\vspace{0.5cm}
 
 The Whitham system can now be written in the form

$${\partial k^{\alpha} \over \partial U^{\nu}} \,
U^{\nu}_{T} \,\, = \,\,
{\partial \omega^{\alpha} \over \partial U^{\nu}} \,
U^{\nu}_{X} \,\,\, , \,\,\,\,\, \alpha \, = \, 1, \dots, m $$
\begin{equation}
\label{wsyst1}
C^{(q)}_{\nu} ({\bf U}) \, U^{\nu}_{T} \,\, = \,\,
D^{(q)}_{\nu} ({\bf U}) \, U^{\nu}_{X} \,\,\, , \,\,\,\,\,
q \, = \, 1, \dots, m+r
\end{equation}
where $rank ||\partial k^{\alpha}/\partial U^{\nu}|| = m$
according to our assumption above. In the generic case the
derivatives $U^{\nu}_{T}$ can be expressed through $U^{\mu}_{X}$
and the Whitham system (\ref{wsyst1}) can be written in the form
(\ref{HTsyst}).

\vspace{0.5cm}

 Let us say that the method described above is not the only one
to get the Whitham system for the system (\ref{dynsyst}). 
In particular, the method of the averaging of conservation 
laws (\cite{whith,ffm,dn1,krichev1,krichev2,dn2,krichev3,dn3,dm})
gives also another way to get the system for the slow
modulations of parameters ${\bf U}(X,T)$. It can be shown that
both these methods give the equivalent systems (\ref{HTsyst})
for the parameters ${\bf U}(X,T)$ (in regular situation).
Thus the averaged conservations laws give then the additional
conservations laws for the system (\ref{wsyst1}).

 We will get here the Symplectic representation of the
conditions of compatibility of the system (\ref{firstappr})
which is also equivalent to (\ref{wsyst1}) in the generic
case. In general we can state that the system (\ref{wsyst1})
admits the averaged Symplectic structure in the sense
discussed above.

\vspace{0.5cm}

 Let us put now some special conditions connected with
"invariant tori" corresponding to quasiperiodic solutions
(\ref{qps}) which we will need for the averaging of the
Symplectic Structure (\ref{conssympform}). Namely, we will
require that we have $m$ linearly independent local flows

\begin{equation}
\label{commfl}
\varphi^{i}_{t^{\alpha}} \, = \, Q^{i}_{(\alpha)}   
(\bm{\varphi}, \bm{\varphi}_{x}, \dots)
\end{equation}  
(which can contain the system (\ref{dynsyst})) which commute
with (\ref{dynsyst}) and admit the same Symplectic Structure
(\ref{conssympform}) with some local Hamiltonian functions
$F_{(\alpha)}[\bm{\varphi}]$, i.e.

$$ \int_{-\infty}^{+\infty} \Omega_{ij}(x,y) \,
Q^{i}_{(\alpha)} (\bm{\varphi}, \bm{\varphi}_{y}, \dots) \,
dy \,\, \equiv \,\,
{\delta \over \delta \varphi^{i}(x)} F_{(\alpha)}$$
where

$$F_{(\alpha)}[\bm{\varphi}] \,\, = \,\,
\int_{-\infty}^{+\infty} f_{(\alpha)}
(\bm{\varphi}, \bm{\varphi}_{x}, \dots) \, dx $$

 This means automatically that the functionals
$H^{(s)} [\bm{\varphi}]$ should give the conservation laws
for the systems (\ref{commfl}) also and we can write

\begin{equation}
\label{jsa}   
h^{(s)}_{t^{\alpha}} \,\, \equiv \,\,
\partial_{x} \, J^{(s)}_{\alpha}
(\bm{\varphi}, \bm{\varphi}_{x}, \dots)
\end{equation}
for some functions
$J^{(s)}_{\alpha} (\bm{\varphi}, \bm{\varphi}_{x}, \dots)$.

 We will require that the flows (\ref{commfl}) generate the
"linear shifts" of the angles $\theta_{0}^{\beta}$ on the
solutions (\ref{qps}) with some frequences
$\omega^{\beta}_{(\alpha)} ({\bf U})$ such that the matrix
$||\omega^{\beta}_{(\alpha)}||$ is non-degenerate, i.e. we
have 

\begin{equation}
\label{commphasesyst}
\omega^{\beta}_{(\alpha)} ({\bf U}) \,
\Phi^{i}_{\theta^{\beta}} \,\, = \,\,
Q^{i}_{(\alpha)} \left(\bm{\Phi}, \, k^{\delta}({\bf U}) \,
\bm{\Phi}_{\theta^{\delta}}, \dots \right)
\end{equation}
with
$det \, ||\omega^{\beta}_{(\alpha)} ({\bf U})|| \, \neq \, 0$.

 Let us denote $||\gamma^{\beta}_{\alpha}||$ the inverse
matrix $||\omega^{\beta}_{(\alpha)}||^{-1}$ such that

\begin{equation}
\label{gammamatr} 
\gamma^{\delta}_{\alpha} ({\bf U}) \,
\omega^{\beta}_{(\delta)} ({\bf U}) \,\,\, = \,\,\,
\delta^{\beta}_{\alpha}
\end{equation}

 We can write also

\begin{equation}
\label{phaserel}
\Phi^{i}_{\theta^{\alpha}} \,\, = \,\,
\gamma^{\beta}_{\alpha} ({\bf U}) \,\,
Q^{i}_{(\beta)} \left(\bm{\Phi}, \, k^{\delta}({\bf U}) \,
\bm{\Phi}_{\theta^{\delta}}, \dots \right)
\end{equation}
on the family (\ref{qps}).

\section{The extended phase space and some technical Lemmas.}
\setcounter{equation}{0}

 In this chapter we will prove some technical Lemmas concerning
the form (\ref{conssympform}) on the "extended" functional
space. As we said already, we consider the form
(\ref{conssympform}) on the loop space ${\cal W}_{0}$
of functions $\varphi^{i}(x)$ rapidly decreasing or
approaching some fixed constants $C^{i}$ for
$x \rightarrow \pm \infty$. Let us define now the extended
space ${\hat {\cal W}}_{0}$ of smooth functions
$\varphi^{i}(\bm{\theta},x)$
($\bm{\theta} = (\theta^{1}, \dots, \theta^{m})$),
$2\pi$-periodic w.r.t. each $\theta^{\alpha}$ and
approaching the same constants $C^{i}$ at each $\bm{\theta}$
for $x \rightarrow \pm \infty$. We define the "extended"
Symplectic Form
${\tilde \Omega}_{ij}(\theta, \theta^{\prime}, x, y)$
by the formula

$${\tilde \Omega}_{ij}(\bm{\theta},\bm{\theta}^{\prime}, x, y)
\, = \, \sum_{k \geq 0} \, \omega^{(k)}_{ij}
(\bm{\varphi}(\bm{\theta},x),
\bm{\varphi}_{x}(\bm{\theta},x), \dots) \,
\delta^{(k)}(x-y) \,
\delta (\bm{\theta} - \bm{\theta}^{\prime}) \, + $$
\begin{equation}
\label{extform}
+ \, \sum_{s=1}^{g} \, e_{s} \,
{\delta {\tilde H}^{(s)} \over
\delta \varphi^{i}(\bm{\theta},x)} \, \nu (x-y) \,
\delta (\bm{\theta} - \bm{\theta}^{\prime}) \,
{\delta {\tilde H}^{(s)} \over
\delta \varphi^{j}(\bm{\theta}^{\prime},y)} \,\,\,\,\, ,
\,\,\,\,\,\,\,\, i,j \, = \, 1, \dots, n
\end{equation}
where the functionals ${\tilde H}^{(s)}$ are defined on 
${\hat {\cal W}}_{0}$ by the formula
\footnote{We can always normalize the densities $h^{(s)}$
such that $h^{(s)}({\bf C},0,\dots) = 0$.}

$${\tilde H}^{(s)}[\bm{\varphi}] \, = \,
\int_{-\infty}^{+\infty}\int_{0}^{2\pi}\!\!\!\dots\int_{0}^{2\pi}
h^{(s)} (\bm{\varphi}(\bm{\theta},x),
\bm{\varphi}_{x}(\bm{\theta},x), \dots) \,
{d^{m} \theta \over (2\pi)^{m}} \, dx $$

 Let us note also that we normalize
$\delta (\bm{\theta}^{\prime} - \bm{\theta})$ such that

$$\int_{0}^{2\pi}\!\!\!\dots\int_{0}^{2\pi} \,
\delta (\bm{\theta}^{\prime} - \bm{\theta}) \,
{d^{m} \theta^{\prime} \over (2\pi)^{m}} \, = \, 1 $$

 Easy to see that (\ref{extform}) gives the closed 2-form
on ${\hat {\cal W}}_{0}$. Let us prove now the first
technical Lemma which we will need later.

\vspace{0.5cm} 

{\bf Lemma 2.}

{\it For any $\alpha, \beta \, = \, 1, \dots, m$ we have

$$C_{\alpha\beta}[\bm{\varphi}] \, = $$
$$=\, \int_{-\infty}^{+\infty}\int_{-\infty}^{+\infty}
\int_{0}^{2\pi}\!\!\!\dots\int_{0}^{2\pi} \,
\varphi_{\theta^{\alpha}}^{i}(\bm{\theta},x) \,
{\tilde \Omega}_{ij}
(\bm{\theta}, \bm{\theta}^{\prime}, x, y) \,
\varphi_{\theta^{\prime\beta}}^{j}(\bm{\theta}^{\prime},x)
\, dx \, dy \, {d^{m} \theta \over (2\pi)^{m}} \,
{d^{m} \theta^{\prime} \over (2\pi)^{m}} \,\,\, \equiv
\,\,\, 0 $$
on ${\hat {\cal W}}_{0}$.
}
 
\vspace{0.5cm}

Proof.

Let us first prove the relation

$${\delta C_{\alpha\beta}[\bm{\varphi}] \over
\delta \varphi^{i}(\bm{\theta},x)} \, \equiv \, 0 $$

We will use the infinite-dimensional form of the relation

$${\partial \over \partial x^{i}} \langle
\bm{\xi} \omega \bm{\eta} \rangle \,\,\, = \,\,\,
\left[
{\cal L}_{\bm{\xi}} \langle \omega \bm{\eta} \rangle
\right]_{i} \, - \,
\left[
{\cal L}_{\bm{\eta}} \langle \omega \bm{\xi} \rangle  
\right]_{i} \, - \,
\langle \omega \left[\bm{\xi}, \bm{\eta}\right]
\rangle_{i}$$
which is valid for the closed form $\omega_{ij}(x)$
on a manifold and any vector fields $\xi^{i}(x)$ and
$\eta^{k}(x)$. The notations
$\langle \bm{\xi} \omega \bm{\eta} \rangle$,
$\langle \omega \bm{\xi} \rangle$ and
$\langle \omega \bm{\eta} \rangle$ mean here the function
$\xi^{j} \, \omega_{jk} \, \eta^{k}$ and the 1-forms
$\omega_{jk} \, \xi^{k}$ and $\omega_{jk} \, \eta^{k}$
respectively. The operators ${\cal L}_{\bm{\xi}}$ and
${\cal L}_{\bm{\eta}}$ are the Lie derivatives w.r.t.
vector fields ${\bm{\xi}}$ and ${\bm{\eta}}$ and
$\left[\bm{\xi}, \bm{\eta}\right]$ is the commutator of
${\bm{\xi}}$ and ${\bm{\eta}}$.

 Indeed, we have for any closed $\omega_{ij}(x)$:

$${\partial \over \partial x^{i}} \left(
\xi^{j} \, \omega_{jk} \, \eta^{k} \right) \, = \,
{\partial \xi^{j} \over \partial x^{i}}
\, \omega_{jk} \, \eta^{k} \, + \,
\xi^{j} \, {\partial \omega_{jk} \over \partial x^{i}} \,
\eta^{k} \, + \, \xi^{j} \, \omega_{jk} \,
{\partial \eta^{k} \over \partial x^{i}} \, = $$   
$$= \, {\partial \xi^{j} \over \partial x^{i}}
\, \omega_{jk} \, \eta^{k} \, + \,
\xi^{j} \, \omega_{jk} \,
{\partial \eta^{k} \over \partial x^{i}} \, - \,
\xi^{j} \, \left(
{\partial \omega_{ki} \over \partial x^{j}} \, + \, 
{\partial \omega_{ij} \over \partial x^{k}} \right) \,
\eta^{k} \, = $$
$$= \, {\partial \xi^{j} \over \partial x^{i}}
\, \omega_{jk} \, \eta^{k} \, + \,
\xi^{j} \, {\partial \over \partial x^{j}} \left[
\omega_{ik} \, \eta^{k} \right] \, - \,
\xi^{j} \, \omega_{jk} \,
{\partial \eta^{k} \over \partial x^{i}} \, - \, 
\eta^{k} \, {\partial \over \partial x^{k}}
\left[ \omega_{ij} \, \xi^{j} \right] \, - \,
\omega_{ik} \, \xi^{j}
{\partial \eta^{k} \over \partial x^{j}} \, + \,
\omega_{ij} \, \eta^{k} \,
{\partial \xi^{j} \over \partial x^{k}} \, = $$
$$= \, \left[
{\cal L}_{\bm{\xi}} \langle \omega \bm{\eta} \rangle
\right]_{i} \, - \,
\left[
{\cal L}_{\bm{\eta}} \langle \omega \bm{\xi} \rangle
\right]_{i} \, - \,
\langle \omega \left[\bm{\xi}, \bm{\eta}\right]
\rangle_{i}$$
(we assume summation over the repeated indices).

 In our case $\partial/\partial x^{i}$ should be replaced
by $\delta/\delta \varphi^{i} (\bm{\theta},x)$ and we can
define the vector fields

$$\xi^{i} (\bm{\theta},x) [\bm{\varphi}] \, = \,
\varphi_{\theta^{\alpha}}^{i} \,\,\,\,\, ,
\,\,\,\,\,\,\,\,
\eta^{i} (\bm{\theta},x) [\bm{\varphi}] \, = \,
\varphi_{\theta^{\beta}}^{i}$$
and the corresponding dynamical systems on
${\hat {\cal W}}_{0}$

$$\varphi^{i}_{t_{1}} \, = \,
\varphi_{\theta^{\alpha}}^{i}
\,\,\,\,\, , \,\,\,\,\,\,\,\,
\varphi^{i}_{t_{2}} \, = \,
\varphi_{\theta^{\beta}}^{i}$$
(let us remind that $x$ and $\bm{\theta}$ play now the
role of "indices" also).

 Easy to see that the fields $\bm{\xi} [\bm{\varphi}]$
and $\bm{\eta} [\bm{\varphi}]$ commute with each other.
 
 The expression $C_{\alpha\beta}[\bm{\varphi}]$ can now  
be written as
$\langle \bm{\xi} {\tilde \Omega} \bm{\eta} \rangle$
and we can write

$${\delta C_{\alpha\beta}[\bm{\varphi}] \over
\delta \varphi^{i}(\bm{\theta},x)} \, = \,
\left[{\cal L}_{\bm{\xi}}
\langle {\tilde \Omega} \bm{\eta} \rangle \right]_{i}
(\bm{\theta},x) \, - \,
\left[{\cal L}_{\bm{\eta}}
\langle {\tilde \Omega} \bm{\xi} \rangle \right]_{i}
(\bm{\theta},x)$$
where
$$\langle {\tilde \Omega} \bm{\xi} \rangle_{i}
(\bm{\theta},x) \, = \, \int_{-\infty}^{+\infty}
\int_{0}^{2\pi}\!\!\!\dots\int_{0}^{2\pi} \,
{\tilde \Omega}_{ij}
(\bm{\theta}, \bm{\theta}^{\prime}, x, y) \,
\varphi_{\theta^{\alpha}}^{j}(\bm{\theta}^{\prime}, y) \,
{d^{m} \theta^{\prime} \over (2\pi)^{m}} \, dy \, =$$  
\begin{equation}
\label{omxi}
= \, \int_{-\infty}^{+\infty} \,
\Omega_{ij} (\bm{\theta}, x, y) \,
\varphi_{\theta^{\alpha}}^{j}(\bm{\theta}, y) \, dy
\end{equation}

 Also
\begin{equation}
\label{ometa}
\langle {\tilde \Omega} \bm{\eta} \rangle_{i}
(\bm{\theta},x) \, = \, \int_{-\infty}^{+\infty} \,
\Omega_{ij} (\bm{\theta}, x, y) \,
\varphi_{\theta^{\beta}}^{j}(\bm{\theta}, y) \, dy
\end{equation}
where $\varphi^{i}(\bm{\theta}, x)$,
$\varphi^{j}(\bm{\theta}, y)$ are considered just as the
functions of $x$ and $y$ at any fixed $\bm{\theta}$.

 The operations of Lie derivatives
$\left[{\cal L}_{\bm{\xi}}
{\bf q}\right]_{i}(\bm{\theta},x)$ and
$\left[{\cal L}_{\bm{\eta}}
{\bf q}\right]_{i}(\bm{\theta},x)$ for any 1-form
$q_{i}(\bm{\theta},x)$ can be written as

$$\left[{\cal L}_{\bm{\xi}}
{\bf q}\right]_{i}(\bm{\theta},x) \, = \,
\int_{-\infty}^{+\infty}\int_{0}^{2\pi}\!\!\!\dots\int_{0}^{2\pi}
\, \varphi^{k}_{\theta^{\prime\alpha}}
(\bm{\theta}^{\prime}, z) \,
{\delta \over \delta  \varphi^{k}
(\bm{\theta}^{\prime}, z)} \, q_{i}(\bm{\theta},x) \,
{d^{m} \theta^{\prime} \over (2\pi)^{m}} \, dz \,\, + $$
$$+ \, \int_{-\infty}^{+\infty}   
\int_{0}^{2\pi}\!\!\!\dots\int_{0}^{2\pi}
\, q_{k}(\bm{\theta^{\prime}},z) \,
{\delta \varphi_{\theta^{\prime\alpha}}^{k}
(\bm{\theta}^{\prime},z)
\over \delta
\varphi^{i} (\bm{\theta},x)} \,
{d^{m} \theta^{\prime} \over (2\pi)^{m}} \, dz $$
where

$${\delta \varphi_{\theta^{\prime\alpha}}^{k}
(\bm{\theta}^{\prime},z)
\over \delta
\varphi^{i} (\bm{\theta},x)} \, = \,
\delta^{k}_{i} \, \delta_{\theta^{\prime\alpha}}
(\bm{\theta}^{\prime} - \bm{\theta}) \, \delta (z-x) $$

 We have so

$$\left[{\cal L}_{\bm{\xi}}
{\bf q}\right]_{i}(\bm{\theta},x) \, = \, -
{\partial \over \partial \theta^{\alpha}}
q_{i}(\bm{\theta},x) \, + \, \int_{-\infty}^{+\infty}
\int_{0}^{2\pi}\!\!\!\dots\int_{0}^{2\pi}
\, \varphi^{k}_{\theta^{\prime\alpha}}
(\bm{\theta}^{\prime}, z) \,
{\delta q_{i}(\bm{\theta},x) \over \delta  \varphi^{k}
(\bm{\theta}^{\prime}, z)} \,
{d^{m} \theta^{\prime} \over (2\pi)^{m}} \, dz $$
which is zero if $q_{i}(\bm{\theta},x)$ does not contain
the explicit dependence on $\bm{\theta}$. (The same for
$\left[{\cal L}_{\bm{\eta}}
{\bf q}\right]_{i}(\bm{\theta},x)$).  

 Using the expressions (\ref{omxi}), (\ref{ometa})
we see that both the forms
$\langle {\tilde \Omega} \bm{\xi} \rangle_{i} 
(\bm{\theta},x)$ and
$\langle {\tilde \Omega} \bm{\eta} \rangle_{i}
(\bm{\theta},x)$
do not depend explicitly on $\bm{\theta}$ so we get
$\delta C_{\alpha\beta}[\bm{\varphi}]/
\delta \varphi^{i}(\bm{\theta},x) \, \equiv \, 0 $
on ${\hat {\cal W}}_{0}$.

 Using now the fact that
$C_{\alpha\beta}[\bm{\varphi}] \, \equiv \, 0 $
on the functions $\varphi^{i}(\bm{\theta},x)$ which
are constants w.r.t. $\bm{\theta}$ at any given $x$ we
get the proof of the Lemma.  

{\hfill Lemma 2 is proved.}

\vspace{0.5cm}

 Let us introduce the nonlocal functionals

$$W^{(s)}(\bm{\theta},x) [\bm{\varphi}] \, = \,
D^{-1} h^{(s)} (\bm{\varphi}, \bm{\varphi}_{x}, \dots)
\, = $$
\begin{equation}
\label{WS}
= \, {1 \over 2} \int_{-\infty}^{x}
h^{(s)} (\bm{\varphi}(\bm{\theta},y), 
\bm{\varphi}_{y}(\bm{\theta},y), \dots) \, dy \, - \,
{1 \over 2} \int_{x}^{+\infty}
h^{(s)} (\bm{\varphi}(\bm{\theta},y),
\bm{\varphi}_{y}(\bm{\theta},y), \dots) \, dy
\end{equation}

 Easy to see that for any $\bm{\varphi}(\bm{\theta},x)$
the functions $W^{(s)}(\bm{\theta},x)$ are
$2\pi$-periodic w.r.t. each $\theta^{\alpha}$ and we
have also

\begin{equation}
\label{antsym}
W^{(s)}(\bm{\theta},-\infty) \, = \, - \, 
W^{(s)}(\bm{\theta},+\infty)
\end{equation}
on ${\hat {\cal L}}_{0}$ at any fixed $\bm{\theta}$.  

 We will need also the following simple Proposition:

\vspace{0.5cm}

{\bf Proposition 1.}

{\it The expressions

$$h^{(s)}_{\theta^{\alpha}} \, - \,
{\delta {\tilde H}^{(s)} \over
\delta \varphi^{i}(\bm{\theta},x)}
\varphi_{\theta^{\alpha}}^{i}$$
can be written as total derivatives w.r.t. $x$ of the
local functions
$T_{\alpha}^{(s)}(\bm{\varphi}, \bm{\varphi}_{x}, \dots)$,
i.e.

\begin{equation}
\label{xderivative}
h^{(s)}_{\theta^{\alpha}} \, - \,
{\delta {\tilde H}^{(s)} \over
\delta \varphi^{i}(\bm{\theta},x)}
\varphi_{\theta^{\alpha}}^{i} \, \equiv \,
{d \over dx}
T_{\alpha}^{(s)}(\bm{\varphi}, \bm{\varphi}_{x}, \dots)
\end{equation}
where

\begin{equation}
\label{tsa}
T_{\alpha}^{(s)}(\bm{\varphi}, \bm{\varphi}_{x}, \dots)
\, = \, \sum_{k \geq 1} \, \sum_{p=0}^{k-1} \, (-1)^{p} \,
\left(
{\partial h^{(s)} \over \partial \varphi^{i}_{kx} }
\right)_{px} \, \varphi^{i}_{\theta^{\alpha},(k-p-1)x}
\end{equation}
}

\vspace{0.5cm} 

Proof.

Using the formulas

$$h^{(s)}_{\theta^{\alpha}} \, = \,
{\partial h^{(s)} \over \partial \varphi^{i} }
\varphi^{i}_{\theta^{\alpha}} \, + \,
{\partial h^{(s)} \over \partial \varphi^{i}_{x} }
\varphi^{i}_{\theta^{\alpha},x} \, + \, \dots $$

$${\delta {\tilde H}^{(s)} \over
\delta \varphi^{i}(\bm{\theta},x)} \, = \,
{\partial h^{(s)} \over \partial \varphi^{i} }
\, - \, \left(
{\partial h^{(s)} \over \partial \varphi^{i}_{x} }
\right)_{x} \, + \, \dots $$
we get the required statement just by direct calculation.

{\hfill Proposition 1 is proved.}

\vspace{0.5cm} 

 Let us prove now another important Lemma.

\vspace{0.5cm}

{\bf Lemma 3.}

{\it 1) For any Symplectic Form (\ref{conssympform}) we have
the relations

$$\varphi^{i}_{\theta^{\alpha}} \sum_{k \geq 0}
\omega^{(k)}_{ij} (\bm{\varphi}, \bm{\varphi}_{x}, \dots) \,
\varphi^{j}_{\theta^{\beta},kx} + \sum_{s=1}^{g} \, e_{s} \,
\left( h^{(s)}_{\theta^{\beta}} T^{(s)}_{\alpha} -
h^{(s)}_{\theta^{\alpha}} T^{(s)}_{\beta} +
\left(T^{(s)}_{\alpha}\right)_{x} T^{(s)}_{\beta} \right)
\,\,\, \equiv $$
\begin{equation}
\label{divform}
\equiv \,\,\, {\partial \over \partial \theta^{\gamma}}
Q^{\gamma}_{\alpha\beta} (\bm{\varphi}, \dots) +
{\partial \over \partial x}
A_{\alpha\beta} (\bm{\varphi}, \dots)
\end{equation}
for some local functions
$Q^{\gamma}_{\alpha\beta} (\bm{\varphi}, \dots)$,
$A_{\alpha\beta} (\bm{\varphi}, \dots)$ (summation over
the repeated indices).

 2) The functions $A_{\alpha\beta} (\bm{\varphi}, \dots)$
(defined modulo the constant functions) can be normalized   
in such a way that
$A_{\alpha\beta} (\bm{\varphi}, \dots) \equiv 0$ for any
$\varphi(\bm{\theta}, x)$ depending on $x$ only (and constant
with respect to $\bm{\theta}$ at every fixed $x$).
}

\vspace{0.5cm}

Proof.

1) Let us consider the values

$$F_{\alpha\beta}(\bm{\theta},x) \, = \,
\varphi^{i}_{\theta^{\alpha}}(\bm{\theta},x) \,
\int_{-\infty}^{+\infty} 
\int_{0}^{2\pi}\!\!\!\dots\int_{0}^{2\pi}
\tilde{\Omega}_{ij}(\bm{\theta},\bm{\theta}^{\prime},x,y) \,
\varphi^{j}_{\theta^{\prime\beta}}(\bm{\theta}^{\prime},y) \,
{d^{m} \theta^{\prime} \over (2\pi)^{m}} \, dy $$

 We have according to Lemma 2:

$$\int_{-\infty}^{+\infty} 
\int_{0}^{2\pi}\!\!\!\dots\int_{0}^{2\pi}
F_{\alpha\beta}(\bm{\theta},x) \,
{d^{m} \theta \over (2\pi)^{m}} \, dx \,\,\, \equiv \,\,\, 0$$

 We have from the other hand

$$F_{\alpha\beta}(\bm{\theta},x) \, = \,
\varphi^{i}_{\theta^{\alpha}} \sum_{k \geq 0}
\omega^{(k)}_{ij} (\bm{\varphi}, \bm{\varphi}_{x}, \dots) \,
\varphi^{j}_{\theta^{\beta},kx} \, + $$
$$+ \, \sum_{s=1}^{g} \, e_{s} \,
\left( \varphi^{i}_{\theta^{\alpha}} \,
{\delta {\tilde H}^{(s)} \over \delta
\varphi^{i}(\bm{\theta},x)} \,
\int_{-\infty}^{+\infty} \nu (x-y) \,   
{\delta {\tilde H}^{(s)} \over \delta
\varphi^{j}(\bm{\theta},y)} \,
\varphi^{j}_{\theta^{\beta}} \, dy \right)$$

 According to Proposition 1 we can write

$$F_{\alpha\beta}(\bm{\theta},x) \, = \,
\varphi^{i}_{\theta^{\alpha}} \sum_{k \geq 0}
\omega^{(k)}_{ij} (\bm{\varphi}, \bm{\varphi}_{x}, \dots) \,  
\varphi^{j}_{\theta^{\beta},kx} \, + $$
$$+ \, \sum_{s=1}^{g} \, e_{s} \,
\left( h^{(s)}_{\theta^{\alpha}} -
\left( T^{(s)}_{\alpha} \right)_{x} \right)
\int_{-\infty}^{+\infty} \nu (x-y) \,
\left( h^{(s)}_{\theta^{\beta}} -
\left( T^{(s)}_{\beta} \right)_{y} \right) \, dy \, = $$
$$= \, \varphi^{i}_{\theta^{\alpha}} \sum_{k \geq 0}
\omega^{(k)}_{ij} \, \varphi^{j}_{\theta^{\beta},kx} \, +
\sum_{s=1}^{g} \, e_{s} \, \left(
W^{(s)}_{\theta^{\alpha}x} W^{(s)}_{\theta^{\beta}} \, - \,
W^{(s)}_{\theta^{\alpha}x} T^{(s)}_{\beta} \, - \,
W^{(s)}_{\theta^{\beta}} \left( T^{(s)}_{\alpha} \right)_{x}
\, + \, \left( T^{(s)}_{\alpha} \right)_{x} T^{(s)}_{\beta}
\right) $$
(we used here the relations (\ref{antsym}) at infinity).

 We can rewrite now $F_{\alpha\beta}(\bm{\theta},x)$
in the following form

$$F_{\alpha\beta}(\bm{\theta},x) \, = \,
\varphi^{i}_{\theta^{\alpha}} \sum_{k \geq 0}
\omega^{(k)}_{ij} \, \varphi^{j}_{\theta^{\beta},kx} \, + \,  
\sum_{s=1}^{g} \, e_{s} \, \left[
W^{(s)}_{\theta^{\beta}x} \, T^{(s)}_{\alpha} \, - \,
W^{(s)}_{\theta^{\alpha}x} \, T^{(s)}_{\beta} \, + \,
\left( T^{(s)}_{\alpha} \right)_{x} \, T^{(s)}_{\beta}
\right] \, + $$
$$+ \, \sum_{s=1}^{g} \, e_{s} \, \left[
{1 \over 2} \left( W^{(s)}_{x} \, W^{(s)}_{\theta^{\beta}}
\right)_{\theta^{\alpha}} \, - \,
{1 \over 2} \left( W^{(s)}_{x} \, W^{(s)}_{\theta^{\alpha}}
\right)_{\theta^{\beta}} \, + \, 
{1 \over 2} \left( W^{(s)}_{\theta^{\alpha}} \,
W^{(s)}_{\theta^{\beta}} \right)_{x}  \, - \,
\left( W^{(s)}_{\theta^{\beta}} \,
T^{(s)}_{\alpha} \right)_{x} \right] $$

 Easy to see that

$$\sum_{s=1}^{g} \, e_{s} \,
\int_{-\infty}^{+\infty} 
\int_{0}^{2\pi}\!\!\!\dots\int_{0}^{2\pi}
{1 \over 2} \left[
\left( W^{(s)}_{x} \, W^{(s)}_{\theta^{\beta}}
\right)_{\theta^{\alpha}} \, - \,
\left( W^{(s)}_{x} \, W^{(s)}_{\theta^{\alpha}}
\right)_{\theta^{\beta}} \right] \,
{d^{m} \theta \over (2\pi)^{m}} \, dx \, \equiv \, 0$$
view the periodicity of $W^{(s)} (\bm{\theta},x)$
w.r.t. all $\theta^{\alpha}$ and

$$\int_{-\infty}^{+\infty}
\int_{0}^{2\pi}\!\!\!\dots\int_{0}^{2\pi}
\left[ {1 \over 2}  \left( W^{(s)}_{\theta^{\alpha}} \,
W^{(s)}_{\theta^{\beta}} \right)_{x}  \, - \,
\left( W^{(s)}_{\theta^{\beta}} \,
T^{(s)}_{\alpha} \right)_{x} \right] \,
{d^{m} \theta \over (2\pi)^{m}} \, dx \, = $$
\begin{equation}
\label{twoterms}
= \, \int_{0}^{2\pi}\!\!\!\dots\int_{0}^{2\pi}
\left[ {1 \over 2} W^{(s)}_{\theta^{\alpha}} \,
W^{(s)}_{\theta^{\beta}} |_{x = -\infty}^{x = +\infty}
\, - \, W^{(s)}_{\theta^{\beta}} \, T^{(s)}_{\alpha}
|_{x = -\infty}^{x = +\infty} \right] \,
{d^{m} \theta \over (2\pi)^{m}}
\end{equation}

 Both terms in(\ref{twoterms}) are zero view (\ref{antsym})
and $T^{(s)}_{\alpha} \rightarrow 0$ for
$x \rightarrow \pm \infty$ on ${\hat {\cal W}}_{0}$.  

 Using now the relations
$W^{(s)}_{\theta^{\alpha}x} \, = \,
h^{(s)}_{\theta^{\alpha}}$,
$W^{(s)}_{\theta^{\beta}x} \, = \, 
h^{(s)}_{\theta^{\beta}}$ we have

$$\int_{-\infty}^{+\infty}
\int_{0}^{2\pi}\!\!\!\dots\int_{0}^{2\pi} \left[
\varphi^{i}_{\theta^{\alpha}} \sum_{k \geq 0}
\omega^{(k)}_{ij} \, \varphi^{j}_{\theta^{\beta},kx} \, + \,
\sum_{s=1}^{g} \, e_{s} \,
\left( h^{(s)}_{\theta^{\beta}} \, T^{(s)}_{\alpha}
\, - \, h^{(s)}_{\theta^{\alpha}} \, T^{(s)}_{\beta}
\, + \, \left( T^{(s)}_{\alpha} \right)_{x} T^{(s)}_{\beta}
\right) \right] \, \times $$
$$ \times \, {d^{m} \theta \over (2\pi)^{m}}
\, dx
\, \equiv \, 0$$
so we get (\ref{divform}).
 
2) We can normalize now the functions   
$A_{\alpha\beta}(\bm{\varphi}, \dots)$ such that
$A_{\alpha\beta} \, = \, 0$ for
$\varphi^{i}(\bm{\theta},x) \, \equiv \, const \, = \, C^{i}$.
Now for any function $\varphi^{i}(\bm{\theta},x)$
depending only on $x$ we have
$\partial/\partial x A_{\alpha\beta}(\bm{\varphi}, \dots)
\, = \, 0$ according to the relation (\ref{divform}).
Using the fact that
$A_{\alpha\beta}(\bm{\theta},\pm \infty) \, = \, 0$
on ${\hat {\cal W}}_{0}$ we get the part (2) of the Lemma
on ${\hat {\cal W}}_{0}$. Now using the fact that
$A_{\alpha\beta}(\bm{\varphi}, \dots)$ is a local
expression of $\bm{\varphi} (\bm{\theta},x)$ and it's
derivatives we get in fact that
$A_{\alpha\beta}(\bm{\theta},\pm \infty) \, \equiv \, 0$
for any $\bm{\varphi} (\bm{\theta},x)$ depending on $x$
only for this normalization of $A_{\alpha\beta}$ which
actually does not depend on the constants $C^{i}$.

{\hfill Lemma 3 is proved.}

\vspace{0.5cm}

 We will need also the following technical Lemma.

\vspace{0.5cm}

{\bf Lemma 4.}

{\it For any Symplectic Form (\ref{conssympform}) we have

$$- \, \int_{0}^{2\pi}\!\!\!\dots\int_{0}^{2\pi}
A_{\alpha\beta} (\bm{\varphi}(\bm{\theta},y),
\bm{\varphi}_{y}(\bm{\theta},y), \dots) \,
{d^{m} \theta \over (2\pi)^{m}} \, + $$
$$+ \, \int_{0}^{2\pi}\!\!\!\dots\int_{0}^{2\pi} \sum_{k \geq 1}
\sum_{p=1}^{k} \, C_{k}^{p} \, (-1)^{p-1} \, \left(
\varphi^{i}_{\theta^{\alpha}}(\bm{\theta},y) \,
\omega^{(k)}_{ij}(\bm{\varphi}(\bm{\theta},y), \dots)  
\, \varphi^{j}_{\theta^{\beta},(k-p)y} \right)_{(p-1)y}
\, {d^{m} \theta \over (2\pi)^{m}} \, -$$
$$- \, {1 \over 2} \int_{0}^{2\pi}\!\!\!\dots\int_{0}^{2\pi}
\sum_{s=1}^{g} \, e_{s} \, \left[
W^{(s)}_{\theta^{\alpha}}(\bm{\theta},y) \,
W^{(s)}_{\theta^{\beta}}(\bm{\theta},y) \, - \,
W^{(s)}_{\theta^{\alpha}}(\bm{\theta}, +\infty) \,
W^{(s)}_{\theta^{\beta}}(\bm{\theta}, +\infty) \right] \,
{d^{m} \theta \over (2\pi)^{m}} \, +$$
$$+ \, \int_{0}^{2\pi}\!\!\!\dots\int_{0}^{2\pi}
\sum_{s=1}^{g} \, e_{s} \,
T^{(s)}_{\alpha}(\bm{\theta},y) \,
W^{(s)}_{\theta^{\beta}}(\bm{\theta},y) \,
{d^{m} \theta \over (2\pi)^{m}} \, +$$
$$+ \, \int_{0}^{2\pi}\!\!\!\dots\int_{0}^{2\pi}
\sum_{s=1}^{g} \, e_{s} \,
\left( W^{(s)}_{\theta^{\alpha}}(\bm{\theta},y) \, - \,
T^{(s)}_{\alpha} (\bm{\theta},y) \right)  
\left( W^{(s)}_{\theta^{\beta}}(\bm{\theta},y) \, - \,
T^{(s)}_{\beta} (\bm{\theta},y) \right) \,
{d^{m} \theta \over (2\pi)^{m}} \,\, \equiv $$

\vspace{0.5cm}

$$\equiv \, \, \int_{0}^{2\pi}\!\!\!\dots\int_{0}^{2\pi}
A_{\beta\alpha} (\bm{\varphi}(\bm{\theta},y),
\bm{\varphi}_{y}(\bm{\theta},y), \dots) \,
{d^{m} \theta \over (2\pi)^{m}} \, + $$
$$+ \, {1 \over 2} \int_{0}^{2\pi}\!\!\!\dots\int_{0}^{2\pi}
\sum_{s=1}^{g} \, e_{s} \, \left[
W^{(s)}_{\theta^{\alpha}}(\bm{\theta},y) \,
W^{(s)}_{\theta^{\beta}}(\bm{\theta},y) \, - \,
W^{(s)}_{\theta^{\alpha}}(\bm{\theta}, +\infty) \,
W^{(s)}_{\theta^{\beta}}(\bm{\theta}, +\infty) \right] \,
{d^{m} \theta \over (2\pi)^{m}} \, -$$
\begin{equation}
\label{lemma4}
- \, \int_{0}^{2\pi}\!\!\!\dots\int_{0}^{2\pi}  
\sum_{s=1}^{g} \, e_{s} \,
W^{(s)}_{\theta^{\alpha}}(\bm{\theta},y) \,
T^{(s)}_{\beta}(\bm{\theta},y) \,
{d^{m} \theta \over (2\pi)^{m}}
\end{equation}
where the values $A_{\alpha\beta}$ (normalized in "right"
way), $W^{(s)}$ and $T^{(s)}_{\alpha}$ are introduced
in (\ref{divform}), (\ref{WS}) and (\ref{tsa}) respectively.
}

\vspace{0.5cm}

 Proof.

 Let us consider the quantities  

$$E_{\alpha\beta}(y) \, = $$
$$= \, \int_{-\infty}^{+\infty} \int_{-\infty}^{+\infty}
\int_{0}^{2\pi}\!\!\!\dots\int_{0}^{2\pi}
\varphi^{i}_{\theta^{\alpha}}(\bm{\theta},z) \,
{\tilde \Omega}_{ij}(\bm{\theta},\bm{\theta}^{\prime},z,w) \,
\varphi^{j}_{\theta^{\prime\beta}}(\bm{\theta}^{\prime},w) \,
\nu (w-y) \, dz \, dw \, {d^{m} \theta \over (2\pi)^{m}} \,
{d^{m} \theta^{\prime} \over (2\pi)^{m}} $$

 We have

$$E_{\alpha\beta}(y) \, = \, \int \,
\varphi^{i}_{\theta^{\alpha}}(\bm{\theta},z) \,
\sum_{k \geq 0} \omega^{(k)}_{ij}
(\bm{\varphi}, \bm{\varphi}_{w}, \dots) \,
\delta^{(k)}(z-w) \,
\varphi^{j}_{\theta^{\beta}}(\bm{\theta},w) \,
\nu (w-y) \, dz \, dw \, {d^{m} \theta \over (2\pi)^{m}}
\,\,\, +$$
$$+ \, \int \sum_{s=1}^{g} \, e_{s} \,
\left( W^{(s)}_{\theta^{\alpha}z} \, - \,
\left(T^{(s)}_{\alpha}\right)_{z} \right) \, \nu (z-w) \,
\left( W^{(s)}_{\theta^{\beta}w} \, - \,
\left(T^{(s)}_{\beta}\right)_{w} \right) \, \nu (w-y) \,
dz \, dw \, {d^{m} \theta \over (2\pi)^{m}} $$

 We can calculate these values in two ways:

\vspace{0.5cm}

 I) Let us first make the integration with respect to $z$.
We have
 
$$E_{\alpha\beta}(y) \, = \,
\int \sum_{k \geq 0} (-1)^{k} \left(
\varphi^{i}_{\theta^{\alpha}}(\bm{\theta},w) \,
\omega^{(k)}_{ij}(\bm{\varphi}, \bm{\varphi}_{w}, \dots) \,
\right)_{kw} \,
\varphi^{j}_{\theta^{\beta}}(\bm{\theta},w) \,
\nu (w-y) \, dw \, {d^{m} \theta \over (2\pi)^{m}}
\,\,\, - $$
$$- \, \int \sum_{s=1}^{g} \, e_{s} \,
\left( W^{(s)}_{\theta^{\alpha}} \, - \,
T^{(s)}_{\alpha}(z) \right) \left(
W^{(s)}_{\theta^{\beta}w} \, - \,
\left(T^{(s)}_{\beta}\right)_{w} \right) \, \nu (w-y) \,
dw \, {d^{m} \theta \over (2\pi)^{m}} \, =$$

\vspace{0.5cm}
 
$$= \, \int \left[ \sum_{k \geq 0} (-1)^{k} \left(
\varphi^{i}_{\theta^{\alpha}} \,
\omega^{(k)}_{ij} \right)_{kw} \,
\varphi^{j}_{\theta^{\beta}} \, - \,
\sum_{s=1}^{g} \, e_{s} \, \left(
h^{(s)}_{\theta^{\alpha}} \, T^{(s)}_{\beta} \, - \,
h^{(s)}_{\theta^{\beta}} \, T^{(s)}_{\alpha} \, + \,
\left( T^{(s)}_{\beta} \right)_{w} \, T^{(s)}_{\alpha}
\right) \right] \times $$
$$\times \,\,\, \nu (w-y) \, dw \,
{d^{m} \theta \over (2\pi)^{m}} \,\,\, - $$
$$- \int \sum_{s=1}^{g}  e_{s}  \left[ 
{1 \over 2} \left( W^{(s)}_{\theta^{\alpha}} \,   
W^{(s)}_{\theta^{\beta}} \right)_{w}  - 
{1 \over 2} \left( W^{(s)}_{\theta^{\beta}} \,
W^{(s)}_{w} \right)_{\theta^{\alpha}}  + 
{1 \over 2} \left( W^{(s)}_{\theta^{\alpha}} \,
W^{(s)}_{w} \right)_{\theta^{\beta}}  - 
\left( W^{(s)}_{\theta^{\alpha}} \,
T^{(s)}_{\beta} \right)_{w} \right]  \times $$
$$\times \,\,
\nu (w-y) \, dw \, {d^{m} \theta \over (2\pi)^{m}} $$
 
 Using now the skew-symmetry of the form
$\tilde{\Omega}_{ij}(\bm{\theta},\bm{\theta}^{\prime},z,w)$
and the relations (\ref{divform}) we can write

$$E_{\alpha\beta}(y) \, = \, - \, \int
\left( A_{\beta\alpha}(\bm{\theta},w) \right)_{w} \,
\nu (w-y) \, dw \,
{d^{m} \theta \over (2\pi)^{m}} \,\,\, - $$
$$- \, \int \sum_{s=1}^{g} \, e_{s} \,
\left[ {1 \over 2} \left( W^{(s)}_{\theta^{\alpha}} \,
W^{(s)}_{\theta^{\beta}} \right)_{w} \, - \,
\left( W^{(s)}_{\theta^{\alpha}} \,
T^{(s)}_{\beta} \right)_{w} \right] \,
\nu (w-y) \, dw \, {d^{m} \theta \over (2\pi)^{m}} \, =$$

\vspace{0.5cm}

$$= \, \int \, A_{\beta\alpha} (\bm{\theta}, y) \,
{d^{m} \theta \over (2\pi)^{m}} \,\,\, - \,\,\, \int
W^{(s)}_{\theta^{\alpha}} (\bm{\theta}, y) \,   
T^{(s)}_{\beta} (\bm{\theta}, y) \,
{d^{m} \theta \over (2\pi)^{m}} \,\,\, +$$
$$+ \, {1 \over 2} \int \sum_{s=1}^{g} \, e_{s} \,
W^{(s)}_{\theta^{\alpha}} (\bm{\theta}, y) \,
W^{(s)}_{\theta^{\beta}} (\bm{\theta}, y) \,
{d^{m} \theta \over (2\pi)^{m}} \, - $$
$$- \, {1 \over 4} \int
\left[ W^{(s)}_{\theta^{\alpha}} (\bm{\theta}, +\infty) \,
W^{(s)}_{\theta^{\beta}} (\bm{\theta}, +\infty) \, + \,
W^{(s)}_{\theta^{\alpha}} (\bm{\theta}, -\infty) \,
W^{(s)}_{\theta^{\beta}} (\bm{\theta}, -\infty) \right]
{d^{m} \theta \over (2\pi)^{m}} $$
(we used the relation
$A_{\beta\alpha} (\bm{\theta}, \pm \infty) \, = \, 0$
on ${\hat {\cal W}}_{0}$.)

\vspace{0.5cm}

 II) Let now make first the integration with respect to
$w$. We have

$$E_{\alpha\beta}(y) \, = \int
\varphi^{i}_{\theta^{\alpha}}(\bm{\theta},z) \,
\sum_{k \geq 0}
\omega^{(k)}_{ij}(\bm{\varphi}, \bm{\varphi}_{z}, \dots) \,
\varphi^{j}_{\theta^{\beta},kz}(\bm{\theta},z) \,
\nu (z-y) \, dz \, {d^{m} \theta \over (2\pi)^{m}} \, +$$
$$+ \, \int \varphi^{i}_{\theta^{\alpha}}(\bm{\theta},z) \,
\sum_{k \geq 1} \sum_{p=1}^{k} C^{p}_{k} \,
\omega^{(k)}_{ij}(\bm{\varphi}, \bm{\varphi}_{z}, \dots) \,
\varphi^{j}_{\theta^{\beta},(k-p)z}(\bm{\theta},z) \,
\delta^{(p-1)}(z-y) \, dz \,
{d^{m} \theta \over (2\pi)^{m}} \, +$$
$$+ \, \int \sum_{s=1}^{g} \, e_{s} \,
\left[ W^{(s)}_{\theta^{\alpha}z}(\bm{\theta},z) \, - \,
\left( T^{(s)}_{\alpha} \right)_{z}(\bm{\theta},z)
\right] \, \times $$
$$\times \, \left[
W^{(s)}_{\theta^{\beta}}(\bm{\theta},z) \, - \,
T^{(s)}_{\beta}(\bm{\theta},z) \, + \, {1 \over 4}
W^{(s)}_{\theta^{\beta}}(\bm{\theta},-\infty) \, - \,
{1 \over 4}
W^{(s)}_{\theta^{\beta}}(\bm{\theta},+\infty) \right]
\, \nu (z-y) \, dz \,
{d^{m} \theta \over (2\pi)^{m}} \,\,\, -$$
$$- \, \int \sum_{s=1}^{g} \, e_{s} \,
\left[ W^{(s)}_{\theta^{\alpha}z}(\bm{\theta},z) \, - \,
\left( T^{(s)}_{\alpha}(\bm{\theta},z) \right)_{z} \right] 
\, \nu (z-y) \,
\left[ W^{(s)}_{\theta^{\beta}}(\bm{\theta},y) \, - \,   
T^{(s)}_{\beta}(\bm{\theta},y)  \right] \, dz \,
{d^{m} \theta \over (2\pi)^{m}} \, = $$

\vspace{0.5cm}

$$= \, \int \left( A_{\alpha\beta} (\bm{\theta},z)
\right)_{z} \, \nu (z-y) \, dz \,
{d^{m} \theta \over (2\pi)^{m}} \, + $$
$$+ \, \int \sum_{k \geq 1} \sum_{p=1}^{k} C^{p}_{k} \,
(-1)^{p-1} \, \left(
\varphi^{i}_{\theta^{\alpha}}(\bm{\theta},y) \,
\omega^{(k)}_{ij}(\bm{\varphi}, \bm{\varphi}_{y}, \dots) \,
\varphi^{j}_{\theta^{\beta},(k-p)y}(\bm{\theta},y)
\right)_{(p-1)y} \,
{d^{m} \theta \over (2\pi)^{m}} \,\,\, + $$
$$+ \, \int \sum_{s=1}^{g} \, e_{s} \,
{1 \over 2} \left[ \left( W^{(s)}_{\theta^{\alpha}} \,
W^{(s)}_{\theta^{\beta}} \right)_{z} \, - \,
\left( W^{(s)}_{z} \,
W^{(s)}_{\theta^{\alpha}} \right)_{\theta^{\beta}} \, + \,
\left( W^{(s)}_{z} \, W^{(s)}_{\theta^{\beta}}
\right)_{\theta^{\alpha}} \right] \, \nu (z-y) \,
dz \,{d^{m} \theta \over (2\pi)^{m}} \,\,\, -$$
$$- \, \int \sum_{s=1}^{g} \, e_{s} \,
\left[ \left( T^{(s)}_{\alpha} \,
W^{(s)}_{\theta^{\beta}} \right)_{z} \, + \,
{1 \over 2} W^{(s)}_{\theta^{\beta}}(\bm{\theta},-\infty)
\left( W^{(s)}_{\theta^{\alpha}} \, - \, T^{(s)}_{\alpha}
\right)_{z} \right] \, \nu (z-y) \,
dz \,{d^{m} \theta \over (2\pi)^{m}} \,\,\, + $$
$$+ \, \int \sum_{s=1}^{g} \, e_{s} \, 
\left( W^{(s)}_{\theta^{\alpha}}(\bm{\theta},y) \, - \,
T^{(s)}_{\alpha}(\bm{\theta},y) \right)
\left( W^{(s)}_{\theta^{\beta}} (\bm{\theta},y) \, - \,
T^{(s)}_{\beta}(\bm{\theta},y) \right) \,
{d^{m} \theta \over (2\pi)^{m}} \, = $$

\vspace{0.5cm}

$$= \, - \, \int A_{\alpha\beta} (\bm{\theta},y) \,   
{d^{m} \theta \over (2\pi)^{m}} \, + $$
$$+ \, \int \sum_{k \geq 1} \sum_{p=1}^{k} C^{p}_{k} \,
(-1)^{p-1} \, \left(
\varphi^{i}_{\theta^{\alpha}}(\bm{\theta},y) \,
\omega^{(k)}_{ij}(\bm{\varphi}, \bm{\varphi}_{y}, \dots) \,
\varphi^{j}_{\theta^{\beta},(k-p)y}(\bm{\theta},y)
\right)_{(p-1)y} \,
{d^{m} \theta \over (2\pi)^{m}} \,\,\, - $$
$$- \,\,\, {1 \over 2} \int \sum_{s=1}^{g} \, e_{s} \,
W^{(s)}_{\theta^{\alpha}}(\bm{\theta},y) \,
W^{(s)}_{\theta^{\beta}} (\bm{\theta},y) \,
{d^{m} \theta \over (2\pi)^{m}} \,\,\, +$$
$$+ \, {1 \over 4}
\int \sum_{s=1}^{g} \, e_{s} \, \left[ 
W^{(s)}_{\theta^{\alpha}}(\bm{\theta},-\infty) \,
W^{(s)}_{\theta^{\beta}} (\bm{\theta},-\infty) \, + \,
W^{(s)}_{\theta^{\alpha}}(\bm{\theta},+\infty) \,
W^{(s)}_{\theta^{\beta}} (\bm{\theta},+\infty)
\right] \, {d^{m} \theta \over (2\pi)^{m}} \, +$$
$$+ \, \int \sum_{s=1}^{g} \, e_{s} \,
T^{(s)}_{\alpha} (\bm{\theta},y) \,
W^{(s)}_{\theta^{\beta}} (\bm{\theta},y) \,
{d^{m} \theta \over (2\pi)^{m}} \, +$$
$$+ \, \int \sum_{s=1}^{g} \, e_{s} \, 
\left( W^{(s)}_{\theta^{\alpha}}(\bm{\theta},y) \, - \,
T^{(s)}_{\alpha}(\bm{\theta},y) \right)
\left( W^{(s)}_{\theta^{\beta}} (\bm{\theta},y) \, - \,
T^{(s)}_{\beta}(\bm{\theta},y) \right) \,
{d^{m} \theta \over (2\pi)^{m}}$$

 Comparing (I) and (II) and using (\ref{antsym}) we get
now the statement of the Lemma.

{\hfill Lemma 4 is proved.}

\section{The averaging of the weakly nonlocal Symplectic
Structures.}
\setcounter{equation}{0}

 Let us now make the change $X = \epsilon x$, $T = \epsilon t$.
We can define again a symplectic form in new coordinates which can
be written as

$${\hat \Omega}_{ij}(\bm{\theta},\bm{\theta}^{\prime},X,Y)
\,\, = \,\, \sum_{k \geq 0} \omega^{(k)}_{ij}
(\bm{\varphi}(\bm{\theta},X),
\epsilon \, \bm{\varphi}_{X}(\bm{\theta},X), \dots) \,
\epsilon^{k} \, \delta^{(k)} (X-Y) \,\,
\delta (\bm{\theta} - \bm{\theta}^{\prime}) \,\,\, +$$   
\begin{equation}
\label{omXY}
+ \,\,\, {1 \over \epsilon} \sum_{s=1}^{g} \, e_{s} \,
{\delta {\hat H}^{(s)} \over \delta \varphi^{i}(\bm{\theta},X)} \,
\nu (X-Y) \,\, \delta (\bm{\theta} - \bm{\theta}^{\prime}) \,
{\delta {\hat H}^{(s)} \over \delta \varphi^{j}(\bm{\theta}^{\prime},Y)}
\,\,\, , \,\,\,\,\, i,j \, = \, 1, \dots, n
\end{equation}
where

$${\hat H}^{(s)} \,\, = \,\, \int_{-\infty}^{+\infty}
\int_{0}^{2\pi}\!\!\!\dots\int_{0}^{2\pi}
h^{s} (\bm{\varphi}(\bm{\theta},X),
\epsilon \, \bm{\varphi}_{X}(\bm{\theta},X), \dots) \,
{d^{m} \theta \over (2\pi)^{m}} \, dX $$
 
 We will assume for simplicity that the family $\Lambda$ of solutions
of (\ref{phasesyst}) contains the solutions corresponding to
$k^{\alpha} \, = \, 0$ for some parameters ${\bf U} = {\bf U}_{0}$
such that
$\Phi^{i}(\bm{\theta}, {\bf U}_{0}) \, = \, C^{i} \, = \, const$
(we should have $Q^{i}({\bf C},0, \dots) \, = \, 0$ in this case).

\vspace{0.5cm}

 Let us introduce the functional "sub-manifold" ${\cal M}_{0}$
in the space of functions $\bm{\varphi}(\bm{\theta},X)$
($2\pi$-periodic w.r.t. each $\theta^{\alpha}$) in the following way

\vspace{0.5cm}

1) We require that the functions $\bm{\varphi}(\bm{\theta},X)$
from ${\cal M}_{0}$ belong to the family $\Lambda$ of solutions
of (\ref{phasesyst}) at any fixed $X$;

2) We put ${\bf U}(X) \, \rightarrow \, {\bf U}_{0}$ 
(i.e. $\varphi^{i}(\bm{\theta},X) \, \rightarrow \, C^{i}$) for
$X \, \rightarrow \, \pm \infty$ (rapidly enough).   

\vspace{0.5cm}
 
 The functions ${\bf U}(X)$, $\bm{\theta}_{0}(X)$ can be taken
as the coordinates on the sub-manifold ${\cal M}_{0}$ such that   
we have  

$$\varphi^{i}_{[{\bf U},\bm{\theta}_{0}]} (\bm{\theta},X)
\, = \, \Phi^{i} (\bm{\theta} + \bm{\theta}_{0}(X), {\bf U}(X)) $$
for the functions belonging to ${\cal M}_{0}$.

\vspace{0.5cm}

 We will consider also the "$\epsilon$-deformations"
${\cal M}_{\epsilon} [\bm{\Psi}_{(1)}]$ of the sub-manifold
${\cal M}_{0}$ defined with the aid of an arbitrary function
$\bm{\Psi}_{(1)} (\bm{\theta},X)$ $2\pi$-periodic w.r.t. each
$\theta^{\alpha}$ and such that

$$\Psi^{i}_{(1)} (\bm{\theta},X) \, \rightarrow \, 0 \,\,\,\,\,
{\rm for} \,\,\,\,\, X \, \rightarrow \, \pm \infty $$

 Namely, we put

$$\varphi^{i}_{[{\bf U},\bm{\theta}_{0}]} (\bm{\theta},X)
\, = \, \Phi^{i} (\bm{\theta} + \bm{\theta}_{0}(X), {\bf U}(X))
\, + \, \epsilon \,
\Psi_{(1)}^{i} (\bm{\theta} + \bm{\theta}_{0}(X), X)$$
which defines the $\epsilon$-deformation of the function
$\bm{\varphi}_{[{\bf U},\bm{\theta}_{0}]}$ corresponding to the   
coordinates ${\bf U}(X)$, $\bm{\theta}_{0}(X)$. Easy to see that the
case $\bm{\Psi}_{(1)} \, = \, 0$ corresponds to the sub-manifold
${\cal M}_{0}$.

\vspace{0.5cm}

 Let us introduce now the new coordinates $\bm{\theta}^{*}_{0}(X)$
on ${\cal M}_{0}$ and ${\cal M}_{\epsilon} [\bm{\Psi}_{(1)}]$
in the following way:

$$\theta^{*\alpha}_{0}(X) \,\,\, = \,\,\,
\theta^{\alpha}_{0}(X) \,\, - \,\, {1 \over \epsilon} \,
S^{\alpha}(X) $$
where
$$S^{\alpha}(X) \,\,\, = \,\,\,
\int_{-\infty}^{+\infty} \nu (X-Y) \,\,
k^{\alpha}({\bf U}(Y)) \, dY $$

 We can write then on any ${\cal M}_{\epsilon} [\bm{\Psi}_{(1)}]$

$$\varphi^{i}_{[{\bf U},\bm{\theta}^{*}_{0}]} (\bm{\theta},X)
\, = \, \Phi^{i} (\bm{\theta} \, + \, \bm{\theta}^{*}_{0}(X) \, + \,
{1 \over \epsilon} \, {\bf S}(X), {\bf U}(X)) \,\, +$$  
\begin{equation}
\label{epsdef}
+ \,\, \epsilon \, \Psi_{(1)}^{i} (\bm{\theta} \, + \,
\bm{\theta}^{*}_{0}(X)\, + \, {1 \over \epsilon} \, {\bf S}(X), X)
\end{equation}

\vspace{0.5cm}
 
 We can see that the functions
$\varphi^{i}_{[{\bf U},\bm{\theta}^{*}_{0}]} (\bm{\theta},X)$
become the rapidly oscillating functions of $X$ (for fixed
$\bm{\theta}$) for any fixed "coordinates"
${\bf U}(X)$, $\bm{\theta}^{*}_{0}(X)$ and $\epsilon \rightarrow 0$.
Easy to see also that (\ref{epsdef}) represents in fact the first
two terms of the expansion of asymptotic solutions
(\ref{whithsol}) for the appropriate $\bm{\Psi}_{(1)}$.

\vspace{0.5cm}

 Let us formulate now the theorem about the restriction of 2-form
${\hat \Omega}_{ij}(\bm{\theta},\bm{\theta}^{\prime},X,Y)$
on the sub-manifolds ${\cal M}_{\epsilon} [\bm{\Psi}_{(1)}]$.

\vspace{0.5cm}

{\bf Theorem 4.}

{\it The restriction of the form
${\hat \Omega}_{ij}(\bm{\theta},\bm{\theta}^{\prime},X,Y)$
to any submanifold ${\cal M}_{\epsilon} [\bm{\Psi}_{(1)}]$
in coordinates $U^{\nu}(X)$,
$\theta^{*\alpha}_{0}(X)$ can be written as
 
$$\Omega^{rest} \,\, = \,\, \int_{-\infty}^{+\infty}
\int_{-\infty}^{+\infty} \Omega^{1}_{\nu\mu}(X,Y) \,\,
\delta U^{\nu}(X) \,\, \delta U^{\mu}(Y) \,\, dX \, dY \, + $$
$$+ \, \int_{-\infty}^{+\infty}
\int_{-\infty}^{+\infty} \Omega^{2}_{\nu\alpha}(X,Y) \,\,
\delta U^{\nu}(X) \,\, \delta \theta^{*\alpha}_{0} (Y) \,\,
dX \, dY  \, + $$
$$+ \, \int_{-\infty}^{+\infty} \int_{-\infty}^{+\infty}
\Omega^{3}_{\alpha\nu}(X,Y) \,\, \delta \theta^{*\alpha}_{0} (X) 
\,\, \delta U^{\nu}(Y) \,\, dX \, dY \, + $$
$$+ \,
\int_{-\infty}^{+\infty} \int_{-\infty}^{+\infty}
\Omega^{4}_{\alpha\beta}(X,Y) \,\,
\delta \theta^{*\alpha}_{0} (X) \,\,
\delta \theta^{*\beta}_{0} (Y) \,\, dX \, dY $$
where

\vspace{0.5cm}  

I) The weak limit\footnote{We mean here the limit in sense of
functionals 
$\int \Omega^{1}_{\nu\mu} (X,Y) \, \xi^{\nu}(X) \, \eta^{\mu}(Y) 
\, dX \, dY$ for any fixed smooth $\xi^{\nu}(X)$, 
$\eta^{\mu}(Y)$.} $\Omega^{1(wl)}_{\nu\mu}(X,Y)$
of the form $\Omega^{1}_{\nu\mu}(X,Y)$ can be written as  

$$\Omega^{1(wl)}_{\nu\mu}(X,Y) \,\, = \,\,
{1 \over \epsilon} \, \sum_{\alpha = 1}^{m} \left(
{\partial k^{\alpha} \over \partial U^{\nu}}(X) \, \nu (X-Y) \,
{\partial I_{\alpha} \over \partial U^{\mu}}(Y) +   
{\partial I_{\alpha} \over \partial U^{\nu}}(X) \, \nu (X-Y) \,
{\partial k^{\alpha} \over \partial U^{\mu}}(Y) \right) \, + $$
$$+ \, {1 \over \epsilon} \, \sum_{s=1}^{g} e_{s}
{\partial \langle h^{(s)} \rangle \over \partial U^{\nu}}(X)
\, \nu (X-Y) \,
{\partial \langle h^{(s)} \rangle \over \partial U^{\mu}}(Y)
\,\,\, + \,\,\, {o(1) \over \epsilon} $$
where the expression $\langle h^{(s)} \rangle ({\bf U})$
are the averaged densities
$h^{(s)} (\bm{\varphi}, \bm{\varphi}_{x}, \dots)$ and the functions
$I^{\alpha}({\bf U})$ are defined through the formulas

$${\partial I_{\alpha} \over \partial U^{\nu}} \,\,\, = \,\,\, - \,
{\partial k^{\beta} \over \partial U^{\nu}} \,\,
\langle A_{\alpha\beta} \rangle \,\,\, + $$

$$+ \,\,\, {\partial k^{\beta} \over \partial U^{\nu}} \,\,
\sum_{s=1}^{g} e_{s} \left[ \gamma^{\delta}_{\alpha}
\left( \langle T^{(s)}_{\beta} J^{(s)}_{\delta} \rangle -
\langle T^{(s)}_{\beta} \rangle \langle J^{(s)}_{\delta} \rangle
\right) \, - \, {1 \over 2} \,
\gamma^{\delta}_{\alpha} \gamma^{\zeta}_{\beta}
\left( \langle J^{(s)}_{\delta} J^{(s)}_{\zeta} \rangle -
\langle J^{(s)}_{\delta} \rangle \langle J^{(s)}_{\zeta} \rangle
\right) \right] + $$

$$+ \,\, \langle \Phi^{i}_{U^{\nu}} \sum_{k \geq 0}
\omega^{(k)}_{ij} (\varphi, \dots) \varphi^{j}_{\theta^{\alpha},kx}
\rangle \,\, - \,\, \sum_{s=1}^{g} e_{s} \langle \Phi^{i}_{U^{\nu}}
{\delta H^{(s)} \over \delta \varphi^{i}(x)} T^{(s)}_{\alpha}
\rangle \,\, + $$

\begin{equation}
\label{actvar}
+ \,\, \sum_{s=1}^{g} e_{s} \gamma^{\beta}_{\alpha} \left( \langle 
\Phi^{i}_{U^{\nu}} {\delta H^{(s)} \over \delta \varphi^{i}(x)}
J^{(s)}_{\beta} \rangle - \langle
\Phi^{i}_{U^{\nu}} {\delta H^{(s)} \over \delta \varphi^{i}(x)}
\rangle \langle J^{(s)}_{\beta} \rangle \right) 
\end{equation}
(the functions $A_{\alpha\beta}$ are normalized according to
Lemma 3);

\vspace{0.5cm}

II) The forms $\Omega^{2}_{\nu\alpha} (X,Y)$,
$\Omega^{3}_{\alpha\nu} (X,Y)$ have the order $O(1)$ for
$\epsilon \rightarrow 0$ on
${\cal M}_{\epsilon} [\bm{\Psi}_{(1)}]$;

\vspace{0.5cm}

The form $\Omega^{4}_{\alpha\beta} (X,Y)$ has the order
$O(\epsilon)$ for $\epsilon \rightarrow 0$ on
${\cal M}_{\epsilon} [\bm{\Psi}_{(1)}]$.

}

\vspace{0.5cm}

Proof.

Let us first rewrite the relations (\ref{divform}) and
(\ref{lemma4}) in the variables $\bm{\theta}$, $X$, i.e.

$$\varphi^{i}_{\theta^{\alpha}} \sum_{k \geq 0} \omega^{(k)}_{ij}
(\bm{\varphi}, \epsilon \, \bm{\varphi}_{X}, \dots) \,
\epsilon^{k} \,
\varphi^{j}_{\theta^{\beta},kX} \, + \, \sum_{s=1}^{g} \, e_{s} \,
\left( h^{(s)}_{\theta^{\beta}} T^{(s)}_{\alpha} \, - \,
h^{(s)}_{\theta^{\alpha}}
T^{(s)}_{\beta} \, + \, \epsilon \,
\left(T^{(s)}_{\alpha}\right)_{X} T^{(s)}_{\beta} \right)
\,\,\, \equiv $$
\begin{equation}
\label{epsdivform}
\equiv \,\,\, {\partial \over \partial \theta^{\gamma}}
Q^{\gamma}_{\alpha\beta}
(\bm{\varphi}, \epsilon \, \bm{\varphi}_{X}, \dots) \, + \,
\epsilon \, {\partial \over \partial X} 
A_{\alpha\beta}
(\bm{\varphi},  \epsilon \, \bm{\varphi}_{X}, \dots)
\end{equation}
and

\vspace{0.5cm}

\vspace{0.5cm}

$$- \, \int_{0}^{2\pi}\!\!\!\dots\int_{0}^{2\pi}
A_{\alpha\beta} (\bm{\varphi}(\bm{\theta},Y),
\epsilon \, \bm{\varphi}_{Y}(\bm{\theta},Y), \dots) \,
{d^{m} \theta \over (2\pi)^{m}} \, + $$
$$+ \! \int_{0}^{2\pi}\!\!\!\dots\int_{0}^{2\pi}\!\sum_{k \geq 1}
\sum_{p=1}^{k} C_{k}^{p} (-1)^{p-1}
\epsilon^{k-1} \! \left(
\varphi^{i}_{\theta^{\alpha}}(\bm{\theta},Y) \,
\omega^{(k)}_{ij}(\bm{\varphi}(\bm{\theta},Y), \dots)
\, \varphi^{j}_{\theta^{\beta},(k-p)Y} (\bm{\theta},Y)
\right)_{(p-1)Y} {d^{m} \theta \over (2\pi)^{m}} -$$
$$- \, {1 \over 2} \, \int_{0}^{2\pi}\!\!\!\dots\int_{0}^{2\pi}
\sum_{s=1}^{g} \, e_{s} \, \left(
W^{(s)}_{\theta^{\alpha}}(\bm{\theta},Y) \,
W^{(s)}_{\theta^{\beta}}(\bm{\theta},Y) \, - \,
W^{(s)}_{\theta^{\alpha}}(\bm{\theta}, +\infty) \,
W^{(s)}_{\theta^{\beta}}(\bm{\theta}, +\infty) \right) \,  
{d^{m} \theta \over (2\pi)^{m}} \, +$$  
$$+ \, \int_{0}^{2\pi}\!\!\!\dots\int_{0}^{2\pi}
\sum_{s=1}^{g} \, e_{s} \,
T^{(s)}_{\alpha}(\bm{\theta},Y) \,
W^{(s)}_{\theta^{\beta}}(\bm{\theta},Y) \,
{d^{m} \theta \over (2\pi)^{m}} \, +$$
$$+ \, \int_{0}^{2\pi}\!\!\!\dots\int_{0}^{2\pi}
\sum_{s=1}^{g} \, e_{s} \,
\left( W^{(s)}_{\theta^{\alpha}}(\bm{\theta},Y) 
\, - \, T^{(s)}_{\alpha} (\bm{\theta},Y) \right)
\left( W^{(s)}_{\theta^{\beta}}(\bm{\theta},Y)
\, - \, T^{(s)}_{\beta} (\bm{\theta},Y) \right) \,
{d^{m} \theta \over (2\pi)^{m}} \,\, \equiv $$

\vspace{0.5cm}

$$\equiv \, \, \int_{0}^{2\pi}\!\!\!\dots\int_{0}^{2\pi}
A_{\beta\alpha} (\bm{\varphi}(\bm{\theta},Y),
\epsilon \, \bm{\varphi}_{Y}(\bm{\theta},Y), \dots) \,
{d^{m} \theta \over (2\pi)^{m}} \, + $$
$$+ \, {1 \over 2} \, \int_{0}^{2\pi}\!\!\!\dots\int_{0}^{2\pi}
\sum_{s=1}^{g} \, e_{s} \, \left[
W^{(s)}_{\theta^{\alpha}}(\bm{\theta},Y) \,
W^{(s)}_{\theta^{\beta}}(\bm{\theta},Y) \, - \,   
W^{(s)}_{\theta^{\alpha}}(\bm{\theta}, +\infty) \,
W^{(s)}_{\theta^{\beta}}(\bm{\theta}, +\infty) \right] \,
{d^{m} \theta \over (2\pi)^{m}} \, -$$
\begin{equation}
\label{epslemma4}
- \, \int_{0}^{2\pi}\!\!\!\dots\int_{0}^{2\pi}
\sum_{s=1}^{g} \, e_{s} \,
W^{(s)}_{\theta^{\alpha}}(\bm{\theta},Y) \,
T^{(s)}_{\beta}(\bm{\theta},Y) \,
{d^{m} \theta \over (2\pi)^{m}}
\end{equation}

\vspace{0.5cm}

 We can write on ${\cal M}_{\epsilon} [\bm{\Psi}_{(1)}]$

$${\delta \varphi^{i} (\bm{\theta},X) \over \delta U^{\nu} (Y)}
\,\,\, = \,\,\, {1 \over \epsilon} \,\, \Phi^{i}_{\theta^{\alpha}}
\left(\bm{\theta} + \bm{\theta}^{*}_{0}(X) + {1 \over \epsilon}
\, {\bf S} (X) , \,\, {\bf U} (X) \right) \,\,
\nu (X-Y) \,\, {\partial k^{\alpha} \over \partial U^{\nu}} (Y)
\,\,\,\,\, + $$
$$+ \,\,\, \Psi^{i}_{(1)\theta^{\alpha}}
\left(\bm{\theta} + \bm{\theta}^{*}_{0}(X) + {1 \over \epsilon}
\, {\bf S} (X) , \,\, X \right) \,\,
\nu (X-Y) \,\, {\partial k^{\alpha} \over \partial U^{\nu}} (Y)
\,\,\,\,\, + $$
$$+ \,\,\,\,\, \Phi^{i}_{U^{\nu}}
\left(\bm{\theta} + \bm{\theta}^{*}_{0}(X) + {1 \over \epsilon}
\, {\bf S} (X) , \,\, {\bf U} (X) \right) \,\, \delta (X-Y) $$

\vspace{0.5cm}

$$= \,\, {1 \over \epsilon} \,\, \varphi^{i}_{\theta^{\alpha}}
(\bm{\theta} , X ) \,\,
\nu (X-Y) \,\, {\partial k^{\alpha} \over \partial U^{\nu}} (Y)
\,\, + \,\, \Phi^{i}_{U^{\nu}}
\left(\bm{\theta} + \bm{\theta}^{*}_{0}(X) + {1 \over \epsilon}
\, {\bf S} (X) , \,\, {\bf U} (X) \right) \,\, \delta (X-Y) $$
and

$${\delta \varphi^{i} (\bm{\theta},X) \over
\delta \theta^{*\alpha}_{0} (Y)} \,\,\, = \,\,\,
\varphi^{i}_{\theta^{\alpha}} (\bm{\theta} , X ) \,\,
\delta (X-Y) $$

 We have so

$$\Omega^{1}_{\nu\mu}(X,Y) \, =
\int \!\! \left( {1 \over \epsilon} \,
{\partial k^{\alpha} \over \partial U^{\nu}} (X) \,
\nu (Z-X) \, \varphi^{i}_{\theta^{\alpha}}
(\bm{\theta}, Z) \, + \,
\delta (Z-X) \, \Phi^{i}_{U^{\nu}}
(\bm{\theta} + \dots, {\bf U}(Z)) \right)  \times $$
$$ \times \, {\tilde \Omega}_{ij}
(\bm{\theta},\bm{\theta}^{\prime},Z,W)
\left( {1 \over \epsilon} \, \varphi^{j}_{\theta^{\beta}}
(\bm{\theta}^{\prime}, W) \, \nu (W-Y) \,
{\partial k^{\beta} \over \partial U^{\mu}} (Y) \, + \,
\delta (W-Y) \, \Phi^{i}_{U^{\mu}}
(\bm{\theta}^{\prime} + \dots, {\bf U}(W)) \right) \times $$   
$$\hspace{9cm} \times \, {d^{m} \theta \over (2\pi)^{m}} \,   
{d^{m} \theta^{\prime} \over (2\pi)^{m}} \,dZ \, dW $$
   
 We can write then

$$\Omega^{1}_{\nu\mu}(X,Y) \, = \, - \, \int
{1 \over \epsilon^{2}} \,
{\partial k^{\alpha} \over \partial U^{\nu}} (X) \,\,
\nu (X-Z) \,\, \sum_{k\geq 0} \varphi^{i}_{\theta^{\alpha}}
(\bm{\theta}, Z) \,\,
\omega^{(k)}_{ij} (\bm{\varphi}(\bm{\theta},Z), \dots)
\,\,\, \times $$
$$\hspace{3cm}
\times \,\,\, \epsilon^{k} \, \varphi^{j}_{\theta^{\beta},kZ}
(\bm{\theta}, Z) \,\, \nu (Z-Y) \,\,
{\partial k^{\beta} \over \partial U^{\mu}} (Y) \,\, dZ \,
{d^{m} \theta \over (2\pi)^{m}} \,\,\, - $$

$$- \, {1 \over \epsilon^{2}} \, \int \,
{\partial k^{\alpha} \over \partial U^{\nu}} (X)
\left[\nu (X-Y) \,\, \sum_{k \geq 1} \sum_{p=1}^{k} C^{p}_{k}
(-1)^{p-1} \epsilon^{k} \, \varphi^{i}_{\theta^{\alpha}}
(\bm{\theta}, Y) \right. \,\, \times
\hspace{2cm}$$
$$\hspace{2cm} \times \,\, \left.
\omega^{(k)}_{ij} (\bm{\varphi}(\bm{\theta},Y), \dots) \,\,   
\varphi^{j}_{\theta^{\beta},(k-p)Y} (\bm{\theta}, Y)  
\right]_{(p-1)Y} \,
{\partial k^{\beta} \over \partial U^{\mu}} (Y) \,
{d^{m} \theta \over (2\pi)^{m}} \,\,\, + $$

$$+ \,\, {1 \over \epsilon} \int \Phi^{i}_{U^{\nu}}
(\bm{\theta} + \dots, {\bf U}(X)) \sum_{k \geq 0} \epsilon^{k} \,
\omega^{(k)}_{ij} (\bm{\varphi}(\bm{\theta},X), \dots) \, \times
\hspace{5cm} $$
$$\hspace{5cm} \times \,\, \left[ \varphi^{j}_{\theta^{\beta}}
(\bm{\theta}, X) \,\, \nu (X-Y) \right]_{kX}
{\partial k^{\beta} \over \partial U^{\mu}} (Y) \,
{d^{m} \theta \over (2\pi)^{m}} \,\,\, -  $$

$$- \,\,\, {1 \over \epsilon} \int
{\partial k^{\alpha} \over \partial U^{\nu}} (X)
\sum_{k \geq 0} (-1)^{k} \, \epsilon^{k} \, \left[
\nu (X-Y) \,\, \varphi^{i}_{\theta^{\alpha}}
(\bm{\theta}, Y) \,\,
\omega^{(k)}_{ij} (\bm{\varphi}(\bm{\theta},Y), \dots)
\right]_{kY} \,\, \times $$
$$\hspace{8cm} \times \,\,
\Phi^{j}_{U^{\mu}} (\bm{\theta} + \dots, {\bf U}(Y))
\, {d^{m} \theta \over (2\pi)^{m}} \,\,\, + $$

$$+ \int \sum_{k \geq 0} \epsilon^{k}
\Phi^{i}_{U^{\nu}} (\bm{\theta} + \dots, {\bf U}(X)) \,\,
\omega^{(k)}_{ij} (\bm{\varphi}(\bm{\theta},X), \dots) \,\,
\times \hspace{3cm} $$
$$\hspace{5cm} \times \,\, \left[
\Phi^{j}_{U^{\mu}} (\bm{\theta} + \dots, {\bf U}(X)) \,
\delta (X-Y) \right]_{kX} \,
{d^{m} \theta \over (2\pi)^{m}} \, -$$

$$- \,\, {1 \over \epsilon^{3}} \, \int \,
\sum_{s=1}^{g} \, e_{s} \,
{\partial k^{\alpha} \over \partial U^{\nu}} (X) \,\,
\nu (X-Z) \,\, \left[
\epsilon \, W^{(s)}_{\theta^{\alpha}Z} (\bm{\theta},Z)
\, - \, \epsilon \, T^{(s)}_{\alpha,Z} (\bm{\theta},Z) \right] \,\,
\nu (Z-W) \,\, \times \hspace{1cm}$$
$$\hspace{2cm}
\times \,\, \left[
\epsilon \, W^{(s)}_{\theta^{\beta}W} (\bm{\theta},W)
\, - \, \epsilon \, T^{(s)}_{\beta,W} (\bm{\theta},W) \right] \,\,
\nu (W-Y) \, {\partial k^{\beta} \over \partial U^{\mu}} (Y) \,
dZ \, dW \, {d^{m} \theta \over (2\pi)^{m}} \,\,\, + $$

$$+ \,\, {1 \over \epsilon^{2}} \, \int \,
\sum_{s=1}^{g} \, e_{s} \,
\Phi^{i}_{U^{\nu}} (\bm{\theta} + \dots, {\bf U}(X)) \,\,
{\delta {\hat H}^{(s)} \over \delta \varphi^{i}(\bm{\theta},X)} \,\,
\nu (X-W) \,\, \times \hspace{5cm}$$  
$$\hspace{2cm} \times \,\, \left[
\epsilon \, W^{(s)}_{\theta^{\beta}W} (\bm{\theta},W)
\, - \, \epsilon \, T^{(s)}_{\beta,W} (\bm{\theta},W) \right] 
\,\, \nu (W-Y) \,\,
{\partial k^{\beta} \over \partial U^{\mu}} (Y) \,\,
dW \, {d^{m} \theta \over (2\pi)^{m}} \,\,\, - $$

$$- \,\,\, {1 \over \epsilon^{2}} \, \int \,
\sum_{s=1}^{g} \, e_{s} \,
{\partial k^{\alpha} \over \partial U^{\nu}} (X) \,\,
\nu (X-Z) \,\, \left[
\epsilon \, W^{(s)}_{\theta^{\alpha}Z} (\bm{\theta},Z)
\, - \, \epsilon \, T^{(s)}_{\alpha,Z} (\bm{\theta},Z) \right] \,\,
\nu (Z-Y) \,\, \times \hspace{1cm} $$
$$\hspace{7cm} \times \,\,
{\delta {\hat H}^{(s)} \over \delta \varphi^{j}(\bm{\theta},Y)} \,\,
\Phi^{j}_{U^{\mu}} (\bm{\theta} + \dots, {\bf U}(Y)) \,\,
dZ \, {d^{m} \theta \over (2\pi)^{m}} \,\,\, + $$

$$+ \, {1 \over \epsilon} \, \int \, \sum_{s=1}^{g} \, e_{s} \,
\Phi^{i}_{U^{\nu}} (\bm{\theta} + \dots, {\bf U}(X)) \,\,
{\delta {\hat H}^{(s)} \over \delta \varphi^{i}(\bm{\theta},X)} \,\,
\nu (X-Y) \,\, \times \hspace{5cm}$$
$$\hspace{7cm} \times \,\,
{\delta {\hat H}^{(s)} \over \delta \varphi^{j}(\bm{\theta},Y)} \,\,
\Phi^{j}_{U^{\mu}} (\bm{\theta} + \dots, {\bf U}(Y)) \,\,
{d^{m} \theta \over (2\pi)^{m}} $$

 We should substitute now the functions $\varphi^{i}$ in the form
(\ref{epsdef}) and
we are interested here in the terms of $\epsilon$-expansion
of $\Omega^{1}_{\nu\mu}(X,Y)$ containing $1/\epsilon$ and omit all
the terms of order $O(1)$ for $\epsilon \rightarrow 0$. We can see
then that we can omit the differentiation of the function
$\nu (X-Y)$ in the second, the third, and the fourth terms of the
expression for $\Omega^{1}_{\nu\mu}(X,Y)$ since they appear only in
regular terms for $\epsilon \rightarrow 0$. By the same reason
we can omit the whole fifth term in the same expression
which is regular for $\epsilon \rightarrow 0$. The whole expression
for $\Omega^{1}_{\nu\mu}(X,Y)$ can then be rewritten (after some
calculation) in the following form

$$\Omega^{1}_{\nu\mu}(X,Y) \,\, = \,\,
- \, {1 \over \epsilon^{2}} \, \int
{\partial k^{\alpha} \over \partial U^{\nu}}(X) \,
\nu (X-Z) \, \sum_{k\geq 0}
\varphi^{i}_{\theta^{\alpha}} (\bm{\theta}, Z) \, \epsilon^{k} \,
\omega^{(k)}_{ij} \left( \bm{\varphi}(\bm{\theta},Z), \dots \right) 
\,\, \times $$
$$\hspace{5cm} \times \,\,
\varphi^{j}_{\theta^{\beta},kZ} (\bm{\theta}, Z) \,
\nu (Z-Y) \, {\partial k^{\beta} \over \partial U^{\mu}}(Y) \,
dZ \, {d^{m} \theta \over (2\pi)^{m}} \,\, -$$
 
$$-  {1 \over \epsilon^{2}} \,
{\partial k^{\alpha} \over \partial U^{\nu}}(X) \,
\nu (X-Y) \, \int \sum_{k\geq 1} \sum_{p=1}^{k} C_{k}^{p} \,
(-1)^{p-1} \, \epsilon^{k} \,\, \times \hspace{5cm} $$
$$\times  \left[ \varphi^{i}_{\theta^{\alpha}} (\bm{\theta}, Y) \,
\omega^{(k)}_{ij} ( \bm{\varphi}(\bm{\theta},Y), \dots)
\, \varphi^{j}_{\theta^{\beta},(k-p)Y}
(\bm{\theta}, Y) \, \right]_{(p-1)Y}
\, {\partial k^{\beta} \over \partial U^{\mu}}(Y) \,   
{d^{m} \theta \over (2\pi)^{m}} +$$

$$+ \,\, {1 \over \epsilon} \, \int \Phi^{i}_{U^{\nu}}
\left( \bm{\theta} + \dots , {\bf U}(X) \right) \, \sum_{k\geq 0} \,
\omega^{(k)}_{ij} \left( \bm{\varphi}(\bm{\theta},X), \dots \right)
\, \epsilon^{k} \,\, \times \hspace{5cm} $$
$$\hspace{5cm} \times \,\, \varphi^{j}_{\theta^{\beta},kX}
(\bm{\theta}, X) \, \nu (X-Y) \,
{\partial k^{\beta} \over \partial U^{\mu}}(Y) \,
{d^{m} \theta \over (2\pi)^{m}} \,\, -$$

$$- \,\, {1 \over \epsilon} \,
{\partial k^{\alpha} \over \partial U^{\nu}}(X) \, 
\nu (X-Y) \, \int \sum_{k\geq 0} (-1)^{k} \, \epsilon^{k} \,\,\,
\times \hspace{7cm}$$
$$\hspace{1cm} \times \,\, \left[
\varphi^{i}_{\theta^{\alpha}} (\bm{\theta}, Y) \,
\omega^{(k)}_{ij} \left( \bm{\varphi}(\bm{\theta},Y), \dots \right)
\right]_{kY} \, \Phi^{j}_{U^{\mu}}
\left( \bm{\theta} + \dots , {\bf U}(Y) \right) \,
{d^{m} \theta \over (2\pi)^{m}} \,\, -$$

$$-  {1 \over \epsilon} \, \int \,
{\partial k^{\alpha} \over \partial U^{\nu}}(X) \,
\nu (X-Z) \,\,\, \times \hspace{9cm}$$
$$\times \sum_{s=1}^{g} \, e_{s} \, \left[
W^{(s)}_{\theta^{\alpha}Z}(\bm{\theta},Z) \,
W^{(s)}_{\theta^{\beta}}(\bm{\theta},Z) \, - \, {1 \over 2} \,
\left( W^{(s)}_{\theta^{\alpha}Z}(\bm{\theta},Z) \, - \,
T^{(s)}_{\alpha,Z} (\bm{\theta},Z) \right) \,
W^{(s)}_{\theta^{\beta}} (\bm{\theta},+\infty) \right. -$$
$$- \, \left( T^{(s)}_{\alpha} (\bm{\theta},Z) \,
W^{(s)}_{\theta^{\beta}} (\bm{\theta},Z) \right)_{Z} \, + \,
{1 \over \epsilon} \, h^{(s)}_{\theta^{\beta}} (\bm{\theta},Z) \,
T^{(s)}_{\alpha} (\bm{\theta},Z) \, - \,
{1 \over \epsilon} \, h^{(s)}_{\theta^{\alpha}} (\bm{\theta},Z) \,
T^{(s)}_{\beta} (\bm{\theta},Z) \, + $$
$$+ \left.
T^{(s)}_{\alpha,Z} (\bm{\theta},Z) \,
T^{(s)}_{\beta} (\bm{\theta},Z) \right] \,
\nu (Z-Y) \, {\partial k^{\beta} \over \partial U^{\mu}}(Y) \,\,
dZ \, {d^{m} \theta \over (2\pi)^{m}} \,\, +$$   

$$+ \,\, {1 \over \epsilon} \, \int \,
\sum_{s=1}^{g} \, e_{s} \,
{\partial k^{\alpha} \over \partial U^{\nu}}(X) \,
\left[ W^{(s)}_{\theta^{\alpha}}(\bm{\theta},X) \, - \,
{1 \over 2} \, W^{(s)}_{\theta^{\alpha}}(\bm{\theta},+\infty)
\, - \, T^{(s)}_{\alpha} (\bm{\theta},X) \right] \,\,
\times \hspace{2cm} $$
$$\hspace{3cm} \times \,\, \nu (X-Y) \, \left[
W^{(s)}_{\theta^{\beta}}(\bm{\theta},Y) \, - \,
T^{(s)}_{\beta} (\bm{\theta},Y) \right] \,
{\partial k^{\beta} \over \partial U^{\mu}}(Y) \,
{d^{m} \theta \over (2\pi)^{m}} \,\, -$$

$$- \,\, {1 \over \epsilon} \, \int \,
\sum_{s=1}^{g} \, e_{s} \,
{\partial k^{\alpha} \over \partial U^{\nu}}(X) \,
\nu (X-Y) \, \left[
W^{(s)}_{\theta^{\alpha}}(\bm{\theta},Y) \, - \,
T^{(s)}_{\alpha} (\bm{\theta},Y) \right] \,\,
\times \hspace{5cm} $$
$$\hspace{5cm} \times \,\, \left[
W^{(s)}_{\theta^{\beta}}(\bm{\theta},Y) \, - \,
T^{(s)}_{\beta} (\bm{\theta},Y) \right] \,
{\partial k^{\beta} \over \partial U^{\mu}}(Y) \,
{d^{m} \theta \over (2\pi)^{m}} \,\, +$$

$$+ \,\, {1 \over \epsilon} \, \int \,
\sum_{s=1}^{g} \, e_{s} \, \Phi^{i}_{U^{\nu}}
\left( \bm{\theta} + \dots , {\bf U}(X) \right) \,
{\delta {\hat H}^{(s)} \over \delta \varphi^{i}(\bm{\theta},X)}
\,\, \times \hspace{7cm} $$
$$\hspace{2cm} \times \,\, \left[
W^{(s)}_{\theta^{\beta}}(\bm{\theta},X) \, - \, {1 \over 2} \,
W^{(s)}_{\theta^{\beta}}(\bm{\theta},+\infty) \, - \,
T^{(s)}_{\beta} (\bm{\theta},X) \right] \,
\nu (X-Y) \, {\partial k^{\beta} \over \partial U^{\mu}}(Y) \,
{d^{m} \theta \over (2\pi)^{m}} \,\, -$$

$$- \,\, {1 \over \epsilon} \, \int \,
\sum_{s=1}^{g} \, e_{s} \, \Phi^{i}_{U^{\nu}}
\left( \bm{\theta} + \dots , {\bf U}(X) \right) \,
{\delta {\hat H}^{(s)} \over \delta \varphi^{i}(\bm{\theta},X)}
\, \nu (X-Y) \,\, \times \hspace{5cm}$$
$$\hspace{5cm} \times \,\,
\left[ W^{(s)}_{\theta^{\beta}}(\bm{\theta},Y)
\, - \, T^{(s)}_{\beta} (\bm{\theta},Y) \right] \,
{\partial k^{\beta} \over \partial U^{\mu}}(Y) \,
{d^{m} \theta \over (2\pi)^{m}} \,\, -$$  

$$- \,\, {1 \over \epsilon} \, \int \,  
\sum_{s=1}^{g} \, e_{s} \,
{\partial k^{\alpha} \over \partial U^{\nu}}(X) \, \left[
W^{(s)}_{\theta^{\alpha}}(\bm{\theta},X) \, - \, {1 \over 2} \,
W^{(s)}_{\theta^{\alpha}}(\bm{\theta},+\infty) \, - \,
T^{(s)}_{\alpha} (\bm{\theta},X) \right] \,\,
\times \hspace{2cm} $$
$$\hspace{5cm} \times \,\, \nu (X-Y) \,
{\delta {\hat H}^{(s)} \over \delta \varphi^{j}(\bm{\theta},Y)} \,\,
\Phi^{j}_{U^{\mu}}
\left( \bm{\theta} + \dots , {\bf U}(Y) \right) \,
{d^{m} \theta \over (2\pi)^{m}} \,\, +$$

$$+ \,\, {1 \over \epsilon} \, \int \,
\sum_{s=1}^{g} \, e_{s} \, \Phi^{i}_{U^{\nu}}
\left( \bm{\theta} + \dots , {\bf U}(X) \right) \,
{\delta {\hat H}^{(s)} \over \delta \varphi^{i}(\bm{\theta},X)} \,
\nu (X-Y) \,\, \times \hspace{5cm} $$
$$\hspace{5cm} \times \,\,
{\delta {\hat H}^{(s)} \over \delta \varphi^{j}(\bm{\theta},Y)} \,\,
\Phi^{j}_{U^{\mu}}
\left( \bm{\theta} + \dots , {\bf U}(Y) \right) \,
{d^{m} \theta \over (2\pi)^{m}} \,\,\,\,\,\,\,\, + \,\,\, O(1)$$

\vspace{0.5cm}

 Let us now consider specially the functions
$W^{(s)}_{\theta^{\alpha}} (\bm{\theta},X)$. We first consider the
sub-manifold ${\cal M}_{0}$ and represent the functions
$\bm{\varphi}_{[{\bf U},\bm{\theta}^{*}]}$ in the form

\begin{equation}
\label{fm0}
\varphi^{i} (\bm{\theta},X) \,\,\, = \,\,\,
\Phi^{i} \left(\bm{\theta} + \bm{\theta}^{*}_{0}(X) +
{1 \over \epsilon} \, {\bf S} (X) , \,\, {\bf U} (X) \right)
\end{equation}

Let us recall the commuting flows (\ref{commfl})
for the system (\ref{dynsyst}) and the corresponding relations
(\ref{jsa}) for the functions
$h^{(s)} (\bm{\varphi}, \bm{\varphi}_{x}, \dots)$. We can write   
in the new "slow" variables $X,T$:   

$$\epsilon \, \varphi^{i}_{T^{\alpha}} \,\,\, = \,\,\,
Q^{i}_{(\alpha)} (\bm{\varphi},\epsilon \, \bm{\varphi}_{X},\dots)$$
and the relations (\ref{jsa}) become now

$$h^{(s)}_{T^{\alpha}} \,\,\, \equiv \,\,\, \partial_{X} \,
J^{(s)}_{\alpha} (\bm{\varphi},\epsilon \, \bm{\varphi}_{X},\dots)$$

 Let us represent the operator $\epsilon \, \partial_{X}$ on the
functions (\ref{fm0}) in the following form:

$$\epsilon \, \partial_{X} \,\, = \,\, \partial^{I}_{X} \, + \,
\epsilon \, \partial^{II}_{X}$$
where

$$\partial^{I}_{X} \,\, = \,\, S^{\alpha}_{X} \,
{\partial \over \partial \theta^{\alpha}} \,\, = \,\,
k^{\alpha}(X) \, {\partial \over \partial \theta^{\alpha}}  
\,\,\,\,\, , \,\,\,\,\,\,\,\,
\partial^{II}_{X} \,\, = \,\, U^{\nu}_{X} \,
{\partial \over \partial U^{\nu}} \, + \,
\theta^{*\alpha}_{0X} \,
{\partial \over \partial \theta^{\alpha}}$$

 We can write on the manifold ${\cal M}_{0}$

$$\omega^{\eta}_{(\alpha)} ({\bf U}(X)) \,
{\partial \over \partial \theta^{\eta}} \,
h^{(s)} (\bm{\varphi}, \partial^{I}_{X} \bm{\varphi}, \dots)
\,\,\, = \,\,\, \partial^{I}_{X} \, J^{(s)}_{\alpha}
(\bm{\varphi}, \partial^{I}_{X} \bm{\varphi}, \dots) $$
or using the relations (\ref{gammamatr}):

$${\partial \over \partial \theta^{\alpha}} \,
h^{(s)} (\bm{\varphi}, \partial^{I}_{X} \bm{\varphi}, \dots)
\,\,\, = \,\,\, \gamma^{\delta}_{\alpha} ({\bf U}(X)) \,\,\,
\partial^{I}_{X} J^{(s)}_{\delta}
(\bm{\varphi}, \partial^{I}_{X} \bm{\varphi}, \dots)
\,\,\, = $$
$$= \,\,\, \gamma^{\delta}_{\alpha} ({\bf U}(X)) \, \left[
\epsilon \, \partial_{X} J^{(s)}_{\delta}
(\bm{\varphi}, \partial^{I}_{X} \bm{\varphi}, \dots) \,\, - \,\,
\epsilon \, \partial^{II}_{X} J^{(s)}_{\delta}
(\bm{\varphi}, \partial^{I}_{X} \bm{\varphi}, \dots) \right] $$

 We have then on ${\cal M}_{0}$

$$h^{(s)}_{\theta^{\alpha}}
(\bm{\varphi}, \epsilon \, \bm{\varphi}_{X}, \dots) \,\, = \,\,
\epsilon \, \left[ \gamma^{\delta}_{\alpha} ({\bf U}) \,\,
J^{(s)}_{\delta} (\bm{\varphi},\partial^{I}_{X}\bm{\varphi},\dots)
\right]_{X} \,\, - $$
$$- \,\, \epsilon \, \left[
\left( \gamma^{\delta}_{\alpha} ({\bf U}) \right)_{X} \,
J^{(s)}_{\delta} (\bm{\varphi},\partial^{I}_{X}\bm{\varphi},\dots)
\, + \, \gamma^{\delta}_{\alpha} ({\bf U}) \,\,
\partial^{II}_{X} J^{(s)}_{\delta}
(\bm{\varphi}, \partial^{I}_{X} \bm{\varphi}, \dots) \right]
\,\, +$$
\begin{equation}
\label{htarel0}
+ \,\, \epsilon \, {\partial \over \partial \theta^{\alpha}} \,
\delta h^{(s)} (\bm{\varphi}, \dots) \,\, + \,\,
O(\epsilon^{2})
\end{equation}
where

$$\delta h^{(s)} (\bm{\varphi}, \dots) \,\, = \,\,
{\partial h^{(s)} \over \partial \varphi^{i}_{x}}
(\bm{\varphi}, \partial^{I}_{X} \bm{\varphi}, \dots) \,
\partial^{II}_{X} \varphi^{i} \,\, + \,\,
{\partial h^{(s)} \over \partial \varphi^{i}_{xx}}
(\bm{\varphi}, \partial^{I}_{X} \bm{\varphi}, \dots) \,
(\partial^{I}_{X} \partial^{II}_{X} \, + \,
\partial^{II}_{X} \partial^{I}_{X}) \varphi^{i} \,\, + \,\,
\dots $$
and the functions $\bm{\varphi}(\bm{\theta},X)$ have the form
(\ref{fm0}).

 Let us now come back to the sub-manifolds
${\cal M}_{\epsilon}[\bm{\Psi}_{(1)}]$ and consider the functions
$\bm{\varphi}_{[{\bf U},\bm{\theta}^{*}]}$ having the form
(\ref{epsdef}). We can see that the relations (\ref{htarel0})
can be rewritten then in the form

$$h^{(s)}_{\theta^{\alpha}}
(\bm{\varphi}, \epsilon \, \bm{\varphi}_{X}, \dots) \,\, = \,\,
\epsilon \, \left[ \gamma^{\delta}_{\alpha} ({\bf U}) \,\,
J^{(s)}_{\delta} (\bm{\varphi},\partial^{I}_{X}\bm{\varphi},\dots)
\right]_{X} \,\, - $$
$$- \,\, \epsilon \, \left[
\left( \gamma^{\delta}_{\alpha} ({\bf U}) \right)_{X} \,
J^{(s)}_{\delta} (\bm{\varphi},\partial^{I}_{X}\bm{\varphi},\dots)
\, + \, \gamma^{\delta}_{\alpha} ({\bf U}) \,\,  
\partial^{II}_{X} J^{(s)}_{\delta}
(\bm{\varphi}, \partial^{I}_{X} \bm{\varphi}, \dots) \right]
\,\, +$$
\begin{equation}
\label{htarel1}
+ \,\, \epsilon \, {\partial \over \partial \theta^{\alpha}} \,
{\tilde {\delta h}}^{(s)} (\bm{\varphi}, \dots) \,\, + \,\,
O(\epsilon^{2})
\end{equation}
where
 
$${\tilde {\delta h}}^{(s)} \,\, = \,\, \delta h^{(s)} \, + \,   
{\partial h^{(s)} \over \partial \varphi^{i}} \Psi^{i}_{(1)} \, + \,
{\partial h^{(s)} \over \partial \varphi^{i}_{x}}
\epsilon \Psi^{i}_{(1)X} \, + \, \dots $$

 We can write now on ${\cal M}_{\epsilon}[\bm{\Psi}_{(1)}]$

$$W^{(s)}_{\theta^{\alpha}} (\bm{\theta},X) \,\, = \,\,   
{1 \over \epsilon} \, \int_{-\infty}^{+\infty} \nu (X-W) \,
h^{(s)}_{\theta^{\alpha}} (\bm{\varphi},W) \, dW \,\, =$$
$$= \,\, \gamma^{\delta}_{\alpha} ({\bf U}(X)) \,\,
J^{(s)}_{\delta}
(\bm{\varphi}, \partial^{I}_{X} \bm{\varphi}, \dots) \,\, - $$
$$- \,\, \int_{-\infty}^{+\infty} \nu (X-W) \,\, \partial^{II}_{W}
\left( \gamma^{\delta}_{\alpha} ({\bf U}(W)) \,\,
J^{(s)}_{\delta}
(\bm{\varphi}, \partial^{I}_{W} \bm{\varphi}, \dots) \right) \,
dW \,\, + $$
\begin{equation}
\label{wform}
+ \,\, \int_{-\infty}^{+\infty} \nu (X-W) \,\,
{\partial \over \partial \theta^{\alpha}} \,
{\tilde {\delta h}}^{(s)} (\bm{\theta},W) \, dW \,\,\, + \,\,\,
O(\epsilon)
\end{equation}
(we use here the operator $\partial^{II}_{W}$ also as
$\partial_{W}$ for the functions
$\gamma^{\delta}_{\alpha} ({\bf U})$
depending on ${\bf U}$ only and assume the normalization of
$J^{(s)}_{\delta} (\bm{\varphi},\dots)$ such that
$J^{(s)}_{\delta} (\bm{\theta}, \pm \infty) = 0$ on ${\cal M}$).

 We can see that the quantities $W^{(s)}_{\theta^{\alpha}}$
have the order $O(1)$ for $\epsilon \rightarrow 0$ and
the fixed coordinates ${\bf U}(x)$, $\bm{\theta}_{0}(X)$ 
on ${\cal M}_{\epsilon}[\bm{\Psi}_{(1)}]$.

 We evidently have also

\begin{equation}
\label{intw}
\int_{0}^{2\pi}\!\!\!\dots\int_{0}^{2\pi}
W^{(s)}_{\theta^{\alpha}} (\bm{\theta},X) \,
{d^{m} \theta \over (2\pi)^{m}} \,\, = \,\, 0
\end{equation}

 Let us consider now in main order of $\epsilon$ the arbitrary
value of the form

$$\int_{0}^{2\pi}\!\!\!\dots\int_{0}^{2\pi}
V (\bm{\theta},X) \, W^{(s)}_{\theta^{\alpha}} (\bm{\theta},Y) \,
{d^{m} \theta \over (2\pi)^{m}} $$
where $V (\bm{\theta},X)$ is arbitrary smooth and periodic w.r.t.
$\bm{\theta}$ function (we can have in particular $X = Y$).

 We have

$$\int_{0}^{2\pi}\!\!\!\dots\int_{0}^{2\pi}
V (\bm{\theta},X) \, W^{(s)}_{\theta^{\alpha}} (\bm{\theta},Y) \,
{d^{m} \theta \over (2\pi)^{m}} \,\, = \,\,
\gamma^{\delta}_{\alpha}(Y) \,
\int_{0}^{2\pi}\!\!\!\dots\int_{0}^{2\pi}
V (\bm{\theta},X) J^{(s)}_{\delta}(\bm{\theta},Y)
{d^{m} \theta \over (2\pi)^{m}} \,\, - $$
$$- \,\, \int_{0}^{2\pi}\!\!\!\dots\int_{0}^{2\pi}
V (\bm{\theta},X) \, \int_{-\infty}^{+\infty} \nu (Y-W) \,\,
\partial^{II}_{W} \left( \gamma^{\delta}_{\alpha}(W) \,
J^{(s)}_{\delta}
(\bm{\varphi}, \partial^{I}_{W} \bm{\varphi}, \dots) \right) \,
dW \, {d^{m} \theta \over (2\pi)^{m}} \,\, + $$
\begin{equation}
\label{vwint}
+ \,\, \int_{0}^{2\pi}\!\!\!\dots\int_{0}^{2\pi}
V (\bm{\theta},X) \, \int_{-\infty}^{+\infty} \nu (Y-W) \,\,
{\partial \over \partial \theta^{\alpha}} \,\,
{\tilde {\delta h}}^{(s)} (\bm{\varphi}(\bm{\theta},W),  \dots) \,
dW \, {d^{m} \theta \over (2\pi)^{m}} \,\,\, + \,\,\, O(\epsilon)
\end{equation}

 The expressions
$J^{(s)}_{\delta} (\bm{\varphi}(\bm{\theta},W),  \dots)$
and $\delta \, h^{(s)} (\bm{\varphi}(\bm{\theta},W),  \dots)$
are the rapidly oscillating functions of $W$ due to the fast
change of the phase according to (\ref{epsdef}). It's not difficult
to show that in the main order of $\epsilon$ the expression
(\ref{vwint}) is given by the independent integration w.r.t.
$\theta$ at the points
$X$ and $W$ integrated then w.r.t. $W$ for smooth generic
${\bf S}(W)$. We can see then that the third term in (\ref{vwint})
disappears in fact in the main order of $\epsilon$.
After that we can also replace in the main order of $\epsilon$
the integration w.r.t.
$\bm{\theta}$ just by the averaging on the family $\Lambda$
in the first two terms of (\ref{vwint})
since the $\epsilon \bm{\Psi}_{(1)}$-corrections give
there just the values of order $O(\epsilon)$. We can write then
on ${\cal M}_{\epsilon} [\bm{\Psi}_{(1)}]$ in the
main order of $\epsilon$

$$\int_{0}^{2\pi}\!\!\!\dots\int_{0}^{2\pi}   
V (\bm{\theta},X) \, W^{(s)}_{\theta^{\alpha}} (\bm{\theta},Y) \, 
{d^{m} \theta \over (2\pi)^{m}} \,\, = \,\,
\gamma^{\delta}_{\alpha}(Y) \, \langle V (\bm{\theta},X)
J^{(s)}_{\delta} (\bm{\theta},Y) \rangle
\,\, - $$
$$- \,\, \langle V (\bm{\theta},X) \rangle
\int_{-\infty}^{+\infty} \nu (Y-W) \,\,
\partial_{W} \left( \gamma^{\delta}_{\alpha}(W) \,
\langle J^{(s)}_{\delta} (\bm{\theta},W) \rangle \right) \,
d W \,\, + \,\, o(1) \,\, = $$
\begin{equation}
\label{vjaver}
= \,\, \gamma^{\delta}_{\alpha}(Y) \, \left[
\langle V (\bm{\theta},X) \, J^{(s)}_{\delta} (\bm{\theta},Y)
\rangle \, - \, \langle V (\bm{\theta},X) \rangle \,
\langle J^{(s)}_{\delta} (\bm{\theta},Y) \rangle \right]
\,\, + \,\, o(1)
\end{equation}

 We can write also the following relation

$$\int_{0}^{2\pi}\!\!\!\dots\int_{0}^{2\pi} V (\bm{\theta},X) \,
W^{(s)}_{\theta^{\alpha}} (\bm{\theta},\pm \infty) \,
{d^{m} \theta \over (2\pi)^{m}} \,\, = \,\, o(1) $$
for $\epsilon \rightarrow 0$ which follows from the formula
(\ref{vjaver}) when we use
$J^{(s)}_{\delta} (\bm{\theta},\pm \infty) \, = \, 0$ on
${\cal M}$.

 Looking now at the expression for $\Omega^{1}_{\nu\mu}(X,Y)$
we can see that all the terms containing the values like
$W^{(s)}_{\theta^{\alpha}} (\bm{\theta},\pm \infty)$ can be actually
omitted in the main ($1/\epsilon$) order of
$\Omega^{1}_{\nu\mu}(X,Y)$ according to the remark above.  

 Using the formula (\ref{epsdivform}) we can write now

$$ - \, {1 \over \epsilon^{2}} \int
{\partial k^{\alpha} \over \partial U^{\nu}}(X) \,
\nu (X-Z) \,\, \sum_{k \geq 0} \varphi^{i}_{\theta^{\alpha}}
(\bm{\theta}, Z) \, \epsilon^{k} \,
\omega^{(k)}_{ij} (\bm{\varphi}(\bm{\theta},Z), \dots ) \,\,
\times \hspace{2cm}$$
$$\hspace{5cm} \times \,\, \varphi^{j}_{\theta^{\beta},kZ}
(\bm{\theta}, Z) \, \nu (Z-Y) \,\,
{\partial k^{\beta} \over \partial U^{\mu}}(Y) \,\, dZ \,
{d^{m} \theta \over (2\pi)^{m}} \,\, -$$

$$- \, {1 \over \epsilon^{2}} \int
{\partial k^{\alpha} \over \partial U^{\nu}}(X) \,
\nu (X-Z) \,\, \sum_{s=1}^{g} \, e_{s} \,
\left[ h^{(s)}_{\theta^{\beta}} (\bm{\theta},Z) \,
T^{(s)}_{\alpha} (\bm{\theta},Z) \, - \,
h^{(s)}_{\theta^{\alpha}} (\bm{\theta},Z) \,
T^{(s)}_{\beta} (\bm{\theta},Z) \,\, + \right. $$
$$\hspace{2cm} \left. + \,\, \epsilon \,
T^{(s)}_{\alpha,Z} (\bm{\theta},Z) \,
T^{(s)}_{\beta} (\bm{\theta},Z) \right] \, \nu (Z-Y) \,\,
{\partial k^{\beta} \over \partial U^{\mu}}(Y) \,\, dZ \,  
{d^{m} \theta \over (2\pi)^{m}} \,\, =$$
 
$$= \,\, - \, {1 \over \epsilon} \int
{\partial k^{\alpha} \over \partial U^{\nu}}(X) \,
\nu (X-Z) \,\, \left[ A_{\alpha\beta}
(\bm{\varphi}(\bm{\theta},Z), \dots ) \right]_{Z} \,\,
\nu (Z-Y) \,\, {\partial k^{\beta} \over \partial U^{\mu}}(Y)
\,\, dZ \, {d^{m} \theta \over (2\pi)^{m}} \,\, =$$

$$= \,\, - \, {1 \over \epsilon} \int
{\partial k^{\alpha} \over \partial U^{\nu}}(X) \,
A_{\alpha\beta} (\bm{\varphi}(\bm{\theta},Z), \dots ) \,\,
\nu (X-Y) \,\, {\partial k^{\beta} \over \partial U^{\mu}}(Y)
\,\, {d^{m} \theta \over (2\pi)^{m}} \,\, +$$

$$+ \,\, {1 \over \epsilon} \int
{\partial k^{\alpha} \over \partial U^{\nu}}(X) \,\,
\nu (X-Y) \,\,
A_{\alpha\beta} (\bm{\varphi}(\bm{\theta},Y), \dots ) \,\,
{\partial k^{\beta} \over \partial U^{\mu}}(Y)
\,\, {d^{m} \theta \over (2\pi)^{m}} $$

 We will also use the identity

$$W^{(s)}_{\theta^{\alpha},Z} (\bm{\theta},Z) \,
W^{(s)}_{\theta^{\beta}} (\bm{\theta},Z) \,\, = \,\,
{1 \over 2} \, \left( W^{(s)}_{\theta^{\alpha}} \,
W^{(s)}_{\theta^{\beta}} \right)_{Z} \, + \, {1 \over 2} \,
\left( W^{(s)}_{Z} \, W^{(s)}_{\theta^{\beta}}
\right)_{\theta^{\alpha}} \, - \, {1 \over 2} \, \left(
W^{(s)}_{Z} \, W^{(s)}_{\theta^{\alpha}}
\right)_{\theta^{\beta}} $$
and so

$$- \, {1 \over \epsilon} \int
{\partial k^{\alpha} \over \partial U^{\nu}}(X) \,
\nu (X-Z) \,\, \sum_{s=1}^{g} \, e_{s} \, \left[
W^{(s)}_{\theta^{\alpha},Z} (\bm{\theta},Z) \,
W^{(s)}_{\theta^{\beta}} (\bm{\theta},Z) \, - \,
{1 \over 2} \, W^{(s)}_{\theta^{\alpha},Z} (\bm{\theta},Z) \,
W^{(s)}_{\theta^{\beta}} (\bm{\theta},+\infty) \, - \right. $$
$$\left. - \,\, \left(
T^{(s)}_{\alpha} (\bm{\theta},Z) \,
W^{(s)}_{\theta^{\beta}} (\bm{\theta},Z) \right)_{Z} \right] \,\,
\nu (Z-Y) \,\, {\partial k^{\beta} \over \partial U^{\mu}}(Y)
\,\, dZ \, {d^{m} \theta \over (2\pi)^{m}} \,\, =$$

$$= \,\, - \, {1 \over \epsilon} \int
{\partial k^{\alpha} \over \partial U^{\nu}}(X) \,
\sum_{s=1}^{g} \, e_{s} \, \left[ {1 \over 2} \,
W^{(s)}_{\theta^{\alpha}} (\bm{\theta},X) \,
W^{(s)}_{\theta^{\beta}} (\bm{\theta},X) \, - \, {1 \over 2} \,
W^{(s)}_{\theta^{\alpha}} (\bm{\theta},X) \,
W^{(s)}_{\theta^{\beta}} (\bm{\theta},+\infty) \, + \right. $$
$$\left. + \, {1 \over 4} \, W^{(s)}_{\theta^{\alpha}} 
(\bm{\theta},+\infty) \,
W^{(s)}_{\theta^{\beta}} (\bm{\theta},+\infty) \, - \,
T^{(s)}_{\alpha} (\bm{\theta},X) \,
W^{(s)}_{\theta^{\beta}} (\bm{\theta},X) \right] \,
\nu (X-Y) \,\, {\partial k^{\beta} \over \partial U^{\mu}}(Y)
\,\, {d^{m} \theta \over (2\pi)^{m}} \,\, +$$

$$ + \,\, {1 \over \epsilon} \int
{\partial k^{\alpha} \over \partial U^{\nu}}(X) \,
\nu (X-Y) \,\, \sum_{s=1}^{g} \, e_{s} \,
\left[ {1 \over 2} \, W^{(s)}_{\theta^{\alpha}} (\bm{\theta},Y) \,
W^{(s)}_{\theta^{\beta}} (\bm{\theta},Y) \, - \right. 
\hspace{3cm} $$
$$\hspace{2cm} \left. - \, {1 \over 2} \,
W^{(s)}_{\theta^{\alpha}} (\bm{\theta},Y) \,
W^{(s)}_{\theta^{\beta}} (\bm{\theta},+\infty) \, - \,  
T^{(s)}_{\alpha} (\bm{\theta},Y) \,
W^{(s)}_{\theta^{\beta}} (\bm{\theta},Y) \right] \,
{\partial k^{\beta} \over \partial U^{\mu}}(Y)
\,\, {d^{m} \theta \over (2\pi)^{m}} $$

 Using also (\ref{epslemma4}) and the remark above we can write 
then

$$\Omega^{1}_{\nu\mu}(X,Y) \,\, = \,\, - \, 
{1 \over \epsilon} \int
{\partial k^{\alpha} \over \partial U^{\nu}}(X) \,
A_{\alpha\beta} (\bm{\varphi}(\bm{\theta},X), \dots) \,\,
\nu (X-Y) \,\, {\partial k^{\beta} \over \partial U^{\mu}}(Y)
\,\, {d^{m} \theta \over (2\pi)^{m}} \,\, +$$

$$+ \,\, {1 \over \epsilon} \int
{\partial k^{\alpha} \over \partial U^{\nu}}(X) \,
\sum_{s=1}^{g} \, e_{s} \, T^{(s)}_{\alpha} (\bm{\theta},X) \,
W^{(s)}_{\theta^{\beta}} (\bm{\theta},X) \,\, \nu (X-Y) \,\,
{\partial k^{\beta} \over \partial U^{\mu}}(Y)
\,\, {d^{m} \theta \over (2\pi)^{m}} \,\, -$$

$$- \,\, {1 \over \epsilon} \int
{\partial k^{\alpha} \over \partial U^{\nu}}(X) \,
\sum_{s=1}^{g} \, e_{s} \, {1 \over 2} \,
W^{(s)}_{\theta^{\alpha}} (\bm{\theta},X) \,
W^{(s)}_{\theta^{\beta}} (\bm{\theta},X) \,\, \nu (X-Y) \,\,
{\partial k^{\beta} \over \partial U^{\mu}}(Y)
\,\, {d^{m} \theta \over (2\pi)^{m}} \,\, -$$ 

$$- \,\, {1 \over \epsilon} \int
{\partial k^{\alpha} \over \partial U^{\nu}}(X) \,\,
\nu (X-Y) \,\, A_{\beta\alpha} (\bm{\varphi}(\bm{\theta},Y), \dots)
\,\, {\partial k^{\beta} \over \partial U^{\mu}}(Y)
\,\, {d^{m} \theta \over (2\pi)^{m}} \,\, +$$

$$+ \,\, {1 \over \epsilon} \int
{\partial k^{\alpha} \over \partial U^{\nu}}(X) \,\,
\nu (X-Y) \,\, \sum_{s=1}^{g} \, e_{s} \,
T^{(s)}_{\beta} (\bm{\theta},Y) \,
W^{(s)}_{\theta^{\alpha}} (\bm{\theta},Y) \,
{\partial k^{\beta} \over \partial U^{\mu}}(Y)
\,\, {d^{m} \theta \over (2\pi)^{m}} \,\, -$$

$$- \,\, {1 \over \epsilon} \int
{\partial k^{\alpha} \over \partial U^{\nu}}(X) \,\,
\nu (X-Y) \,\, \sum_{s=1}^{g} \, e_{s} \, {1 \over 2} \,
W^{(s)}_{\theta^{\alpha}} (\bm{\theta},Y) \,
W^{(s)}_{\theta^{\beta}} (\bm{\theta},Y) \,
{\partial k^{\beta} \over \partial U^{\mu}}(Y)
\,\, {d^{m} \theta \over (2\pi)^{m}} \,\, +$$

$$+ \,\, {1 \over \epsilon} \int
\Phi^{i}_{U^{\nu}} (\bm{\theta} + \dots, {\bf U}(X)) \,\,
\sum_{k \geq 0} \omega^{(k)}_{ij}
(\bm{\varphi}(\bm{\theta},X), \dots) \, \epsilon^{k} \,\,
\times \hspace{5cm}$$
$$\hspace{5cm} \times \,\, \Phi^{i}_{\theta^{\beta},kX}
(\bm{\theta} + \dots, {\bf U}(X)) \,\, \nu (X-Y) \,\,
{\partial k^{\beta} \over \partial U^{\mu}}(Y)
\,\, {d^{m} \theta \over (2\pi)^{m}} \,\, -$$

$$- \,\, {1 \over \epsilon} \int
{\partial k^{\alpha} \over \partial U^{\nu}}(X) \,\,
\nu (X-Y) \,\, \sum_{k \geq 0} (-1)^{k} \, \epsilon^{k} \, \left[
\Phi^{i}_{\theta^{\alpha}} (\bm{\theta} + \dots, {\bf U}(Y)) \,\,
\omega^{(k)}_{ij} (\bm{\varphi}(\bm{\theta},Y), \dots)
\right]_{kY} \,\, \times $$
$$\hspace{7cm} \times \,\,
\Phi^{j}_{U^{\mu}} (\bm{\theta} + \dots, {\bf U}(Y))
\,\, {d^{m} \theta \over (2\pi)^{m}} \,\, +$$

$$+ \,\, {1 \over \epsilon} \int \sum_{s=1}^{g} \, e_{s} \,
\Phi^{i}_{U^{\nu}} (\bm{\theta} + \dots, {\bf U}(X)) \,\,
{\delta {\hat H}^{(s)} \over \delta \varphi^{i}(\bm{\theta},X)}
\left( W^{(s)}_{\theta^{\beta}} (\bm{\theta},X) \, - \,
T^{(s)}_{\beta} (\bm{\theta},X) \right) \,\, \times
\hspace{1cm} $$
$$\hspace{5cm} \times \,\, \nu (X-Y) \,\,
{\partial k^{\beta} \over \partial U^{\mu}}(Y)
\,\, {d^{m} \theta \over (2\pi)^{m}} \,\, +$$

$$+ \,\, {1 \over \epsilon} \int \sum_{s=1}^{g} \, e_{s} \,
{\partial k^{\alpha} \over \partial U^{\nu}}(X) \,\, \nu (X-Y) \,\,
\times \hspace{7cm} $$
$$\hspace{2cm} \times \,\, \left(
W^{(s)}_{\theta^{\alpha}} (\bm{\theta},Y) \, - \,
T^{(s)}_{\alpha} (\bm{\theta},Y) \right)
{\delta {\hat H}^{(s)} \over \delta \varphi^{j}(\bm{\theta},Y)} \,\,
\Phi^{j}_{U^{\mu}} (\bm{\theta} + \dots, {\bf U}(Y))
\,\, {d^{m} \theta \over (2\pi)^{m}} \,\, +$$

$$+ \,\, {1 \over \epsilon} \int
\sum_{s=1}^{g} \, e_{s} \,
{\partial k^{\alpha} \over \partial U^{\nu}}(X) \,\, \left(
W^{(s)}_{\theta^{\alpha}} (\bm{\theta},X) \, - \,
T^{(s)}_{\alpha} (\bm{\theta},X) \right) \, \nu (X-Y) \,\,
\times \hspace{3cm}$$
$$\hspace{5cm} \times \,\, \left(
W^{(s)}_{\theta^{\beta}} (\bm{\theta},Y) \, - \,
T^{(s)}_{\beta} (\bm{\theta},Y) \right)
{\partial k^{\beta} \over \partial U^{\mu}}(Y)
\,\, {d^{m} \theta \over (2\pi)^{m}} \,\, -$$

$$- \,\, {1 \over \epsilon} \int \sum_{s=1}^{g} \, e_{s} \,
\Phi^{i}_{U^{\nu}} (\bm{\theta} + \dots, {\bf U}(X)) \,\,
{\delta {\hat H}^{(s)} \over \delta \varphi^{i}(\bm{\theta},X)}
\,\, \nu (X-Y) \,\, \times \hspace{3cm} $$
$$\hspace{3cm} \times \,\, \left(
W^{(s)}_{\theta^{\beta}} (\bm{\theta},Y) \, - \,
T^{(s)}_{\beta} (\bm{\theta},Y) \right)
{\partial k^{\beta} \over \partial U^{\mu}}(Y)   
\,\, {d^{m} \theta \over (2\pi)^{m}} \,\, -$$

$$- \,\, {1 \over \epsilon} \int \sum_{s=1}^{g} \, e_{s} \,
{\partial k^{\alpha} \over \partial U^{\nu}}(X) \left(
W^{(s)}_{\theta^{\alpha}} (\bm{\theta},X) \, - \,
T^{(s)}_{\alpha} (\bm{\theta},X) \right) \,\, \nu (X-Y) \,\,
\times \hspace{3cm} $$
$$\hspace{5cm} \times \,\,
{\delta {\hat H}^{(s)} \over \delta \varphi^{j}(\bm{\theta},Y)} \,\,
\Phi^{j}_{U^{\mu}} (\bm{\theta} + \dots, {\bf U}(Y))
\,\, {d^{m} \theta \over (2\pi)^{m}} \,\, +$$

$$+ \,\, {1 \over \epsilon} \int \sum_{s=1}^{g} \, e_{s} \,
\Phi^{i}_{U^{\nu}} (\bm{\theta} + \dots, {\bf U}(X)) \,\,
{\delta {\hat H}^{(s)} \over \delta \varphi^{i}(\bm{\theta},X)}
\,\, \nu (X-Y) \,\, \times \hspace{3cm} $$   
$$\hspace{3cm} \times \,\,
{\delta {\hat H}^{(s)} \over \delta \varphi^{j}(\bm{\theta},Y)} \,\,
\Phi^{j}_{U^{\mu}} (\bm{\theta} + \dots, {\bf U}(Y))
\,\, {d^{m} \theta \over (2\pi)^{m}} \hspace{1cm} +
\hspace{1cm} O(1) $$

 We will investigate now the weak limit
$\Omega^{1(wl)}_{\nu\mu}(X,Y)$ of the form
$\Omega^{1}_{\nu\mu}(X,Y)$, i.e. the limit in sence of
the integrals

$$\int_{-\infty}^{+\infty} \int_{-\infty}^{+\infty}
\xi^{\nu}(X) \,\, \Omega^{1}_{\nu\mu}(X,Y) \,\,
\eta^{\mu}(Y) \,\, dX \, dY $$
for fixed (smooth) $\xi^{\nu}(X)$ and $\eta^{\mu}(Y)$.

 We will use first the formulas (\ref{wform}) for the values like
$W^{(s)}_{\theta^{\alpha}}$, $W^{(s)}_{\theta^{\beta}}$ in the
expression above. It's easy to see then that
$\Omega^{1}_{\nu\mu}(X,Y)$ contains actually just the terms of
order $1/\epsilon$ in the main part.

 We note after that that the integration w.r.t.
$\theta$  in the last four terms
can be done independently at the points $X$ and $Y$ in the weak limit
for the rapidly oscillating functions of $X$ and $Y$ in the full
analogy with the remark before the formula (\ref{vjaver}).
Using then the formula (\ref{intw}) we see that the values like
$W^{(s)}_{\theta^{\alpha}}$, $W^{(s)}_{\theta^{\beta}}$ can be
actually omitted in the order $1/\epsilon$ for the weak limit
of the last four terms of $\Omega^{1}_{\nu\mu}(X,Y)$. We can also
replace in the same terms the integration w.r.t. $\bm{\theta}$
just by the averaging on the quasiperiodic solutions in the
main $(1/\epsilon)$ order of $\epsilon$.

 It's not difficult to prove also the formula

\begin{equation}
\label{hav}
{\partial \langle h^{(s)} \rangle \over \partial U^{\nu}} (X)
\,\,\, = \,\,\, \langle
{\delta {\hat H}^{(s)} \over \delta \varphi^{i}(\bm{\theta},X)} \,
\Phi^{i}_{U^{\nu}} (\bm{\theta},X) \rangle \,\, + \,\, 
{\partial k^{\alpha} \over \partial U^{\nu}}(X)
\langle T^{(s)}_{\alpha} (\bm{\theta},X) \rangle   
\end{equation}
according to the definition (\ref{tsa}) of the functions
$T^{(s)}_{\alpha}$.

 Using the formula (\ref{hav}) and the remarks above we can see
then that the last four terms in the expression for
$\Omega^{1}_{\nu\mu}(X,Y)$ give the terms

$$ {1 \over \epsilon} \, \sum_{s=1}^{g} e_{s}
{\partial \langle h^{(s)} \rangle \over \partial U^{\nu}}(X) 
\, \nu (X-Y) \,
{\partial \langle h^{(s)} \rangle \over \partial U^{\mu}}(Y)  
\,\,\, + \,\,\, {o(1) \over \epsilon} $$
for $\Omega^{1(wl)}_{\nu\mu}(X,Y)$.

 If we introduce now the functions

$$\tau_{\beta\nu} \, = \,
{\partial k^{\alpha} \over \partial U^{\nu}} \left[ - \,
\langle A_{\alpha\beta} \rangle \, + \, \sum_{s=1}^{g} \, e_{s} \,
\langle T^{(s)}_{\alpha} \, W^{(s)}_{\theta^{\beta}} \rangle
\, - \, {1 \over 2} \, \sum_{s=1}^{g} \, e_{s} \,
\langle W^{(s)}_{\theta^{\alpha}} \, W^{(s)}_{\theta^{\beta}} \rangle
\right] \,\, + $$

$$+ \,\, \langle \Phi^{i}_{U^{\nu}} \sum_{k \geq 0}
\omega^{(k)}_{ij} (\bm{\varphi}, \dots) \,
\varphi^{j}_{\theta^{\beta},kx} \rangle \, + \,
\langle \Phi^{i}_{U^{\nu}} \,
{\delta H^{(s)} \over \delta \varphi^{i}(x)} \,
\left( W^{(s)}_{\theta^{\beta}} \, - \,
T^{(s)}_{\beta} \right) \rangle $$
and use the formulas (\ref{vjaver}) we can see that the form
$\Omega^{1(wl)}_{\nu\mu}(X,Y)$ can be written as

$$\Omega^{1(wl)}_{\nu\mu}(X,Y) \,\, = \,\,
{1 \over \epsilon} \, \sum_{\alpha = 1}^{m} \left(
{\partial k^{\alpha} \over \partial U^{\nu}}(X) \, \nu (X-Y) \,
\tau_{\alpha\mu} (Y) \, + \,
\tau_{\alpha\nu} (X) \, \nu (X-Y) \,
{\partial k^{\alpha} \over \partial U^{\mu}}(Y) \right) \, + $$
$$+ \, {1 \over \epsilon} \, \sum_{s=1}^{g} e_{s}
{\partial \langle h^{(s)} \rangle \over \partial U^{\nu}}(X)
\, \nu (X-Y) \,
{\partial \langle h^{(s)} \rangle \over \partial U^{\mu}}(Y)
\,\,\, + \,\,\, {o(1) \over \epsilon} $$
where the values $\tau_{\alpha\nu}({\bf U})$ are given by the
formulas (\ref{actvar}) for the values
$\partial I_{\alpha}/\partial U^{\nu}$.

 Let us prove now that $\tau_{\alpha\nu}({\bf U})$ can be in fact
represented as the derivatives
$\partial I_{\alpha}/\partial U^{\nu}$ for some functions
$I_{\alpha}({\bf U})$. We will assume as we said already that the
gradients $d \, k^{1}$, $\dots$, $d \, k^{m}$ are linearly
independent on ${\cal M}^{N}$. From the closeness of the form
$\Omega^{rest}$ it follows that the form
$\Omega^{1(wl)}_{\nu\mu}(X,Y)$ is also closed on ${\cal M}$.
Easy to see that the part

$$ {1 \over \epsilon} \, \sum_{s=1}^{g} e_{s}
{\partial \langle h^{(s)} \rangle \over \partial U^{\nu}}(X)
\, \nu (X-Y) \,
{\partial \langle h^{(s)} \rangle \over \partial U^{\mu}}(Y) $$
is closed according to Theorem 2. We get then that the form

$${1 \over \epsilon} \, \sum_{\alpha = 1}^{m} \left(
{\partial k^{\alpha} \over \partial U^{\nu}}(X) \, \nu (X-Y) \,
\tau_{\alpha\mu} (Y) \, + \,
\tau_{\alpha\nu} (X) \, \nu (X-Y) \,
{\partial k^{\alpha} \over \partial U^{\mu}}(Y) \right) $$
should also be closed on ${\cal M}$. Using Theorem 2 it's
not difficult to see then that we should have
$\tau_{\alpha\nu}({\bf U}) \, = \,
\partial I_{\alpha}/\partial U^{\nu}$ for some functions
$I_{\alpha}({\bf U})$ in this case.

\vspace{0.5cm}

II) We have
$\Omega^{2}_{\nu\alpha}(X,Y) \, = \, - \,
\Omega^{3}_{\alpha\nu}(Y,X)$ and

$$\Omega^{2}_{\nu\alpha}(X,Y) \, = \, \int \left( - {1 \over
\epsilon} {\partial k^{\gamma} \over \partial U^{\nu}}(X) \,\, \nu
(X-Z) \,\, \varphi^{i}_{\theta^{\gamma}}(\bm{\theta},Z) \, + \,
\delta (X-Z) \, \Phi^{i}_{U^{\nu}} (\bm{\theta} + \dots, {\bf
U}(Z)) \right) \,\, \times $$
$$\times \,\, {\hat \Omega}_{ij}
(\bm{\theta},\bm{\theta}^{\prime},Z,W) \,\,
\varphi^{j}_{\theta^{\alpha}} (\bm{\theta}^{\prime},W) \,
\delta(W-Y) \, {d^{m} \theta \over (2\pi)^{m}} \, {d^{m}
\theta^{\prime} \over (2\pi)^{m}} \, dZ \, dW $$

 Easy to see that we can omit all the terms of order $O(1)$ in this
expression keeping in mind the statement of the Theorem. In particular
we can omit the differentiation of the function $\delta (W-Y)$ in the
local part and write

$$\Omega^{2}_{\nu\alpha}(X,Y) \, = $$

$$= \, - \, {1 \over \epsilon} \, \int
{\partial k^{\gamma} \over \partial U^{\nu}}(X) \, \nu (X-Y) \,
\varphi^{i}_{\theta^{\gamma}}(\bm{\theta},Y) \, \sum_{k\geq 0}
\omega^{(k)}_{ij} (\bm{\varphi}(\bm{\theta},Y),\dots ) \, \epsilon^{k}
\, \varphi^{j}_{\theta^{\alpha},kY} (\bm{\theta},Y) \,
{d^{m} \theta \over (2\pi)^{m}} \, - $$

$$- \, \int \sum_{s=1}^{g} e_{s} \,
{\partial k^{\gamma} \over \partial U^{\nu}}(X) \, \nu (X-Z) \,   
\left( W^{(s)}_{\theta^{\gamma}Z} (\bm{\theta},Z) \, - \,
T^{(s)}_{\gamma,Z} (\bm{\theta},Z) \right) \, \nu (Z-Y) \, \times
\hspace{3cm}$$
$$\hspace{5cm} \times \,
\left( W^{(s)}_{\theta^{\alpha}Y} (\bm{\theta},Y) \, - \,
T^{(s)}_{\alpha,Y} (\bm{\theta},Y) \right) \,
{d^{m} \theta \over (2\pi)^{m}} \, dZ \,\, + $$

$$+ \, \int \sum_{s=1}^{g} e_{s} \,
\Phi^{i}_{U^{\nu}} (\bm{\theta} + \dots, {\bf U}(X)) \,
{\delta {\hat H}^{(s)} \over \delta \varphi^{i} (\bm{\theta},X)} \,   
\nu (X-Y) \, \times \hspace{5cm} $$
$$\hspace{5cm} \times \,
\left( W^{(s)}_{\theta^{\alpha}Y} (\bm{\theta},Y) \, - \,
T^{(s)}_{\alpha,Y} (\bm{\theta},Y) \right) \,
{d^{m} \theta \over (2\pi)^{m}} \,\,\,\,\,\,\,\, +
\,\,\,\,\,\,\,\, O(1) \,\, = $$

$$= \, - \int {1 \over \epsilon} \,
{\partial k^{\gamma} \over \partial U^{\nu}}(X) \, \nu (X-Y) \,
\varphi^{i}_{\theta^{\gamma}}(\bm{\theta},Y) \, \sum_{k\geq 0}
\omega^{(k)}_{ij} (\bm{\varphi}(\bm{\theta},Y),\dots ) \, \epsilon^{k}
\, \varphi^{j}_{\theta^{\alpha},kY} (\bm{\theta},Y) \,
{d^{m} \theta \over (2\pi)^{m}} \, - $$

$$- \, {\partial \over \partial Y} \, \int \sum_{s=1}^{g} e_{s} \,
{\partial k^{\gamma} \over \partial U^{\nu}}(X) \, \nu (X-Z) \,
\left( W^{(s)}_{\theta^{\gamma}Z} (\bm{\theta},Z) \, - \,
T^{(s)}_{\gamma,Z} (\bm{\theta},Z) \right) \, \times \hspace{3cm} $$
$$\hspace{5cm} \times \, \nu (Z-Y) \, \left(   
W^{(s)}_{\theta^{\alpha}} (\bm{\theta},Y) \, - \,
T^{(s)}_{\alpha} (\bm{\theta},Y) \right) \,
{d^{m} \theta \over (2\pi)^{m}} \, dZ \,\, - $$

$$- \, \int \sum_{s=1}^{g} e_{s} \,
{\partial k^{\gamma} \over \partial U^{\nu}}(X) \, \nu (X-Y) \,
\left( W^{(s)}_{\theta^{\gamma}Y} (\bm{\theta},Y) \, - \,
T^{(s)}_{\gamma,Y} (\bm{\theta},Y) \right) \, \times \hspace{5cm} $$
$$\hspace{7cm} \times \,
\left( W^{(s)}_{\theta^{\alpha}} (\bm{\theta},Y) \, - \,
T^{(s)}_{\alpha} (\bm{\theta},Y) \right) \,
{d^{m} \theta \over (2\pi)^{m}} \, + $$

$$+ \, {\partial \over \partial Y} \, \int \sum_{s=1}^{g} e_{s} \,
\Phi^{i}_{U^{\nu}} (\bm{\theta} + \dots, {\bf U}(X)) \,
{\delta {\hat H}^{(s)} \over \delta \varphi^{i} (\bm{\theta},X)} \,
\nu (X-Y) \, \times \hspace{5cm} $$
$$\hspace{7cm} \times \,
\left( W^{(s)}_{\theta^{\alpha}} (\bm{\theta},Y)
\, - \, T^{(s)}_{\alpha} (\bm{\theta},Y) \right) \,
{d^{m} \theta \over (2\pi)^{m}} \, + $$

$$+ \, \int \sum_{s=1}^{g} e_{s} \,
\Phi^{i}_{U^{\nu}} (\bm{\theta} + \dots, {\bf U}(X)) \,
{\delta {\hat H}^{(s)} \over \delta \varphi^{i} (\bm{\theta},X)}
\, \times \hspace{7cm} $$
$$\hspace{5cm} \times \,
\left( W^{(s)}_{\theta^{\alpha}} (\bm{\theta},Y) \, - \,
T^{(s)}_{\alpha} (\bm{\theta},Y) \right) \, \delta (X-Y) \,
{d^{m} \theta \over (2\pi)^{m}} \,\,\,\,\,\,\,\, +
\,\,\,\,\,\,\,\, O(1) $$

 Using the relations
$W^{(s)}_{\theta^{\alpha}} (\bm{\theta},Y) \sim O(1)$,
$\epsilon \rightarrow 0$ we can omit now the last two terms. The
second term can be rewritten in the form   

$$- \, {\partial \over \partial Y} \, \int \sum_{s=1}^{g} e_{s} \,
{\partial k^{\gamma} \over \partial U^{\nu}}(X) \,
\left( W^{(s)}_{\theta^{\gamma}} (\bm{\theta},X) \, - \,
T^{(s)}_{\gamma} (\bm{\theta},X) \, - \, {1 \over 2}
\, - \, W^{(s)}_{\theta^{\gamma}} (\bm{\theta},+\infty) \right)
\, \times \hspace{3cm}$$
$$\hspace{5cm} \times \,
\nu (X-Y) \, \left( W^{(s)}_{\theta^{\alpha}} (\bm{\theta},Y)
\, - \, T^{(s)}_{\alpha} (\bm{\theta},Y) \right) \,
{d^{m} \theta \over (2\pi)^{m}} \, + $$

$$+ \,  {\partial \over \partial Y} \, \int \sum_{s=1}^{g} e_{s} \,
{\partial k^{\gamma} \over \partial U^{\nu}}(X) \, \nu (X-Y) \, 
\left( W^{(s)}_{\theta^{\gamma}} (\bm{\theta},Y) \, - \,
T^{(s)}_{\gamma} (\bm{\theta},Y)  \right) \, \times \hspace{5cm} $$
$$\hspace{7cm} \times \, \left(
W^{(s)}_{\theta^{\alpha}} (\bm{\theta},Y)
\, - \, T^{(s)}_{\alpha} (\bm{\theta},Y) \right) \,
{d^{m} \theta \over (2\pi)^{m}} $$
and can be omitted by the same reason.
 
 We have so

$$\Omega^{2}_{\nu\alpha}(X,Y) \, = $$      

$$= \, - \, {1 \over \epsilon} \, \int
{\partial k^{\gamma} \over \partial U^{\nu}}(X) \, \nu (X-Y) \,
\varphi^{i}_{\theta^{\gamma}}(\bm{\theta},Y) \, \sum_{k\geq 0}
\omega^{(k)}_{ij} (\bm{\varphi}(\bm{\theta},Y),\dots ) \, \epsilon^{k}
\, \varphi^{j}_{\theta^{\alpha},kY} (\bm{\theta},Y) \,
{d^{m} \theta \over (2\pi)^{m}} \, - $$

$$-  \, \int \sum_{s=1}^{g} e_{s} \,
{\partial k^{\gamma} \over \partial U^{\nu}}(X) \, \nu (X-Y) \,
\left( {1 \over 2} \, \left( W^{(s)}_{Y} (\bm{\theta},Y) \,
W^{(s)}_{\theta^{\alpha}} (\bm{\theta},Y) \right)_{\theta^{\gamma}}
\right. \, - $$
$$- {1 \over 2} \left( W^{(s)}_{\theta^{\gamma}} (\bm{\theta},Y)
W^{(s)}_{Y} (\bm{\theta},Y) \right)_{\theta^{\alpha}} + 
{1 \over 2} \left( W^{(s)}_{\theta^{\gamma}} (\bm{\theta},Y)
W^{(s)}_{\theta^{\alpha}} (\bm{\theta},Y) \right)_{Y} -
\left( T^{(s)}_{\gamma} (\bm{\theta},Y)  
W^{(s)}_{\theta^{\alpha}} (\bm{\theta},Y) \right)_{Y} + $$
$$+ \, \left. {1 \over \epsilon} \,
h^{(s)}_{\theta^{\alpha}} (\bm{\theta},Y) \,
T^{(s)}_{\gamma} (\bm{\theta},Y) \, - \,
{1 \over \epsilon} \, h^{(s)}_{\theta^{\gamma}} (\bm{\theta},Y)
\, T^{(s)}_{\alpha} (\bm{\theta},Y) \, + \,
T^{(s)}_{\gamma,Y} (\bm{\theta},Y) \,      
T^{(s)}_{\alpha} (\bm{\theta},Y) \, \right) \,
{d^{m} \theta \over (2\pi)^{m}} \,\,\, + $$
\begin{equation}
\label{om2}
+ \,\,\,\,\,\,\,\, O(1)
\end{equation}

 We can omit now the total derivatives w.r.t. $\theta^{\gamma}$ and
$\theta^{\alpha}$ in the second integral. The term

$$- \, \int \sum_{s=1}^{g} e_{s} \,
{\partial k^{\gamma} \over \partial U^{\nu}}(X) \, \nu (X-Y) \,
\times \hspace{8cm} $$
$$\hspace{3cm} \times \,
\left[ {1 \over 2} \, \left( W^{(s)}_{\theta^{\gamma}} (\bm{\theta},Y)
\, W^{(s)}_{\theta^{\alpha}} (\bm{\theta},Y) \right)_{Y} \, - \,
\left( T^{(s)}_{\gamma} (\bm{\theta},Y) \,
W^{(s)}_{\theta^{\alpha}} (\bm{\theta},Y) \right)_{Y} \right] \,
{d^{m} \theta \over (2\pi)^{m}} $$
can be written as

$$- \, {\partial \over \partial Y} \, \int \sum_{s=1}^{g} e_{s} \,
{\partial k^{\gamma} \over \partial U^{\nu}}(X) \, \nu (X-Y) \,
\times \hspace{8cm} $$
$$\hspace{3cm} \times \,
\left[ {1 \over 2} \, W^{(s)}_{\theta^{\gamma}} (\bm{\theta},Y)
\, W^{(s)}_{\theta^{\alpha}} (\bm{\theta},Y)  \, - \,
T^{(s)}_{\gamma} (\bm{\theta},Y) \,
W^{(s)}_{\theta^{\alpha}} (\bm{\theta},Y) \right] \,
{d^{m} \theta \over (2\pi)^{m}} \, - $$

$$- \, \int \sum_{s=1}^{g} e_{s} \,
{\partial k^{\gamma} \over \partial U^{\nu}}(X) \, \delta (X-Y) \, 
\left[ {1 \over 2} \, W^{(s)}_{\theta^{\gamma}} (\bm{\theta},Y)
\, W^{(s)}_{\theta^{\alpha}} (\bm{\theta},Y)  \, - \,
T^{(s)}_{\gamma} (\bm{\theta},Y) \,
W^{(s)}_{\theta^{\alpha}} (\bm{\theta},Y) \right] \,
{d^{m} \theta \over (2\pi)^{m}} $$
and has also the order $O(1)$ for $\epsilon \rightarrow 0$.

 The rest of the expression (\ref{om2}) can now be written according
to (\ref{epsdivform}) as

$$- \, {1 \over \epsilon} \, \int {\partial k^{\gamma} \over
\partial U^{\nu}}(X) \, \nu (X-Y) \, \left[ {\partial \over
\partial \theta^{\beta}} \, Q^{\beta}_{\gamma\alpha}
(\bm{\varphi}(\bm{\theta},Y),\dots ) \, + \,
\epsilon {\partial \over \partial Y} \,
A_{\gamma\alpha} (\bm{\varphi}(\bm{\theta},Y),\dots
) \right] \, {d^{m} \theta \over (2\pi)^{m}} \, + $$

$$+ \,\,\, O(1) \,\,\, = \,\,\, O(1) $$

 So we get the part (II) of the Theorem.

\vspace{0.5cm}

III) We have

$$\Omega^{4}_{\alpha\beta} (X,Y) \, = \, \int
\varphi^{i}_{\theta^{\alpha}} (\bm{\theta},X) \sum_{k\geq 0}
\omega^{(k)}_{ij} (\bm{\varphi} (\bm{\theta},X), \dots) \,
\epsilon^{k} \, \varphi^{j}_{\theta^{\beta},kX} (\bm{\theta},X)
\, {d^{m} \theta \over (2\pi)^{m}} \, \delta (X-Y) \, + $$
 
$$+ \, \epsilon \int \sum_{s=1}^{g} e_{s} \, \left(
W^{(s)}_{\theta^{\alpha}X} (\bm{\theta},X) -
T^{(s)}_{\alpha,X} (\bm{\theta},X) \right) \, \nu(X-Y) \,\, 
\times \hspace{5cm}$$
$$\hspace{3cm} \times \,\,
\left( W^{(s)}_{\theta^{\beta}Y} (\bm{\theta},Y) -
T^{(s)}_{\beta,Y} (\bm{\theta},Y) \right) \,
{d^{m} \theta \over (2\pi)^{m}} \, \delta (X-Y) \,\,\,\,\,
+ \,\,\,\,\, O(\epsilon) \, = $$

$$= \, \int \varphi^{i}_{\theta^{\alpha}} (\bm{\theta},X)
\sum_{k\geq 0}
\omega^{(k)}_{ij} (\bm{\varphi} (\bm{\theta},X), \dots) \,
\epsilon^{k} \, \varphi^{j}_{\theta^{\beta},kX} (\bm{\theta},X)
\, {d^{m} \theta \over (2\pi)^{m}} \, \delta (X-Y) \, + $$

$$+ \, \epsilon {\partial^{2} \over \partial X \, \partial Y} \,
\int \sum_{s=1}^{g} e_{s} \, \left(
W^{(s)}_{\theta^{\alpha}} (\bm{\theta},X) -
T^{(s)}_{\alpha} (\bm{\theta},X) \right) \, \nu(X-Y) \,
\times \hspace{5cm}$$
$$\hspace{7cm} \times \,
\left( W^{(s)}_{\theta^{\beta}} (\bm{\theta},Y) -  
T^{(s)}_{\beta} (\bm{\theta},Y) \right) \,  
{d^{m} \theta \over (2\pi)^{m}} \, + $$

$$+ \, \epsilon  \int \sum_{s=1}^{g} e_{s} \, \left(
W^{(s)}_{\theta^{\alpha}} (\bm{\theta},X) -
T^{(s)}_{\alpha} (\bm{\theta},X) \right) \, 
\left( W^{(s)}_{\theta^{\beta}} (\bm{\theta},X) -
T^{(s)}_{\beta} (\bm{\theta},X) \right) \,
{d^{m} \theta \over (2\pi)^{m}} \, \delta^{\prime} (X-Y) \, + $$

$$+ \, \epsilon  \int \sum_{s=1}^{g} e_{s} \, \left(
{1 \over 2} \left(W^{(s)}_{X} (\bm{\theta},X) \,
W^{(s)}_{\theta^{\beta}} (\bm{\theta},X) \right)_{\theta^{\alpha}}
\, - \, {1 \over 2} \left( W^{(s)}_{\theta^{\alpha}} (\bm{\theta},X)
\, W^{(s)}_{X} (\bm{\theta},X) \right)_{\theta^{\beta}} \, +
\right. $$
$$+ \,\, {1 \over 2} \left( W^{(s)}_{\theta^{\alpha}} (\bm{\theta},X)
\, W^{(s)}_{\theta^{\beta}} (\bm{\theta},X) \right)_{X} \, - \,
\left( T^{(s)}_{\alpha} (\bm{\theta},X) \,
W^{(s)}_{\theta^{\beta}} (\bm{\theta},X) \right)_{X} \, + $$
$$+ \,\,
{1 \over \epsilon} \, h^{(s)}_{\theta^{\beta}} (\bm{\theta},X) \,
T^{(s)}_{\alpha} (\bm{\theta},X) \, - \,
{1 \over \epsilon} \, h^{(s)}_{\theta^{\alpha}} (\bm{\theta},X) \,
T^{(s)}_{\beta} (\bm{\theta},X) \, + $$
$$+ \, \left. T^{(s)}_{\alpha,X} (\bm{\theta},X) \,
T^{(s)}_{\beta} (\bm{\theta},X) \right) \,
{d^{m} \theta \over (2\pi)^{m}} \, \delta (X-Y) \,\,\,\,\,\,\,\, +
\,\,\,\,\,\,\,\, O(\epsilon) $$

 Using the same arguments as before we can write now

$$\Omega^{4}_{\alpha\beta} (X,Y) \, = \, \int \left[
{\partial \over \partial \theta^{\gamma}} \,
Q^{\gamma}_{\alpha\beta} (\bm{\theta},X) \, \delta (X-Y) \, +
\epsilon \, \left( A_{\alpha\beta} (\bm{\theta},X) \right)_{X} \,
\delta (X-Y) \right] \, {d^{m} \theta \over (2\pi)^{m}} \, +
\, O(\epsilon) $$

 So we get the part (III) of the theorem.

{\hfill Theorem 4 is proved.}

\vspace{1cm}

{\bf Definition 4.}  {\it We call the form

$$\Omega^{av}_{\nu\mu} (X,Y) \, = \, \sum_{\alpha = 1}^{m}
\left( {\partial k^{\alpha} \over \partial U^{\nu}} (X) \,
\nu (X-Y) \, {\partial I_{\alpha} \over \partial U^{\mu}}(Y)
\,\, + \,\, {\partial I_{\alpha} \over \partial U^{\nu}}(X) \,
\nu (X-Y) \, {\partial k^{\alpha} \over \partial U^{\mu}} (Y)
\right) \,\, + $$
\begin{equation}
\label{avsympstr}
+ \,\, \sum_{s=1}^{g} e_{s} \,
{\partial \langle h^{(s)} \rangle \over \partial U^{\nu}}(X) 
\, \nu (X-Y) \,
{\partial \langle h^{(s)} \rangle \over \partial U^{\mu}}(Y)
\end{equation}   
- the averaging of the form (\ref{conssympform}) on the space
of $m$-phase solutions of system (\ref{dynsyst}).

 We call the functions $I^{\alpha}({\bf U})$ defined through the
formulas (\ref{actvar}) the action variables conjugated with the
wave numbers $k^{\alpha}({\bf U})$.
}

\vspace{0.5cm}

 We will prove now that the Symplectic structure (\ref{avsympstr})
can be considered actually as the Symplectic structure for the
Whitham system (\ref{wsyst1}) while the value
$\int \langle h \rangle (X) \, dX$ plays the role of the Hamiltonian
function for this system. Let us prove here the following
Theorem:

\vspace{0.5cm}

{\bf Theorem 5.}

{\it If the functions

$$\phi_{(1)}^{i} (\bm{\theta},X,T,\epsilon) \, = \,
\Phi^{i} \left({{\bf S}(X,T) \over \epsilon} +
\bm{\theta}^{*}(X,T) + \bm{\theta}, \, {\bf U}(X,T) \right)
\, + \, \epsilon \, \Psi^{i}_{(1)}
\left({{\bf S}(X,T) \over \epsilon} + \bm{\theta}, X,T
\right) $$
satisfy the system (\ref{epssyst}) modulo the terms
$O(\epsilon^{2})$ then the following relation is true

\begin{equation}
\label{sympeq}
\int_{-\infty}^{+\infty} \Omega^{av}_{\nu\mu} (X,Y) \,\,
U^{\mu}_{T} (Y) \,\, dY \,\,\, = \,\,\,
{\partial \langle h \rangle \over \partial U^{\nu}} (X)
\end{equation}
}

\vspace{0.5cm}

 Proof.

 Let us prove first that under the conditions of the Theorem
the following relations hold in the weak limit

$$\int \left( - {1 \over \epsilon}
{\partial k^{\alpha} \over \partial U^{\nu}}(X) \, \nu (X-Z) \,
\phi_{(1)\theta^{\alpha}}^{i} (\bm{\theta},Z,\epsilon) \, + \,
\delta (X-Z) \, \Phi^{i}_{U^{\nu}}
(\bm{\theta} + \dots , {\bf U}(Z)) \right) \,\, \times $$
$$\times \,\, {\hat \Omega}_{ij} \, [\bm{\phi}_{(1)}] \,
(\bm{\theta},\bm{\theta}^{\prime},Z,W) \,\, \times $$
$$\times \,\, \left( \phi_{(1)\theta^{\prime\beta}}^{j}
(\bm{\theta}^{\prime},W,\epsilon) \, \left(
S^{\beta}_{T}(W) \, + \, \epsilon \, \theta^{*\beta}_{T}(W) \right)
\, + \, \epsilon \, \Phi^{j}_{U^{\mu}}
(\bm{\theta}^{\prime} + \dots , {\bf U}(W)) \,\, U^{\mu}_{T}(W)
\right) \,\, \times $$
$$\times \,\, dZ \, dW \, {d^{m} \theta \over (2\pi)^{m}} \,
{d^{m} \theta^{\prime} \over (2\pi)^{m}} \,\,\, = $$   

$$= \,\,\, \int \left( - {1 \over \epsilon}
{\partial k^{\alpha} \over \partial U^{\nu}}(X) \, \nu (X-Z) \,
\phi_{(1)\theta^{\alpha}}^{i} (\bm{\theta},Z,\epsilon) \, + \,
\delta (X-Z) \, \Phi^{i}_{U^{\nu}}
(\bm{\theta} + \dots , {\bf U}(Z)) \right) \,\, \times $$
\begin{equation}
\label{eq1}
\times \,\,
{\delta {\hat H} \over \delta \varphi^{i}(\bm{\theta},Z)} \,
\left( \phi_{(1)}^{i} (\bm{\theta},Z,\epsilon), \dots \right) \,
dZ \, {d^{m} \theta \over (2\pi)^{m}}
\,\,\,\,\,\,\,\, + \,\,\,\,\,\,\,\, o(1)
\end{equation}
$\epsilon \rightarrow 0$.

 Easy to see that the expression

$$\phi_{(1)\theta^{\prime\beta}}^{j}
(\bm{\theta}^{\prime},W,\epsilon) \, \left(
S^{\beta}_{T}(W) \, + \, \epsilon \, \theta^{*\beta}_{T}(W) \right)
\, + \, \epsilon \, \Phi^{j}_{U^{\mu}}
(\bm{\theta}^{\prime} + \dots , {\bf U}(W)) \,\, U^{\mu}_{T}(W) $$
gives actually the value
$\epsilon \, \phi_{(1)T}^{j} (\bm{\theta}^{\prime},W,\epsilon)$
up to the terms of order $O(\epsilon^{2})$.

 We can write then

$$\phi_{(1)\theta^{\prime\beta}}^{j}
(\bm{\theta}^{\prime},W,\epsilon) \, \left(
S^{\beta}_{T}(W) \, + \, \epsilon \, \theta^{*\beta}_{T}(W) \right)
\, + \, \epsilon \, \Phi^{j}_{U^{\mu}}
(\bm{\theta}^{\prime} + \dots , {\bf U}(W)) \,\, U^{\mu}_{T}(W)
\,\, =$$

$$= \,\, Q^{j} \left( \bm{\phi}_{(1)},
\epsilon \, \bm{\phi}_{(1)W}, \dots \right) \,\, + \,\,
\epsilon^{2} \,  G^{j} (\bm{\theta}^{\prime} + \dots, W)$$
where $G^{j} (\bm{\theta}^{\prime},W)$ are some local
expressions of $\bm{\Phi} (\bm{\theta}^{\prime},{\bf U}(W))$,
$\bm{\Psi}_{(1)} (\bm{\theta}^{\prime},W)$ and their derivatives.

 Let us start with the nonlocal part of the form
${\hat \Omega}_{ij} \, (\bm{\theta},\bm{\theta}^{\prime},Z,W)$.
First we note that

$$\int \left( - {1 \over \epsilon}
{\partial k^{\alpha} \over \partial U^{\nu}}(X) \, \nu (X-Z) \,
\phi_{(1)\theta^{\alpha}}^{i} (\bm{\theta},Z,\epsilon) \, + \, 
\delta (X-Z) \, \Phi^{i}_{U^{\nu}}
(\bm{\theta} + \dots , {\bf U}(Z)) \right) \,\, \times $$
 
$$\times \,\, {1 \over \epsilon} \sum_{s=1}^{g} \, e_{s} \,
{\delta {\hat H}^{(s)} \over \delta \varphi^{i}(\bm{\theta},Z)}
\, [\bm{\phi}_{(1)}] \,
\nu (Z-W) \,\, \delta (\bm{\theta} - \bm{\theta}^{\prime}) \,
{\delta {\hat H}^{(s)} \over
\delta \varphi^{j}(\bm{\theta}^{\prime},W)} \, [\bm{\phi}_{(1)}]
\,\, \times \hspace{3cm}$$
$$\hspace{7cm} \times \,\, \epsilon^{2} \,
G^{j} (\bm{\theta}^{\prime} + \dots, W) \,\,
dZ \, dW \, {d^{m} \theta \over (2\pi)^{m}} \,
{d^{m} \theta^{\prime} \over (2\pi)^{m}} \,\,\, = $$

$$= \,\,\, \int \left(
- {\partial k^{\alpha} \over \partial U^{\nu}}(X) \, \nu (X-Z) \,
\Psi_{(0)\theta^{\alpha}}^{i} (\bm{\theta} + \dots, Z) \right)
\,\, \times \hspace{5cm} $$
$$\times \,\, \sum_{s=1}^{g} \, e_{s} \,
{\delta {\hat H}^{(s)} \over \delta \varphi^{i}(\bm{\theta},Z)}
\, [\bm{\Psi}_{(0)}] \,\, \nu (Z-W) \,\,
{\delta {\hat H}^{(s)} \over
\delta \varphi^{j}(\bm{\theta},W)} \,
[\bm{\Psi}_{(0)}] \,\, \times $$
$$\hspace{7cm} \times \,\,  G^{j} (\bm{\theta} + \dots, W)
\,\, dZ \, dW \, {d^{m} \theta \over (2\pi)^{m}}
\,\,\,\,\,\,\,\, + \,\,\,\,\,\,\,\, O(\epsilon)$$

 Using the same arguments as before we note that the rapidly   
oscillating functions of $Z$ and $W$ should be averaged in the
weak limit separately in the main order of $\epsilon$ (for   
generic ${\bf S}(Z)$, ${\bf S}(W)$) and besides that

$$\langle \Psi_{(0)\theta^{\alpha}}^{i} (\bm{\theta} + \dots, Z) \,
{\delta {\hat H}^{(s)} \over \delta \varphi^{i}(\bm{\theta},Z)}
\, [\bm{\Psi}_{(0)}] \rangle \,\, = \,\,
\langle h^{(s)}_{\theta^{\alpha}} \, - \, \epsilon \,
{\partial \over \partial X} \, T^{(s)}_{\alpha} \rangle \,\, =
\,\, - \, \epsilon \, {\partial \over \partial X} \,
\langle T^{(s)}_{\alpha} \rangle \, = \, O(\epsilon)$$

 We can claim then that the terms consisting $ G^{j}$
can be actually omitted since they do not affect (\ref{eq1})
both in the non-local and local parts of ${\hat \Omega}_{ij}$.

 Let us use now the relations

$${\delta {\hat H}^{(s)} \over
\delta \varphi^{j}(\bm{\theta}^{\prime},W)} \, \epsilon \,
\varphi^{j}_{T} (\bm{\theta}^{\prime},W) \,\, = \,\,
{\delta {\hat H}^{(s)} \over
\delta \varphi^{j}(\bm{\theta}^{\prime},W)} \,  
Q^{j} (\bm{\varphi}, \epsilon \, \bm{\varphi}_{W}, \dots )
\,\, \equiv \,\, \epsilon \, \partial_{W} \, {\bar J}^{(s)}    
(\bm{\varphi}, \epsilon \, \bm{\varphi}_{W}, \dots ) $$
which follows from (\ref{hjrel}) (where the functions
${\bar J}^{(s)}$ are in general different from $J^{(s)}$
introduced in (\ref{hjrel})).

 Using the identity

$$\sum_{k \geq 0} \omega^{(k)}_{ij} (\bm{\theta},Z) \,
\epsilon^{k} \, {\partial^{k} \over \partial Z^{k}}
Q^{j} (\bm{\theta},Z) \, + \, \sum_{s=1}^{g} \, e_{s} \,
{\delta {\hat H}^{(s)} \over \delta \varphi^{i}(\bm{\theta},Z)}
\, {\bar J}^{(s)} (\bm{\theta},Z) \, \equiv \,
{\delta {\hat H} \over \delta \varphi^{i}(\bm{\theta},Z)} $$
(which is the definition of the symplectic structure of the 
system (\ref{dynsyst})) we get (\ref{eq1}).

 Now using the relation

$$S^{\beta}_{T}(W) \,\, = \,\, \int_{-\infty}^{+\infty}
\nu (W-Y) \, {\partial k^{\beta} \over \partial U^{\mu}}(Y) \,\,
U^{\mu}_{T}(Y) \,\, dY $$
we can see that the left-hand part of (\ref{eq1}) can be
written as

$$\epsilon \, \int_{-\infty}^{+\infty}
\Omega^{1}_{\nu\mu} (X,Y) \,\, U^{\mu}_{T}(Y) \,\, dY  
\,\,\, + \,\,\, \epsilon \, \int_{-\infty}^{+\infty} 
\Omega^{2}_{\nu\beta} (X,Y) \,\, \theta^{*\beta}_{T}(Y) 
\,\, dY $$
where $\Omega^{1}_{\nu\mu} (X,Y)$,
$\Omega^{2}_{\nu\beta} (X,Y)$ are the parts of the restriction
of the form ${\hat \Omega}_{ij}$ on the manifold
${\cal M}_{\epsilon} [\bm{\Psi}_{(1)}]$ introduced in the
Theorem 4.

 Using the relation (\ref{xderivative}) and the integration
by parts (w.r.t. $Z$) in the right-hand part of (\ref{eq1})
we can see that the right-hand part of (\ref{eq1}) can be   
written as

$${\partial k^{\alpha} \over \partial U^{\nu}}(X) \,
\langle T_{\alpha} (\bm{\theta},X) \rangle \,\, + \,\,
\langle
{\delta {\hat H} \over \delta \varphi^{i}(\bm{\theta},X)}
\, \Phi^{i}_{U^{\nu}} (\bm{\theta},X) \rangle \,\, + \,\,
O(\epsilon) $$
where $T_{\alpha}$ is the analog of the functions
$T^{(s)}_{\alpha}$ for the functional ${\hat H}$.
So we have that the right-hand part of (\ref{eq1})
is equal to
$\partial \langle h \rangle /\partial U^{\nu} \,\, (X) 
\,\,\, + \,\,\, O(\epsilon)$ according to (\ref{hav}).

 If we consider now the weak limit of the relation (\ref{eq1})
and use the parts (I), (II) of Theorem 4  
we get the relation (\ref{sympeq}) in the main ($O(1)$) order
of $\epsilon$.

{\hfill Theorem 5 is proved.}

\vspace{0.5cm}
 
 As we already said previously, we can consider the system 
(\ref{sympeq}) as the Whitham system for (\ref{dynsyst})    
in the generic situation.

\section{ The weakly nonlocal 1-forms and the averaging of the
weakly non-local Lagrangian functions.}
\setcounter{equation}{0}

 Let us consider now the 1-forms
$\omega_{i}[\bm{\varphi}](x)$ on the space of functions
$\varphi^{i}(x)$, $i = 1, \dots, n$ having the form

\begin{equation}
\label{wnonl1form}
\omega_{i}[\bm{\varphi}](x) \, = \,
c_{i} (\bm{\varphi},\bm{\varphi}_{x}, \dots) \, - \,
{1 \over 2} \, \sum_{s=1}^{g} e_{s} \,
{\delta H^{(s)} \over \delta \varphi^{i}(x)} \,
\int_{-\infty}^{+\infty} \nu (x-y) \,
h^{(s)} (\bm{\varphi},\bm{\varphi}_{y}, \dots) \, dy
\end{equation}
where $H^{(s)}[\bm{\varphi}] \, = \, \int_{-\infty}^{+\infty}
h^{(s)} (\bm{\varphi},\bm{\varphi}_{x}, \dots) \, dx$.

 We can see that the forms (\ref{wnonl1form}) have the purely
local part and the nonlocal "tail" of the fixed form which we
will call weakly nonlocal in this situation. We will call the
form $\omega_{i}[\bm{\varphi}](x)$ purely local if it has the form

$$\omega_{i}[\bm{\varphi}](x) \, = \,
c_{i} (\bm{\varphi},\bm{\varphi}_{x}, \dots) $$
for some functions $c_{i} (\bm{\varphi},\bm{\varphi}_{x}, \dots)$.

 We call the weakly nonlocal form (\ref{wnonl1form}) purely
nonlocal if

$$\omega_{i}[\bm{\varphi}](x) \, = \, \, - \,
{1 \over 2} \, \sum_{s=1}^{g} e_{s} \,
{\delta H^{(s)} \over \delta \varphi^{i}(x)} \,
\int_{-\infty}^{+\infty} \nu (x-y) \,
h^{(s)} (\bm{\varphi},\bm{\varphi}_{y}, \dots) \, dy $$

 The action of the forms $\omega_{i}[\bm{\varphi}](x)$
on the "tangent vectors" $\xi^{i}[\bm{\varphi}](x)$ is defined
in the natural way

$$(\bm{\omega}, \bm{\xi}) [\bm{\varphi}]\, = \,
\int_{-\infty}^{+\infty} \omega_{i}[\bm{\varphi}](x) \,
\xi^{i}[\bm{\varphi}](x) \, dx $$
 
 The forms (\ref{wnonl1form}) are closely connected with the 
weakly nonlocal 2-forms (\ref{conssympform}). Namely, let us 
consider the external derivative of the form
$\omega_{i}[\bm{\varphi}](x)$:

$$\left[ d \bm{\omega} \right]_{ij} (x,y) \, = \,
{\delta \omega_{j}[\bm{\varphi}](y) \over \delta \varphi^{i}(x)}  
\,\, - \,\,
{\delta \omega_{i}[\bm{\varphi}](x) \over \delta \varphi^{j}(y)} $$

\vspace{0.5cm}

{\bf Lemma 5.}

{\it The external derivative
$\left[ d \bm{\omega} \right]_{ij} (x,y)$ is the closed two-form
having the form (\ref{conssympform}) with some local functions
$\omega_{ij}^{(k)} (\bm{\varphi},\bm{\varphi}_{x}, \dots)$.
}

\vspace{0.5cm}

Proof.
 
First we note that the closeness of $d \bm{\omega}$ is a trivial
fact since $d \bm{\omega}$ is exact. Easy to see that the
derivative of the local part of $\omega_{i}$ can be written as

$${\partial c_{j} \over \partial \varphi^{i}}
(\bm{\varphi},\bm{\varphi}_{y}, \dots) \, \delta (y-x) \, + \,    
{\partial c_{j} \over \partial \varphi_{y}^{i}}
(\bm{\varphi},\bm{\varphi}_{y}, \dots) \,
\delta^{\prime} (y-x) \, + \, \dots $$

$$- \, {\partial c_{i} \over \partial \varphi^{j}}
(\bm{\varphi},\bm{\varphi}_{x}, \dots) \, \delta (x-y) \, - \,
{\partial c_{i} \over \partial \varphi_{x}^{j}}
(\bm{\varphi},\bm{\varphi}_{x}, \dots) \,
\delta^{\prime} (x-y) \, - \, \dots $$
and is a purely local 2-form.

 The derivative of the nonlocal part of $\omega_{i}$ can be   
written as

$$- \, {1 \over 2} \, \sum_{s=1}^{g} e_{s} \,
{\delta^{2} H^{(s)} \over \delta \varphi^{i}(x) \,
\delta \varphi^{j}(y)} \,
\int_{-\infty}^{+\infty} \nu (y-z) \,
h^{(s)} (\bm{\varphi},\bm{\varphi}_{z}, \dots) \, dz \,\, -$$   

$$- \,\, {1 \over 2} \, \sum_{s=1}^{g} e_{s} \,
{\delta H^{(s)} \over \delta \varphi^{j}(y)} \,
\int_{-\infty}^{+\infty} \nu (y-z) \,
{\delta h^{(s)} (\bm{\varphi},\bm{\varphi}_{z}, \dots)
\over \delta \varphi^{i}(x)} \, dz \,\, + $$   

$$+ \,\, {1 \over 2} \, \sum_{s=1}^{g} e_{s} \,
{\delta^{2} H^{(s)} \over \delta \varphi^{j}(y) \,
\delta \varphi^{i}(x)} \,
\int_{-\infty}^{+\infty} \nu (x-z) \,
h^{(s)} (\bm{\varphi},\bm{\varphi}_{z}, \dots) \, dz \,\, +$$

$$+ \,\, {1 \over 2} \, \sum_{s=1}^{g} e_{s} \,
{\delta H^{(s)} \over \delta \varphi^{i}(x)} \,
\int_{-\infty}^{+\infty} \nu (x-z) \,
{\delta h^{(s)} (\bm{\varphi},\bm{\varphi}_{z}, \dots)
\over \delta \varphi^{j}(y)} \, dz $$
 
 We have

$${\delta H^{(s)} \over \delta \varphi^{i}(x)} \, = \,
{\partial h^{(s)} \over \partial \varphi^{i}} (x) \, - \,
{\partial \over \partial x}
{\partial h^{(s)} \over \partial \varphi_{x}^{i}} (x) \, + \,
\dots $$
and

$${\delta^{2} H^{(s)} \over \delta \varphi^{i}(x) \,  
\delta \varphi^{j}(y)} \,\, = \,\,
{\delta^{2} H^{(s)} \over \delta \varphi^{j}(y) \,
\delta \varphi^{i}(x)} $$
for smooth functions
$h^{(s)} (\bm{\varphi},\bm{\varphi}_{x}, \dots)$.

 We have also

$${\delta^{2} H^{(s)} \over \delta \varphi^{i}(x) \,
\delta \varphi^{j}(y)} \,\, = \,\,
{\delta^{2} H^{(s)} \over \delta \varphi^{j}(y) \,
\delta \varphi^{i}(x)} \,\, = \,\,
\sum_{k\geq 0} A^{(s)k}_{ij}
(\bm{\varphi},\bm{\varphi}_{x}, \dots) \,
\delta^{(k)}(x-y) $$
for some local functions
$A^{(s)k}_{ij} (\bm{\varphi},\bm{\varphi}_{x}, \dots)$.

 Using the formulas

$$\delta^{(k)}(x-y) \, \nu (y-z) \,\, = \,\,
\delta^{(k)}(x-y) \, \nu (x-z) \, + \, \sum_{p=1}^{k}
C_{k}^{p} \, \delta^{(k-p)}(x-y) \, \delta^{(p-1)}(x-z) $$
we can write then

$$- \, {1 \over 2} \, \sum_{s=1}^{g} e_{s} \,
{\delta^{2} H^{(s)} \over \delta \varphi^{i}(x) \,
\delta \varphi^{j}(y)} \,
\int_{-\infty}^{+\infty} \nu (y-z) \,
h^{(s)} (\bm{\varphi},\bm{\varphi}_{z}, \dots) \, dz \,\, +$$

$$+ \,\, {1 \over 2} \, \sum_{s=1}^{g} e_{s} \,
{\delta^{2} H^{(s)} \over \delta \varphi^{j}(y) \,
\delta \varphi^{i}(x)} \,
\int_{-\infty}^{+\infty} \nu (x-z) \,
h^{(s)} (\bm{\varphi},\bm{\varphi}_{z}, \dots) \, dz \,\, =$$

$$- \, {1 \over 2} \, \sum_{s=1}^{g} e_{s} \,\,
\sum_{k\geq 1} A^{(s)k}_{ij}
(\bm{\varphi},\bm{\varphi}_{x}, \dots) \,\,
\sum_{p=1}^{k} C_{k}^{p} \, \left(
h^{(s)} (\bm{\varphi},\bm{\varphi}_{x}, \dots) \right)_{(p-1)x}
\, \delta^{(k-p)}(x-y) $$
which is a local expression.

 Now we have

$$\int_{-\infty}^{+\infty} \nu (y-z) \,
{\delta h^{(s)} (\bm{\varphi},\bm{\varphi}_{z}, \dots)
\over \delta \varphi^{i}(x)} \, dz \,\, = $$

$$= \,\, \int_{-\infty}^{+\infty} \nu (y-z) \, \left(
{\partial h^{(s)} \over \partial \varphi^{i}}(z) \,
\delta (z-x) \, + \,
{\partial h^{(s)} \over \partial \varphi_{z}^{i}}(z) \,
\delta^{\prime} (z-x) \, + \, \dots \right) \, dz \,\, = $$

$$= \,\, \sum_{p\geq 0} (-1)^{p} \, \left[ \nu (y-x) \,
{\partial h^{(s)} \over \partial \varphi^{i}_{px}}(x)
\right]_{px} \,\,\, = \,\,\, \nu (y-x) \,
{\delta H^{(s)} \over \delta \varphi^{i}(x)} \,\, + \,\,
{\rm (local \,\, part)} $$

 Also

$$\int_{-\infty}^{+\infty} \nu (x-z) \,
{\delta h^{(s)} (\bm{\varphi},\bm{\varphi}_{z}, \dots)
\over \delta \varphi^{j}(y)} \, dz \,\,\, =
\,\,\, \nu (x-y) \,
{\delta H^{(s)} \over \delta \varphi^{j}(y)} \,\, + \,\,
{\rm (local \,\, part)} $$

 We have finally

$$\left[ d \bm{\omega} \right]_{ij} (x,y) \,\, = $$
$$= \, - \, {1 \over 2} \, \sum_{s=1}^{g} e_{s} \,
{\delta H^{(s)} \over \delta \varphi^{j}(y)} \,
\nu (y-x) \,
{\delta H^{(s)} \over \delta \varphi^{i}(x)} \, + \,
{1 \over 2} \, \sum_{s=1}^{g} e_{s} \,
{\delta H^{(s)} \over \delta \varphi^{i}(x)} \,
\nu (x-y) \,
{\delta H^{(s)} \over \delta \varphi^{j}(y)} \,\, + \,\,
{\rm (local \,\, part)} \,\, =$$

$$= \,\, \sum_{s=1}^{g} e_{s} \,
{\delta H^{(s)} \over \delta \varphi^{i}(x)} \,
\nu (x-y) \,
{\delta H^{(s)} \over \delta \varphi^{j}(y)} \,\, + \,\,
{\rm (local \,\, part)} $$

{\hfill Lemma 5 is proved.}

\vspace{0.5cm}

 It's not difficult to prove also (using the analogous
statement for purely local symplectic structures) that
every closed 2-form (\ref{conssympform}) can be locally
represented as the external derivative of some 1-form
(\ref{wnonl1form}) on the space $\bm{\varphi}(x)$.

\vspace{0.5cm}

 We are going to give now the procedure of averaging of
1-forms (\ref{wnonl1form}) connected with the averaging of
the Symplectic structures (\ref{conssympform}). Namely,
we will assume now that the form $\Omega_{ij}(x,y)$ is  
represented as the external derivative of the form
(\ref{wnonl1form}). The corresponding procedure of averaging
of the form (\ref{wnonl1form}) should then give the weakly
nonlocal 1-form of "Hydrodynamic type" which is connected
with the form $\Omega^{av}_{\nu\mu} (X,Y)$ in the same way.

\vspace{0.5cm}

{\bf Definition 5.} {\it We call the form
$\omega_{\nu}[{\bf U}](X)$ on the space of functions
$U^{1}(X), \dots, U^{N}(X)$ the weakly nonlocal 1-form of
Hydrodynamic type if it has the form

\begin{equation}
\label{HT1form}
\omega_{\nu}[{\bf U}](X) \,\, = \,\, - {1 \over 2}
\sum_{s,p=1}^{M} \kappa_{sp} \,
{\partial f^{(s)} \over \partial U^{\nu}} ({\bf U}(X)) \,
\int_{-\infty}^{+\infty} \nu(X-Y) \, f^{(p)}({\bf U}(Y)) \,
dY
\end{equation}
for some functions $f^{(s)}({\bf U})$ and the quadratic form
$\kappa_{sp}$.
}

\vspace{0.5cm}

 It's not difficult to see that the form 
$\Omega_{\nu\mu}(X,Y)$ given by (\ref{HTconsform}) is
connected with (\ref{HT1form}) by the relation

\begin{equation}
\label{drelation}
\Omega_{\nu\mu} (X,Y) \,\, = \,\, 
\left[ d \bm{\omega} \right]_{\nu\mu} (X,Y)
\end{equation}

\vspace{0.5cm}

 As previously, we introduce the extended space of functions
$\bm{\varphi}(\bm{\theta},x)$ $2\pi$-periodic w.r.t. each   
$\theta^{\alpha}$. After the change of coordinate
$X = \epsilon \, x$ we can introduce the 1-form

$${\hat \omega}_{i} (\bm{\theta},X) \,\, = \,\,
c_{i} (\bm{\varphi}(\bm{\theta},X), \epsilon \,
\bm{\varphi}_{X}(\bm{\theta},X),\dots) \,\, - $$

\begin{equation}
\label{hat1form}
- \,\, {1 \over 2\epsilon} \, \sum_{s=1}^{g} e_{s} \,
{\delta {\hat H}^{(s)} \over \delta \varphi^{i}(\bm{\theta},X)}
\, \int_{-\infty}^{+\infty} \nu(X-Y) \,
h^{(s)} (\bm{\varphi}(\bm{\theta},Y), \epsilon \,
\bm{\varphi}_{Y}(\bm{\theta},Y),\dots) \, dY
\end{equation}
where ${\hat H}^{(s)} \, = \, \int_{-\infty}^{+\infty}
h^{(s)} (\bm{\varphi}(\bm{\theta},X), \dots) \, dX$.

 Easy to see that the relation
 
$$\Omega_{ij}(x,y) \,\, = \,\, [d \bm{\omega}]_{ij}(x,y) $$
gives

$${\hat \Omega}_{ij}(\bm{\theta},\bm{\theta}^{\prime},X,Y) \,\,
= \,\, [d {\hat{\bm{\omega}}}]_{ij}
(\bm{\theta},\bm{\theta}^{\prime},X,Y)$$
on the "extended" functional space.

 According to our previous approach we will investigate here the
main term of the restriction of the 1-form
${\hat \omega}_{i} [\bm{\varphi}] (\bm{\theta},X)$ on the
submanifolds ${\cal M}_{\epsilon}[\bm{\Psi}_{(1)}]$
(in coordinates $({\bf U}, \bm{\theta}_{0}^{*})$) in the weak
sense. Let us formulate here the corresponding theorem.

\vspace{0.5cm}

{\bf Theorem 6.}

{\it The restriction of the form
${\hat \omega}_{i} (\bm{\theta},X)$ to any submanifold
${\cal M}_{\epsilon}[\bm{\Psi}_{(1)}]$ in coordinates
$U^{\nu}(X)$, $\theta_{0}^{*\alpha}(X)$ can be written as
 
$$\bm{\omega}^{rest} \,\, = \,\, \int_{-\infty}^{+\infty}  
\omega^{1}_{\nu}(X) \, \delta U^{\nu}(X) \, dX \,\, + \,\,
\int_{-\infty}^{+\infty} \omega^{2}_{\alpha}(X) \,
\delta \theta_{0}^{*\alpha}(X) \, dX $$
where

I) The form $\omega^{1}_{\nu}(X)$ can be written as

$$\omega^{1}_{\nu}(X) \,\, = \,\,
- \, {1 \over \epsilon} \,
{\partial k^{\alpha} \over \partial U^{\nu}}(X) \, 
\int_{-\infty}^{+\infty} \nu(X-Y) \, I_{\alpha}(Y) \, dY
\,\, - $$

$$- \,\, {1 \over 2\epsilon} \, \sum_{s=1}^{g} e_{s} \,
{\partial \langle h^{(s)} \rangle \over \partial U^{\nu}}(X)
\, \int_{-\infty}^{+\infty} \nu(X-Y) \,
\langle h^{(s)} \rangle (Y) \, dY \,\,\,\,\, +
\,\,\,\,\, {o(1) \over \epsilon}$$
(summation over $\alpha = 1,\dots,m$) where

\begin{equation}
\label{formactvar}
I_{\alpha}({\bf U}) \,\, = \,\,
\langle c_{i} \, \varphi^{i}_{\theta^{\alpha}} \rangle \,  
+ \, {1 \over 2} \, \gamma^{\delta}_{\alpha}({\bf U}) \,  
\sum_{s=1}^{g} e_{s} \,
\left[ \langle h^{(s)} \, J^{(s)}_{\delta} \rangle \, - \,
\langle h^{(s)} \rangle \langle J^{(s)}_{\delta} \rangle
\right] \, - \, {1 \over 2} \, \sum_{s=1}^{g} e_{s} \,
\langle h^{(s)} \, T^{(s)}_{\alpha} \rangle
\end{equation}
and the values
$J^{(s)}_{\delta}(\bm{\varphi},\dots)$,
$\gamma^{\delta}_{\alpha}({\bf U})$ and
$T^{(s)}_{\alpha}(\bm{\varphi},\dots)$ are introduced in
(\ref{jsa}), (\ref{gammamatr}) and (\ref{tsa}).

II) The form $\omega^{2}_{\alpha}(X)$ has the order $O(1)$ for
$\epsilon \rightarrow 0$.
}

\vspace{0.5cm}

 Proof.

 We have

$$\omega^{1}_{\nu}(X) \, = \, \int \left( {1 \over \epsilon} \,
{\partial k^{\alpha} \over \partial U^{\nu}}(X) \, \nu (Z-X) \,
\varphi^{i}_{\theta^{\alpha}} (\bm{\theta},Z) \, + \,
\delta (Z-X) \, \Phi^{i}_{U^{\nu}}
(\bm{\theta} + \dots, {\bf U}(Z)) \right) \, \times $$  
$$\hspace{7cm} \times \, {\hat \omega}_{i} (\bm{\theta},Z) \,
{d^{m} \theta \over (2\pi)^{m}} \, dZ \,\, =$$

$$= \, - \, {1 \over \epsilon} \, \int
{\partial k^{\alpha} \over \partial U^{\nu}}(X) \, \nu (X-Z) \,
c_{i} (\bm{\varphi}(\bm{\theta},Z), \dots) \,
\varphi^{i}_{\theta^{\alpha}} (\bm{\theta},Z) \,
{d^{m} \theta \over (2\pi)^{m}} \, dZ \,\, +$$ 

$$+ \, {1 \over 2\epsilon^{2}} \int \sum_{s=1}^{g} e_{s} \,   
{\partial k^{\alpha} \over \partial U^{\nu}}(X) \, \nu (X-Z) \,
\varphi^{i}_{\theta^{\alpha}} (\bm{\theta},Z) \, \,
{\delta {\hat H}^{(s)} \over \delta \varphi^{i} (\bm{\theta},Z)}
\,\, \times \hspace{5cm} $$
$$\hspace{5cm} \times \,\,  \nu (Z-Y) \,
h^{(s)} (\bm{\varphi}(\bm{\theta},Y), \dots) \,
{d^{m} \theta \over (2\pi)^{m}} \, dY \, dZ \,\, -$$

$$- \, {1 \over 2\epsilon} \int \sum_{s=1}^{g} e_{s} \,
\Phi^{i}_{U^{\nu}} (\bm{\theta} + \dots, {\bf U}(X)) \,
{\delta {\hat H}^{(s)} \over \delta \varphi^{i} (\bm{\theta},X)}
\,\, \times \hspace{5cm}$$
$$\hspace{3cm} \times  \nu (X-Y) \,
h^{(s)} (\bm{\varphi}(\bm{\theta},Y), \dots) \,
{d^{m} \theta \over (2\pi)^{m}} \, dY \,\,\,\,\, + \,\,\,\,\,
O(1) \,\,\, = $$

\vspace{0.5cm}

$$= \,\, - \, {1 \over \epsilon} \, \int
{\partial k^{\alpha} \over \partial U^{\nu}}(X) \, \nu (X-Z) \,\,
\langle c_{i} (\bm{\varphi}(\bm{\theta},Z), \dots) \,\,
\varphi^{i}_{\theta^{\alpha}} (\bm{\theta},Z) \rangle \, dZ \,\,
+ $$

$$+ \,\, {1 \over 2\epsilon} \, \int \sum_{s=1}^{g} e_{s} \,   
{\partial k^{\alpha} \over \partial U^{\nu}}(X) \, \nu (X-Z) \,
W^{(s)}_{\theta^{\alpha}Z} (\bm{\theta},Z) \,\, \times
\hspace{5cm}$$
$$\hspace{5cm} \times  \,\, \nu (Z-Y) \,
h^{(s)} (\bm{\varphi}(\bm{\theta},Y), \dots) \,
{d^{m} \theta \over (2\pi)^{m}} \, dY \, dZ \,\, -$$

$$- \,\, {1 \over 2\epsilon} \, \int \sum_{s=1}^{g} e_{s} \,
{\partial k^{\alpha} \over \partial U^{\nu}}(X) \, \nu (X-Z) \,
T^{(s)}_{\alpha,Z} (\bm{\theta},Z) \,\, \times
\hspace{5cm}$$
$$\hspace{5cm} \times \,\, \nu (Z-Y) \,
h^{(s)} (\bm{\varphi}(\bm{\theta},Y), \dots) \,
{d^{m} \theta \over (2\pi)^{m}} \, dY \, dZ \,\, -$$

$$- \, {1 \over 2\epsilon} \int \sum_{s=1}^{g} e_{s} \,
\Phi^{i}_{U^{\nu}} (\bm{\theta} + \dots, {\bf U}(X)) \,
{\delta {\hat H}^{(s)} \over \delta \varphi^{i} (\bm{\theta},X)}
\,\, \times \hspace{5cm}$$
$$\hspace{3cm} \times \,\, \nu (X-Y) \,
h^{(s)} (\bm{\varphi}(\bm{\theta},Y), \dots) \,
{d^{m} \theta \over (2\pi)^{m}} \, dY \,\,\,\,\, + \,\,\,\,\,   
O(1) $$

 Here $\langle \dots \rangle$ means again the averaging on the 
family $\Lambda$ and the functions $W^{(s)}$, $T^{(s)}_{\alpha}$
are the same as in (\ref{WS}), (\ref{tsa}).

 As in the proof of the Theorem 4 we can omit here also
(by the same reason) the averaging with the values like
$W^{(s)}_{\theta^{\alpha}} (\bm{\theta},\pm\infty)$ in the main
order of $\epsilon$. We can write then

$$\omega^{1}_{\nu}(X) \,\, = \,\, - \, {1 \over \epsilon} \, \int
{\partial k^{\alpha} \over \partial U^{\nu}}(X) \,\, \nu (X-Y)
\,\, \langle c_{i} (\bm{\varphi}(\bm{\theta},Y), \dots) \,\,
\varphi^{i}_{\theta^{\alpha}} (\bm{\theta},Y) \rangle \, dY \,\,
+ $$

$$+ \,\, {1 \over 2\epsilon} \, \int \sum_{s=1}^{g} e_{s} \,
{\partial k^{\alpha} \over \partial U^{\nu}}(X) \,
W^{(s)}_{\theta^{\alpha}} (\bm{\theta},X) \,\, \nu (X-Y) \,\,   
h^{(s)} (\bm{\varphi}(\bm{\theta},Y), \dots) \,
{d^{m} \theta \over (2\pi)^{m}} \, dY \,\, - $$

$$- \,\, {1 \over 2\epsilon} \, \int \sum_{s=1}^{g} e_{s} \,    
{\partial k^{\alpha} \over \partial U^{\nu}}(X) \,\, \nu (X-Y)
\,\, W^{(s)}_{\theta^{\alpha}} (\bm{\theta},Y) \,
h^{(s)} (\bm{\varphi}(\bm{\theta},Y), \dots) \,
{d^{m} \theta \over (2\pi)^{m}} \, dY \,\, - $$

$$- \,\, {1 \over 2\epsilon} \, \int \sum_{s=1}^{g} e_{s} \,
{\partial k^{\alpha} \over \partial U^{\nu}}(X) \,
T^{(s)}_{\alpha} (\bm{\theta},X) \,\, \nu (X-Y) \,\,
h^{(s)} (\bm{\varphi}(\bm{\theta},Y), \dots) \,
{d^{m} \theta \over (2\pi)^{m}} \, dY \,\, + $$

$$+ \,\, {1 \over 2\epsilon} \, \int \sum_{s=1}^{g} e_{s} \,
{\partial k^{\alpha} \over \partial U^{\nu}}(X) \,\, \nu (X-Y)   
\,\, T^{(s)}_{\alpha} (\bm{\theta},Y) \,
h^{(s)} (\bm{\varphi}(\bm{\theta},Y), \dots) \,
{d^{m} \theta \over (2\pi)^{m}} \, dY \,\, - $$

$$- \,\, {1 \over 2\epsilon} \, \int \sum_{s=1}^{g} e_{s} \,
\Phi^{i}_{U^{\nu}} (\bm{\theta} + \dots, {\bf U}(X)) \,
{\delta {\hat H}^{(s)} \over \delta \varphi^{i} (\bm{\theta},X)}
\,\, \times \hspace{5cm}$$
$$\hspace{3cm} \times \,\, \nu (X-Y) \,
h^{(s)} (\bm{\varphi}(\bm{\theta},Y), \dots) \,
{d^{m} \theta \over (2\pi)^{m}} \, dY \,\,\,\,\, + \,\,\,\,\,
O(1) $$

 We can use now the same arguments is in the proof of the
Theorem 4 and make in the main order of $\epsilon$ the
independent integration w.r.t. $\theta$ of the rapidly
oscillating functions depending on $X$ and $Y$ before the
integration w.r.t. $Y$. We can omit then the second term of the
expression above in the main order. Using also the relations
(\ref{vjaver}) and (\ref{hav}) we get the statement (I) of the
theorem.

\vspace{0.5cm}

 II) We have

$$\omega^{2}_{\alpha} (X) \,\,\, = \,\,\, \int 
\varphi^{i}_{\theta^{\alpha}} (\bm{\theta},X) \,\,
{\hat {\omega}}_{i} (\bm{\theta},X) \,\,
{d^{m} \theta \over (2\pi)^{m}} \,\,\, = $$

$$= \,\,\, \int c_{i} (\bm{\varphi}(\bm{\theta},X), \dots) \,\, 
\varphi^{i}_{\theta^{\alpha}} (\bm{\theta},X) \,\,
{d^{m} \theta \over (2\pi)^{m}} \,\,\, - $$

$$- \, {1 \over 2} \, \int \sum_{s=1}^{g} e_{s}
\left( W^{(s)}_{\theta^{\alpha}X} (\bm{\theta},X) -
T^{(s)}_{\alpha,X} (\bm{\theta},X) \right) \,\, \nu (X-Y) \,\,
h^{(s)} (\bm{\varphi}(\bm{\theta},Y), \dots) \,
{d^{m} \theta \over (2\pi)^{m}} \, dY $$

 Using the identity

$$- {1 \over 2} \int \sum_{s=1}^{g} e_{s}
\left( W^{(s)}_{\theta^{\alpha}X} (\bm{\theta},X) -
T^{(s)}_{\alpha,X} (\bm{\theta},X) \right) \,\, \nu (X-Y) \,\,
h^{(s)} (\bm{\varphi}(\bm{\theta},Y), \dots) \,
{d^{m} \theta \over (2\pi)^{m}}  dY \, =$$

$$= - {1 \over 2} {\partial \over \partial X}
\int \sum_{s=1}^{g} e_{s}
\left( W^{(s)}_{\theta^{\alpha}} (\bm{\theta},X) -
T^{(s)}_{\alpha} (\bm{\theta},X) \right) \,\, \nu (X-Y) \,\,
h^{(s)} (\bm{\varphi}(\bm{\theta},Y), \dots) \,
{d^{m} \theta \over (2\pi)^{m}}  dY \, +$$ 

$$+ \,\, {1 \over 2} \, \int \sum_{s=1}^{g} e_{s}
\left( W^{(s)}_{\theta^{\alpha}} (\bm{\theta},X) -
T^{(s)}_{\alpha} (\bm{\theta},X) \right) \,
h^{(s)} (\bm{\varphi}(\bm{\theta},X), \dots) \,\,
{d^{m} \theta \over (2\pi)^{m}} $$
we easily get the part (II) of the theorem.

{\hfill Theorem 6 is proved.}

\vspace{0.5cm}
 
{\bf Definition 6.} {\it We call the 1-form

$$\omega^{av}_{\nu}(X) \,\, = \,\, - \,
{\partial k^{\alpha} \over \partial U^{\nu}}(X) \,
\int_{-\infty}^{+\infty} \nu(X-Y) \, I_{\alpha}(Y) \, dY
\,\, - $$

\begin{equation}
\label{av1form}
- \,\, {1 \over 2} \, \sum_{s=1}^{g} e_{s} \,
{\partial \langle h^{(s)} \rangle \over \partial U^{\nu}}(X)
\, \int_{-\infty}^{+\infty} \nu(X-Y) \,
\langle h^{(s)} \rangle (Y) \, dY
\end{equation}
where $I_{\alpha}({\bf U})$ are defined by the formula
(\ref{formactvar}) the averaging of the 1-form
(\ref{wnonl1form}) on the family of $m$-phase solutions of
(\ref{dynsyst}).
}

\vspace{0.5cm}

 As follows from our construction we have the relation

$$\Omega^{av}_{\nu\mu} (X,Y) \,\,\, = \,\,\,
[d \, \bm{\omega}^{av} ]_{\nu\mu} (X,Y) $$
for the forms (\ref{avsympstr}) and (\ref{av1form}).

 Using the remark (\ref{drelation}) it's not difficult
to prove also that the quantities (\ref{formactvar}) give the
action variables defined in (\ref{actvar}).

 We can see that the formulas (\ref{av1form}),
(\ref{formactvar}) give another procedure for the
averaging of 2-forms $\Omega_{ij}(x,y)$ represented in the
form of the external derivatives of weakly-nonlocal 1-forms 
$\omega_{i}(x)$.

 We can also write the formal Lagrangian formalism for the
Whitham equations in the form

$$\delta \,\, \int \int \left[ \omega^{av}_{\nu}(X) \,
U^{\nu}_{T}(X) \, - \, \langle h \rangle ({\bf U}) \right]
\,\, dX \, dT \,\,\, = \,\,\, 0 $$
or using (\ref{av1form})

$$\delta \,\, \int \int \left[ k^{\alpha}_{T}(X) \,
\nu (X-Y) \, I_{\alpha}(Y) \, + \right. \hspace{9cm}$$
\begin{equation}
\label{Lagrang}
\hspace{2cm} + \, {1 \over 2} \left.
\sum_{s=1}^{g} \, e_{s} \, \langle h^{(s)} \rangle_{T}(X) \,
\nu (X-Y) \, \langle h^{(s)} \rangle (Y) \, + \,
\langle h \rangle \right] \, dX \, dY \, dT \, = \, 0 
\end{equation}

\vspace{0.5cm}

  The work was supported by Grant of President of
Russian Federation (MK - 1375.2003.02), Russian Foundation
for Basic Research (grant RFBR 03-01-00368), and Russian
Science Foundation.

\end{document}